\documentclass[12pt]{report}
\usepackage[utf8]{inputenc}
\usepackage[T1]{fontenc}
\usepackage{graphicx}

\usepackage[square,compress,comma,numbers]{natbib}

\usepackage{sidecap}

\usepackage[raggedright]{titlesec}

\usepackage{amsmath,amsfonts,amssymb,bbm, color}
\usepackage{mathrsfs}
\usepackage[colorlinks]{hyperref}
\hypersetup{
	bookmarksnumbered,
	pdfstartview={FitH},
	citecolor={srednigreen},
	linkcolor={darkblue},
	urlcolor={darkblue},
	pdfpagemode={UseOutlines}}
\definecolor{darkgreen}{RGB}{20,100,20}
\definecolor{srednigreen}{RGB}{30,160,30}
\definecolor{darkblue}{RGB}{0,0,130}
\definecolor{darkred}{rgb}{.8,0,0}

\newcommand{\twh}{\textcolor{white}}

\usepackage{fancyhdr}
\renewcommand{\sectionmark}[1]{}
\fancypagestyle{intro}{\rhead{\it INTRODUCTION}}
\fancypagestyle{concl}{\rhead{\it CONCLUSIONS}}

\newcommand{\ket}[1]{|#1 \rangle}
\newcommand{\bra}[1]{\langle #1|}

\newcommand{\avg}[1]{\langle #1 \rangle}
\def\d{^{\dag}}
\def\s{^{\star}}
\newcommand{\Tr}[1]{\mathrm{Tr}(#1)}
\def\I{ \mathbbm{1} }
\renewcommand{\Re}{\text{Re}} 
\renewcommand{\Im}{\text{Im}}
\newcommand\defn{\mathrel{\overset{\makebox[0pt]{\mbox{\normalfont\tiny\sffamily def}}}{=}}}
\newcommand{\x}[1]{\mathrm{#1}}
\def\A{\cal{A}}
\newcommand{\pa}{{\mkern3mu\vphantom{\perp}\vrule depth 0pt\mkern2mu\vrule depth 0pt\mkern3mu}}

\def\mat{\boldsymbol}

\def\wiO{w_{\text{I}\Omega}}
\def\wiiO{w_{\text{II}\Omega}}
\def\wiX{w_{\text{I}\Xi}}
\def\wiiX{w_{\text{II}\Xi}}
\def\biO{\hat{b}_{\text{I}\Omega}}
\def\biiO{\hat{b}_{\text{II}\Omega}}
\def\biX{\hat{b}_{\text{I}\Xi}}
\def\biiX{\hat{b}_{\text{II}\Xi}}
\def\bLO{\hat{b}_{\Lambda\Omega}}

\def\wiOk{w_{\text{I}\Omega{\bf k_\perp}}}
\def\wiiOk{w_{\text{II}\Omega{\bf k_\perp}}}
\def\wiXl{w_{\text{I}\Xi{\bf l_\perp}}}

\def\biOk{\hat{b}_{\text{I}\Omega{\bf k_\perp}}}
\def\biiOk{\hat{b}_{\text{II}\Omega{\bf k_\perp}}}
\def\biXl{\hat{b}_{\text{I}\Xi{\bf l_\perp}}}
\def\biiXl{\hat{b}_{\text{II}\Xi{\bf l_\perp}}}

\newcommand{\ali}{\alpha_\text{I}}
\newcommand{\alii}{\alpha_\text{II}}
\newcommand{\bei}{\beta_\text{I}}
\newcommand{\beii}{\beta_\text{II}}

\newcommand{\sdvac}{\boldsymbol\sigma^{(d)}_\x{vac}}
\newcommand{\sdvel}{\left(\sigma^{(d)}_\x{vac}\right)}

\newcommand{\oIok}{(\psi_\text{I}, w_{\text{I}\Omega{\bf k_\perp}})}
\newcommand{\oIxl}{(\psi_\text{I}, w_{\text{I}\Xi{\bf l_\perp}})}
\newcommand{\oIIxl}{(\psi_\text{II}, w_{\text{II}\Xi{\bf l_\perp}})}
\newcommand{\oIIok}{(\psi_\text{II}, w_{\text{II}\Omega{\bf k_\perp}})}
\newcommand{\oIIxk}{(\psi_\text{II}, w_{\text{II}\Xi{\bf k_\perp}})}
\newcommand{\oIIomk}{(\psi_\text{II}, w_{\text{II}\Omega{\bf -k_\perp}})}
\newcommand{\oIIxmk}{(\psi_\text{II}, w_{\text{II}\Xi{\bf -k_\perp}})}
\newcommand{\olok}{(\psi_\Lambda, w_{\Lambda\Omega{\bf k_\perp}})}

\newcommand{\bi[1]}{\hat{b}_{\text{I}#1}}
\newcommand{\bii[1]}{\hat{b}_{\text{II}#1}}

\newcommand{\pe}{\perp}

\newcommand*\pusta{\newpage\null\thispagestyle{empty}\newpage}

\usepackage{bookmark}
\makeatletter
\usepackage{etoolbox}
\pretocmd{\tableofcontents}{%
  \if@openright\cleardoublepage\else\clearpage\fi
  \pdfbookmark[0]{\contentsname}{toc}%
}{}{}%
\makeatother

\title{
{\bf\Huge Effects of uniform acceleration on quantum states and vacuum entanglement in relativistic qunatum information}
\\\vspace{1em}
{\bf\Large Krzysztof Lorek}\\
{\includegraphics{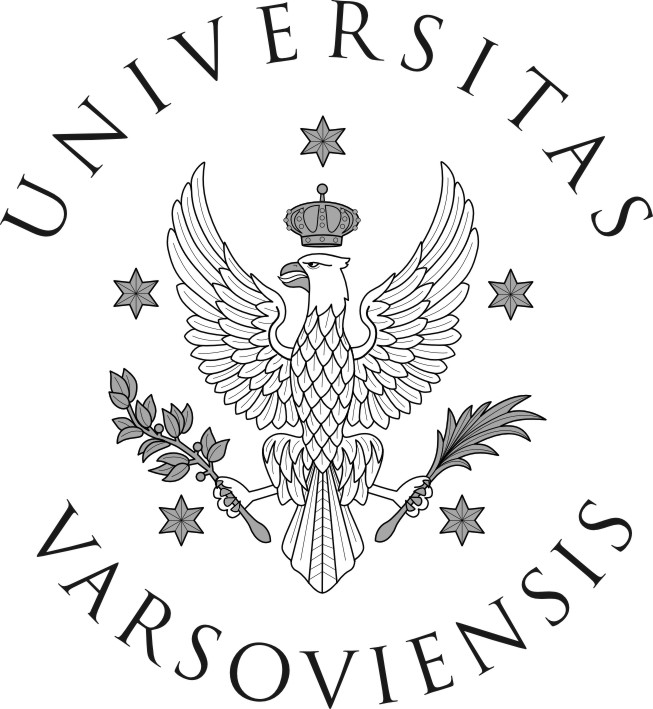}}\\
{\Large Faculty of Physics, University of Warsaw}\\
{\Large August 2017}
}
\date{}

\begin{document}


\thispagestyle{empty}

\begin{center}

\belowpdfbookmark{Title page}{}

{\bf\huge 
Effects of uniform acceleration \\ \vspace{0.00em}
on quantum states \\ \vspace{0.1em}
and vacuum entanglement  \\ \vspace{0.3em}
\mbox{in relativistic quantum information}}
\\ \vspace{3em}
\textbf{\textit{\Large Krzysztof Lorek}}\\ \vspace{3em}
{\includegraphics{logo_lac.jpg}}\\ \vspace{3em}
{\Large Doctoral dissertation under the supervision of}\\ \vspace{0.5em}
\textbf{\Large dr hab. Andrzej Dragan}\\ \vspace{0.5em}
{\Large Institute of Theoretical Physics} \\ \vspace{0.5em}
{\Large Faculty of Physics, University of Warsaw}\\ \vspace{1.9em}
{Warsaw, December 2017}

\end{center}

\pagebreak


\pusta
\thispagestyle{empty}

\begin{flushright}
\begin{large}

$\,$
\vspace{1cm}

To my high school physics teacher, \\
{\it Wojciech Lech}, \\
who taught me to learn.

\vspace{7cm}

Mojemu nauczycielowi fizyki z liceum, \\
{\it Wojciechowi Lechowi}, \\
który nauczył mnie uczyć się.

\end{large}
\end{flushright}


\chapter*{Acknowledgements}
\thispagestyle{empty}

First of all, I would like to thank my supervisor Andrzej Dragan for allowing me to do physics within his group in Poland, for these four years of guidance and countless helpful suggestions. Also, I would like to thank him for letting me present my research at many international conferences, and for helping me to substantially improve this thesis.

Furthermore, I would like to thank my parents and brother for their unconditional love and support, and for keeping up with my subsequent ideas. They provided me with everything I needed to pursue them and gave me freedom to do so.

None of this would have been possible without my exceptional high school teacher Wojciech Lech. During our countless hours with a piece of chalk and a blackboard or with a soldering iron and a mess of wires and components, he not only sparked my interest in science but shaped me as a person in~many ways. He not only taught me physics, but taught me how to learn physics (and other things). Whenever I walk out of his classroom, even now, I am inspired to pursue new paths and to aim high. Let me also acknowledge other teachers from STSG high school in Świnoujście, Wiesław Kaczanowski and Barbara Sadza, whom I learnt much from.

Moreover, I am grateful to all the other people who supported me at any point during the doctoral studies. This includes Ania, Arek, Kamil, Krzysiek, Mateusz, Fr. Rafał and many others. Thank you! Also, I~am grateful to dr~Anna Kaczorowska for inspiring me in a time when I did not expect it.

I would like to acknowledge Susan who answered my countless questions about her mother tongue, which I attempted to write this thesis in. Also, I~am grateful to Mateusz for useful comments about the Introduction.

Finally, I thank the Lord for being my shepherd and leading me through green pastures and through valleys of the shadow of death. 

\begin{flushright} {\it Krzysztof Lorek} \end{flushright}

\chapter*{Podziękowania}
\thispagestyle{empty}

Przede wszystkim chciałbym podziękować mojemu promotorowi, Andrzejowi Draganowi, za umożliwienie mi zajmowania się fizyką w jego grupie, w Polsce, za te cztery lata opieki i wiele cennych sugestii. Chciałbym mu również podziękować za to, że mogłem zaprezentować wyniki moich badań na międzynarodowych konferencjach i za ogromną pomoc w udoskonaleniu tej rozprawy.

Następnie chciałbym podziękować moim rodzicom i bratu za ich bezwarunkową miłość i wsparcie i za to, że nadążali za moimi kolejnymi pomysłami. Oni zapewnili mi wszystko, czego potrzebowałem by je realizować i dali mi wolność, bym mógł to robić.

Żadna z tych rzeczy nie byłaby możliwa bez mojego wyjątkowego nauczyciela z liceum, Wojciecha Lecha. Podczas naszych niezliczonych godzin przy tablicy z kawałkiem kredy, lub z lutownicą przy kłębowisku kabli i elektroniki, nie tylko zainteresował mnie fizyką, ale ukształtował wiele aspektów mojej osoby. Nie tylko nauczył mnie fizyki, ale nauczył mnie jak uczyć się fizyki (i innych rzeczy). Kiedy wychodzę z jego sali, nawet teraz, jestem zainspirowany do wędrowania nowymi ścieżkami i mierzenia wysoko. Chciałbym też podziękować innym nauczycielom z liceum STSG w Świnoujściu, Wiesławowi Kaczanowskiemu i Barbarze Sadzy, od których wiele się nauczyłem.

Jestem też wdzięczny wszystkim innym osobom, które wsparły mnie w różnych momentach podczas studiów doktoranckich, w tym: Ani, Arkowi, Kamilowi, Krzyśkowi, Mateuszowi i ks. Rafałowi. Dziękuję! Jestem też wdzięczny dr Annie Kaczorowskiej za inspirację, gdy się jej nie spodziewałem.

Dziękuję Susan, która odpowiadała na moje niezliczone pytania na temat jej języka ojczystego, w którym próbowałem napisać tą rozprawę. Jestem też wdzięczny Mateuszowi za cenne komentarze dotyczące Wstępu.

Wreszcie dziękuję również Bogu za bycie moim pasterzem i prowadzenie mnie przez zielone pastwiska i przez ciemne doliny.

\begin{flushright} {\it Krzysztof Lorek} \end{flushright}

\chapter*{Abstract}

\noindent Relativistic quantum information is a field at the intersection of general relativity and quantum field theory. It investigates quantum phenomena, in particular those related to quantum information theory, in the presence of gravity, which is treated as a classical spacetime. Effects that are of fundamental importance for relativistic quantum information include the Unruh effect, regarding the presence of particles in quantum vacuum, when observed from an accelerated frame of reference; the Hawking radiation, which is the radiation emitted by a black hole; and the dynamical Casimir effect, which concerns the production of particles in a cavity with a moving boundary. The field is rapidly developing as new effects are predicted, and first empirical tests are performed, such as one of the dynamical Casimir effect.

In this thesis we focus on the influence of uniform acceleration on quantum states and their properties, as well as on quantum entanglement contained in the vacuum. 

First, we analyze a quantum clock measuring time in terms of the decay of an unstable particle. We compare the rate of ticking of a stationary clock to the one of a uniformly accelerating clock. We discover that there exists a deviation from what the discrepancy between those rates is predicted to be by special relativity. This is caused by the fact that the accelerating clock is surrounded by particles arising due to the Unruh effect.

It is known that quantum states have different forms when observed from an inertial and a uniformly accelerated frame. In a further part of this thesis we focus on the family of two-mode Gaussian states of two localized modes of a given quantum field. A general framework is developed for investigating the properties of such states, when observed by two uniformly accelerated observers, having access to another pair of localized modes. In particular, we focus on the vacuum state and examine the amount of quantum entanglement perceived by the observers. The framework is developed for three cases: the quantum scalar field in $1+1$ and $3+1$ dimensions and the Dirac spinor field in $1+1$ dimensions. We find that in general more entanglement is seen when the accelerations of the observers increase and when the size of the modes of the observers and their central frequencies decrease. Also, for the $1+1$-dimensional scalar field we plot the dependence of the amount of entanglement on the separation of the observers and their accelerations independently and discover cases of sudden death of entanglement.

Furthermore, we extend our attention also to tripartite entanglement and investigate its extraction from the quantum vacuum by three particle detectors in a cavity, interacting with a quantum field for a finite time. Detailed maps of amounts of tripartite and bipartite entanglement extracted by the detectors, are produced and compared between two types of boundary conditions. It is found vacuum entanglement is most easily extracted in the case of periodic boundary conditions and it is easier to extract tripartite than bipartite entanglement.

\chapter*{Streszczenie}

Relatywistyczna informacja kwantowa jest dziedziną fizyki z pogranicza ogólnej teorii względności i kwantowej teorii pola. Bada ona przebieg zjawisk kwantowych, w szczególności tych związanych z teorią informacji kwantowej, w obecności pól grawitacyjnych, opisanych jako klasyczna czasoprzestrzeń. Najważniejsze efekty w relatywistycznej informacji kwantowej to efekt Unruha, dotyczący obecności cząstek w kwantowej próżni, gdy obserwowana jest ona z przyspieszonego układu odniesienia; promieniowanie Hawkinga, tj. promieniowanie emitowane przez czarną dziurę; oraz dynamiczny efekt Casimira, który dotyczy produkcji cząstek we wnęce rezonansowej z ruchomą ścianką. Relatywistyczna informacja kwantowa jest szybko rozwijającą się dziedziną, w ramach której formułowane są nowe teoretyczne przewidywania i wykonywane są pierwsze eksperymenty, jak np. weryfikacja dynamicznego efektu Casimira.

Tematem tej rozprawy jest wpływ jednostajnego przyspieszenia na stany kwantowe i ich właściwości, jak również splątanie kwantowe obecne w próżni.



W pierwszej kolejności rozważamy zegar kwantowy, mierzący czas za pomocą czasu połowicznego rozpadu nietrwałej cząstki. Porównujemy tempo tykania takiego zegara, gdy jest on w spoczynku, z sytuacją gdy przyspiesza on jednostajnie. Otrzymana różnica temp odbiega od przewidywań szczególnej teorii względności. Jest to spowodowane faktem, że przyspieszający zegar otaczają cząstki obecne z uwagi na efekt Unruha.


Stany kwantowe mają różne formy, gdy są obserwowane z inercjalnego lub z jednostajnie przyspieszonego układu odniesienia. W dalszej części rozprawy zajmujemy się dwumodowymi stanami gaussowskimi dwóch zlokalizowanych modów pola kwantowego. Opracowujemy ogólną metodę badania właściwości takich stanów, gdy obserwowane są przez dwóch jednostajnie przyspieszających obserwatorów, mających dostęp do kolejnej pary zlokalizowanych modów. W szczególności koncentrujemy się na stanie próżni i badamy ilość splątania kwantowego rejestrowanego przez obserwatorów. Metoda jest opracowana dla trzech przypadków: kwantowe pole skalarne w $1+1$ i $3+1$ wymiarach, oraz pole spinorowe Diraca w $1+1$ wymiarach. Wyniki wykazują, że obserwatorzy rejestrują więcej splątania, gdy ich przyspieszenia rosną, oraz gdy rozmiary modów i ich częstotliwości środkowe maleją. Ponadto dla $1+1$-wymiarowego pola skalarnego wykreślamy zależność ilości splątania od odległości między obserwatorami i od ich przyspieszeń i obserwujemy przypadki nagłego zaniku splątania.


Następnie badamy ekstrakcję trzyciałowego splątania z próżni kwantowej, przez trzy detektory cząstek we wnęce rezonansowej, oddziałujące z polem kwantowym przez skończony czas. Otrzymujemy szczegółowe diagramy przedstawiające ilość wydobytego trzyciałowego i dwuciałowego splątania i porównujemy je dla dwóch typów warunków brzegowych. Wyniki wykazują, że splątanie najłatwiej jest wydobywać przy periodycznych warunkach brzegowych, oraz że łatwiej jest wydobywać trzyciałowe splątanie niż dwuciałowe splątanie.

\tableofcontents 

\pagebreak
\thispagestyle{plain}

{\large\bf This thesis is based on the following publications:}
\vspace{1em}

\begin{enumerate}
\item Krzysztof Lorek, Daniel P\k{e}cak, Eric G. Brown and Andrzej Dragan,
\\ \textit{Extraction of genuine tripartite entanglement from the vacuum,} \cite{lor_trip}
\\ \href{https://link.aps.org/doi/10.1103/PhysRevA.90.032316}{Phys. Rev. A {\bf 90}, 032316 (2014)},

\item Krzysztof Lorek, Jorma Louko and Andrzej Dragan,
\\ \textit{Ideal clocks - a convenient fiction,} \cite{lor_clk}
\\ \href{http://stacks.iop.org/0264-9381/32/i=17/a=175003}{Class. Quantum Grav. {\bf 32}, 175003 (2015)},

\item Mehdi Ahmadi, Krzysztof Lorek, Agata Chęcińska, 
\\ Alexander R. H. Smith, Robert B. Mann and Andrzej Dragan
\\ \textit{Effect of relativistic acceleration on localized two-mode Gaussian quantum states,} \cite{lor_1d}
\\ \href{https://link.aps.org/doi/10.1103/PhysRevD.93.124031}{Phys. Rev. {\bf D} 93, 124031 (2016)},

\item Benedikt Richter, Krzysztof Lorek, Andrzej Dragan and Yasser Omar
\\ \textit{Effect of acceleration on localized fermionic Gaussian states: from vacuum entanglement to maximally entangled states,} \cite{lor_fer}
\\ \href{https://link.aps.org/doi/10.1103/PhysRevD.95.076004}{Phys. Rev. {\bf D} 95, 076004 (2017)},

\item Krzysztof Lorek, Piotr Grochowski, Mehdi Ahmadi and Andrzej Dragan
\\ \textit{Effect of relativistic acceleration on localized two-mode Gaussian quantum states - the $3+1$-dimensional case,} \cite{lor_3d}
\\ (in preparation).
\end{enumerate}

\chapter*{Introduction}
\addcontentsline{toc}{part}{Introduction}


The 20th century was a time of great advances in theoretical physics. At the beginning of the century two theories were developed, that entirely changed the way Nature was thought of. In the 1900s the first quantum models of blackbody radiation~\cite{intro_qm_bb} and of the photoelectric effect~\cite{intro_qm_photo} paved the way towards the theory of quantum mechanics~\cite{dirac_principles, shankar}, which provides the description of phenomena at atomic and subatomic scales. Simultaneously, the theory of special relativity was formulated~\cite{intro_sr_einst}, which constitutes a correct description of relations between inertial frames of reference, moving at any subluminal velocity. It replaced the previously used, so-called Galilean relativity, which turned out to be merely an approximation of special relativity for speeds much lower than the speed of light. Ten years later general relativity~\cite{intro_gr_einst} was formulated, which explains gravity in terms of the geometry of spacetime. In the 1920s and 1930s attempts were made at combining quantum mechanics and special relativity together, which led to the discovery of the Klein-Gordon equation~\cite{intro_qft_klein, intro_qft_gordon} and the Dirac equation~\cite{intro_qft_dirac1, intro_qft_dirac2}. These were the beginnings of quantum field theory~\cite{peskin} which describes Nature at all scales and in any inertial frame of reference, ultimately in terms of particle collisions and decays. Within this framework, in the 1960s the Standard Model of particle physics was developed~\cite{intro_qft_sm1,intro_qft_sm2,intro_qft_sm3}, including three out of four known fundamental interactions, i.e. the electromagnetic, and the strong and weak nuclear interactions, as well as the Higgs mechanism predicted in~\cite{intro_qft_higgs1,intro_qft_higgs2,intro_qft_higgs3}. In order to construct the ultimate theory of Nature, the remaining task was to incorporate gravity into the picture. While general relativity extended the scope of special relativity to include non-inertial frames of reference, it is a classical theory, treating the gravitational interaction fundametally differently from how quantum field theory treats the other forces of Nature. The theory of general relativity and quantum field theory constitute the current state of human knowledge about physics.

In 1970s improtant predictions were made within the so-called quantum field theory in curved spacetime~\cite{birrell_davies}. This framework applies in situations whereby quantum field theory and general relativity have to be used simultaneously, but gravity does not need to be treated as a quantum field. Gravitational forces are treated as a classical curved background spacetime, not interacting with the quantum fields inhabiting it. This framework may be safely applied unless very extreme conditions are considered, such as e.g. during the first instances after the Big Bang, or at very short distances away from a black hole singularity. 

In~\cite{intro_hawking} the existence of the Hawking radiation was deduced. A scattering of plane wave modes off a collapsing star was considered. Modes were assumed to incide from an infinite distance away, and the final state of modes outgoing to infinity was examined. Even if the incoming modes contained no particles, the outgoing ones were found to be in a thermal state. 
The effect leads to the evaporation of black holes, which poses interesting fundamental questions, such as the black hole information paradox.

In~\cite{unruh_effect} the Unruh effect was predicted. In quantum field theory particles are merely excitations of a quantum field. Let us consider a quantum state in an inertial frame of reference. A change of the frame of reference from the inertial to a non-inertial one, results in such a change of the state, that the average particle number on it, becomes different. The Unruh effect concerns a situation, whereby one observer is stationary in quantum vacuum, and hence detects no particles. It is predicted that another observer who accelerates through the same location, may see a thermal state. This effect has not been detected yet because it is predicted to be very weak at accelerations accessible experimentally.

\thispagestyle{intro}

Another important prediction of quantum field theory in curved spacetime, was the dynamical Casimir effect~\cite{intro_dcasimir}. If a quantum field is subject to a moving boundary condition, such as e.g. an oscillating reflecting mirror, the presence of the moving boundary may lead to particle creation.

These are the foundations of the field of relativistic quantum information, which investigates the implications of quantum field theory in curved spacetime, in particular those related to quantum information theory~\cite{qi_horodecki_book, qi_nc}. Topics that have been examined in relativistic quantum information since its birth, include the generation of quantum entanglement in cosmology~\cite{intro_cosmo_ent}, extraction of quantum entanglement from the vacuum state of a quantum field~\cite{reznik1,reznik2}, the relativity of classicality of a quantum state~\cite{lor_witness}, the relativistic aspects of quantum metrology~\cite{intro_rqmetrology} and creating quantum gates with motion~\cite{intro_gates}.

Let us now discuss the empirical verification of the implications of quantum field theory in curved spacetime. First quantum mechanical experiments in the presence of gravity are already performed. In 2001 gravitational bound states of neutrons were observed~\cite{valery}. Also, the dynamical Casimir effect was empirically confirmed in 2011~\cite{intro_dcasimir_conf}. It was realised using a superconducting system, in an analogue experiment. This means that the investigated system was engineered such that its behaviour was governed by the same equations as the setup wherein the effect was predicted. In 2014 and 2016 Hawking radiation was observed in an analogue experiment using a Bose-Einstein condesate~\cite{Steinhauer1, Steinhauer2}.

The research in relativistic quantum information is interesing from the theoretical perspective, since it provides many insights into the properties of quantum states and quantum information protocols under the influence of gravity, without the need of using a full theory of quantized gravity, which is not known as of today. Also, its predictions are becoming testable and may be applicable to the technology of the future. 

In this thesis we investigate the topics in relativistic quantum information, concerning uniformly accelerated frames of reference and the entanglement of the vacuum.

\thispagestyle{intro}

First, we investigate a consequence of the presence of the particles arising due to the Unruh effect. Two quantum clocks, measuring time in terms of a particle decay rate, are considered. One is stationary in an inertial frame of reference and the other one uniformly accelerates with respect to it. The clocks differ only in the fact that the accelerating one is surrounded by particles appearing because of the Unruh effect. It is found that there exists a deviation from what the difference between the readouts of the clocks is predicted to be by special relativity. For a clock travelling along a general accelerated trajectory, it is impossible to correct this deviation without knowing the full trajectory. The Unruh effect proves to fundmentally alter the way time is measured~\cite{lor_clk}. This topic was previously investigated using a different clock model~\cite{clk_andrzej}, and brought similar conclusions. Both models are simplified, and recently a more realistic setup was considered \cite{clk_roberto}.

Further investigation of the consequences of the Unruh effect concerns the question how a general two-mode Gaussian state is perceived by uniformly accelerating observers. Gaussian states are a family of states that includes in particular coherent, squeezed and thermal states. A Gaussian state of two modes of a quantum field, that are stationary in an inertial frame, is considered to be observed by two accelerating observers having access to two accelerated modes. Mathematically, the transformation of the state of the stationary modes into the state of the modes of the observers, can be expressed as a quantum channel. This framework can produce an output state of the channel for any desired input Gaussian state. Previously, this was investigated in~\cite{channel_andrzej} for the case of one accelerated and one inertial observer and considering only a particular input state, i.e. a squeezed state. 
Our work is performed for a scalar field in $1+1$ dimensions~\cite{lor_1d} and $3+1$ dimensions~\cite{lor_3d}, and for a Dirac spinor field in $1+1$ dimensions~\cite{lor_fer}. The expressions for the channel are derived and in particular they are applied to the input vacuum state, to yield insights about the vacuum entanglement. The sudden death of entanglement is seen in certain regimes of parameters.

The structure of the vacuum entanglement proves to be rich and interesting. In the final part of the thesis, not only bipartite but also tripartite entanglement of the vacuum is considered. Bipartite entanglement is quantum entanglement across two degrees of freedom, while tripartite entanglement is quantum entanglement across three degrees of freedom of a given system. The entanglement is probed by three particle detectors, which are modelled by harmonic oscillators, interacting with a field confined in a cavity. From the state of the detectors after the interaction, one can infer that both bipartite and tripartite entanglement exist in the vacuum, and furthermore, the tripartite entanglement seems to be easier to extract~\cite{lor_trip}.

\thispagestyle{intro}

The thesis is structured in the following way: In Part \ref{part1} the necessary theory is introduced. In Chapter \ref{chqm} we discuss elementary notions of quantum mechanics and quantum information, including density matrices, the description of composite systems and quantum entanglement. In Chapter \ref{chgqm} 
we restrict our attention to a less general, but still a very broad family of states, i.e. Gaussian states, and express them using the Gaussian quantum mechanics formalism. This formalism substantially simplifies many calculations. In Chapter~\ref{chgr} we introduce the Minkowski and Rindler coordinate systems which are used throughout the thesis. In Chapter~\ref{chqft} we discuss the description of a quantum scalar field in these coordinate systems, and the Bogolyubov transformation, which allows to transform a given state between various frames of reference. This part of the thesis is concluded with Chapter~\ref{chrqi} in which the Unruh effect is derived and the existence of vacuum entanglement is demonstrated.

In Part \ref{part2} one of the implications of the Unruh effect is discussed. It is the alteration of the reading of a uniformly accelerating clock because of the presence of the Unruh thermal bath. The calculation is performed for an inertial clock, and then repeated analogously for an accelerated clock. The results are compared and the implications are discussed.

In Part \ref{part3} we move on to the computation of the quantum channel, that transforms any Gaussian state of two inertial modes of a quantum field to the state perceived by two accelerating observers having access to two accelerated modes. In Chapter \ref{chb1d} the calculation is performed, wherein the field is a scalar $1+1$-dimensional field. Further, in Chapter \ref{chb3d} the computation is repeated for a scalar field in $3+1$ dimensions. Finally, in Chapter \ref{chf1d} also a $1+1$-dimensional Dirac spinor field is considered. For these three cases we calculate the amount of bipartite entanglement that is extracted by the observers from the vacuum.

In Part \ref{part4} we discuss the work on three particle detectors, extracting bipartite and tripartite entanglement from the vacuum. The calculation is presented, together with the numerical results showing the amounts of extracted entanglement of each kind for two types of boundary conditions.

Finally, the thesis is concluded and further work is discussed, that could be undertaken to extend this research.

\vspace{2em}

\thispagestyle{intro}

\noindent Throughout the thesis we use the following conventions and notation:
\begin{itemize}
\item natural units are used with $c=\hbar=1$;

\item quantum operators are denoted with hats, e.g. $\biO$;

\item creation and annihilation operators and modes related to a free field are denoted with lowercase letters, e.g. $u_k$;

\item creation and annihilation operators and modes related to a cavity field are denoted with capital letters, e.g. $U_n$;

\item Schr\"odinger picture is used;

\item the chosen metric signature is $(+,-)$ in $1+1$ dimensions and $(+,-,-,-)$ in $3+1$ dimensions.

\end{itemize}



\part{Preliminaries} \label{part1}

In the first part of the thesis, we present the theoretical foundations for the further chapters. In Chapter \ref{chqm} we recall basics of quantum mechanics and quantum information theory, i.e. the notion of density matrices, the description of composite systems and quantum entanglement. In Chapter \ref{chgqm} we focus on the family of Gaussian states and formulate their description in terms of the formalism of Gaussian quantum mechanics, which is extensively used in the further parts of the thesis. In Chapter \ref{chgr} we discuss the Minkowski and Rindler coordinate systems. Chapter \ref{chqft} is devoted to the description of a quantum scalar field in the aforementioned coordinate systems. Also, the Bogolyubov transformation is introduced. Finally, in Chapter \ref{chrqi} we derive the Unruh effect and discuss the entanglement contained in the vacuum.


\chapter{Elements of quantum mechanics and quantum information theory} \label{chqm}

In this chapter we discuss topics in quantum mechanics, that are used throughout the thesis. First, in Sec. \ref{sec_qm_density} the formalism of density matrices, with their evolution is briefly outlined, mainly for comparison with the covariance matrix formalism introduced in Chapter \ref{chgqm}. Further, in Sec. \ref{sec_qm_composite} we discuss states of composite systems. Later we move on to quantum entanglement, beginning with bipartite entanglement in Sec. \ref{sec_qm_bipartite} and generalizing to multipartite entanglement in Sec. \ref{sec_qm_multi}. We finish with Sec. \ref{sec_neg} by introducing a particularly useful way of quantifying the amount of entanglement - the logarithmic negativity.

Let us remind the Reader that throughout the thesis the Schr\"odinger picture is used.

\section{Density matrices}       \label{sec_qm_density}

Any quantum state, whether pure or mixed, can be described with a density matrix $\rho$~\cite{shankar}. The density matrix satisfies the following properties:
\begin{equation}      \label{qm_density_prop}
\begin{gathered}
\rho\d = \rho, \\
\Tr{\rho} = 1, \\
\rho \geq 0.
\end{gathered}
\end{equation}
Suppose we express $\rho$ in an orthonormal, discrete basis of states $\{\ket{\psi_n}\}$. The diagonal elements are interpreted as probabilities of the system being in a given basis state:
\begin{equation}
P(\ket{\psi_n}) = \bra{\psi_n} \,\rho\, \ket{\psi_n}.
\end{equation}
Out of these probabilities one can construct probability of the system being in any state.

The average (or expected) value of a Hermitian operator $\hat{O}$ on a state $\rho$ can be calculated as:
\begin{equation}
\avg{\hat{O}} = \Tr{\rho\, \hat{O}}.
\end{equation}

The time evolution of a density matrix starting at $t=0$ is governed by the equation:
\begin{equation}  \label{qm_evo}
\rho(t) = \hat{U}(t) \,\rho\, \hat{U}\d(t) \defn
e^{-i\hat{H}t} \,\rho\, e^{i\hat{H}t},
\end{equation}
where $\hat{H}$ is the Hamiltonian governing the evolution of the state. We have also defined $\hat{U}$, called the time shift operator. 
From the definition of the time shift operator, one may see that it is unitary:
\begin{equation}
\hat{U}\d\hat{U} = \I.
\end{equation}
The time evolution described with a unitary operator is fully reversible, and if $\hat{U}(t)$ evolves a state by $t$ forward in time, then $\hat{U}\d(t)$ evolves it by $t$ backwards.

\section{Composite systems}        \label{sec_qm_composite}

Suppose we have two disjoint systems - system $A$ with an orthonormal basis $\{\ket{\psi_n}\}$ and system $B$ with an orthonormal basis $\{\ket{\phi_m}\}$. Then the two systems together constitute the composite system of $A$ and $B$ with an orthonormal basis $\{\ket{\psi_n} \otimes \ket{\phi_m}\}$~\cite{qi_nc}. Let $\rho_{AB}$ be the density matrix of the composite system. In order to consider the subsystem $A$ only, one needs to calculate the {\it reduced density matrix} of the subsystem $A$, discarding all the information about the subsystem $B$. This operation is called {\it partial trace} over the basis of the subsystem $B$, also colloquially referred to as ``tracing~out'' the subsystem $B$. It is performed as follows:
\begin{equation} \label{qm_pTr}
\rho_{A} = \text{Tr}_B (\rho_{AB}) \defn
\sum\limits_m \bra{\phi_m} \,\rho_{AB}\, \ket{\phi_m}.
\end{equation}
The state $\rho_A$ is called the {\it reduced state} of the system $A$.

In the basis $\{\ket{\psi_n}\}$, the matrix $\rho_{A}$ has dimensions $n \times n$, and in the basis $\{\ket{\phi_m}\}$, the matrix $\rho_{B}$ has dimensions $m \times m$. Likewise, in the basis $\{\ket{\psi_n} \otimes \ket{\phi_m}\}$, the matrix $\rho_{AB}$ has dimensions $nm \times nm$.

A general state consisting of two subsystems $A$ and $B$ may be written as:
\begin{equation}   \label{qm_rhoab_decomp}
\rho_{AB} = \sum\limits_{nmrs} p_{nmrs} \,\ket{\psi_n}_A\bra{\psi_m} \otimes \ket{\phi_r}_B\bra{\phi_s},
\end{equation}
where the basis states are defined as above, and the coefficients $p_{nmrs}$ are chosen such that \eqref{qm_rhoab_decomp} satisfies the properties \eqref{qm_density_prop}. The partial trace of $\rho_{AB}$ with respect to the subsystem $A$ is defined with the indices $n$ and $m$ interchanged:
\begin{equation}
(\rho_{AB})^{\Gamma_A} \defn \sum\limits_{nmrs} p_{mnrs} \,\ket{\psi_n}_A\bra{\psi_m} \otimes \ket{\phi_r}_B\bra{\phi_s}.
\end{equation}
This operation will be used in the definition of the logarithmic negativity in Sec.~\ref{sec_neg}.

The discussion in this section can be generalized to a higher number of subsystems.

\section{Bipartite entanglement}         \label{sec_qm_bipartite}


A quantum composite state $\rho_{AB}$ is {\it separable}, if it can be written as a mixture of tensor products of certain states of the subsystem $A$ and of the subsystem~$B$~\cite{werner_sep}:
\begin{equation} \label{qmsep}
\rho_{AB} =\sum\limits_i p_i \,\rho_{Ai} \otimes \rho_{Bi},
\end{equation}
where $\rho_{Ai}$ are reduced density matrices of the system $A$, $\rho_{Bi}$ are reduced density matrices of the system $B$, and $p_i$ are nonnegative probabilities that add up to unity.
 
If the state of the system $AB$ is separable, it is fully characterized by its reduced state on $A$ and its reduced state on $B$. Also, a measurement performed on one subsystem, does not provide any information on the outcomes of any future measurements performed on the other subsystem.
If the separable state of the system $AB$ is pure, then its reduced state on $A$ and its reduced state on $B$ are also pure.
 
The basic criterion for a composite state $\rho_{AB}$ to be separable is the Peres-Horodecki (PPT) criterion~\cite{peres_ppt,horodecki_ppt}. It states that if $\rho_{AB}$ is separable, then $(\rho_{AB})^{\Gamma_A}\geq 0$. For systems with dimensions $2\times 2$ and $2\times 3$, the condition $(\rho_{AB})^{\Gamma_A}\geq 0$ is necessary and sufficient for $\rho_{AB}$ to be separable.

If the state $\rho_{AB}$ is not separable, we say that it is {\it entangled}~\cite{horodecki_entanglement}. There is a type of entanglement called \textit{bound entanglement} that may exist in the state despite the PPT criterion being satisfied for systems with dimensions higher than $2\times 3$~\cite{horodecki_bound}. 

The entanglement discussed here is also called {\it bipartite} entanglement, because the entanglement exists across a bipartition of the composite system. If the state of the system $AB$ is entangled, its reduced state on $A$ and its reduced state on $B$ do not suffice to characterize it. There is also information in the quantum correlations between its parts. Furthermore, if measurements are performed simultaneously on subsystems $A$ and $B$, their outcomes may be quantum-correlated, regardless of the fact that the subsystems may be localized in space a substantial distance apart. 

If the system $AB$ is in an pure state, and the reduced states of $A$ and $B$ are mixed, then the state of $AB$ is entangled~\cite{qi_nc}.



\section{Multipartite entanglement}          \label{sec_qm_multi}

In a system composed of multiple subsystems, there exist more types of entanglement. Let us consider a system with a certain number $n$ of subsystems $A$, $B$, ..., $Z$ ($n$ is not necessarily 26). A composite state $\rho_{A...Z}$ is fully separable, if it can be written in analogy to Eq. \eqref{qmsep}, as~\cite{horodecki_entanglement}:
\begin{equation} \label{qmsepmulti}
\rho_{A...Z} =\sum\limits_i p_i \,\rho_{Ai} \otimes ... \otimes \rho_{Zi},
\end{equation}
where $\rho_{Ai}$ are reduced states of the system $A$, and similarly for the other subsystems, and $p_i$ are nonnegative probabilities that add up to unity.

If this criterion is not met, the state is entangled. We distinguish however, between different types of entanglement. First of all, there could be bipartite entanglement in the state of certain two subsystems, say $A$ and $B$. In more general, there may be bipartite entanglement between any two subsests of all the subsystems $A$, $B$, ... , $Z$, e.g. across the bipartition $AB|CDE$ of the state of $ABCDE$.

Apart from bipartite entanglement, there might be also tripartite entanglement in the state of e.g. $A$, $B$ and $C$. This means that simultaneous measurements of $A$, $B$ and $C$ may be quantum-correlated. This might be the case in an absence of bipartite entanglement between $A$ and $B$, $B$ and $C$, and $A$ and $C$, meaning that if we perform the partial trace over one subsystem, the other two are separable. An example of such a behaviour is a GHZ state~\cite{ghz}. It could also be the case in a presence of some of those bipartite entanglements, as it is the case e.g. for a W state~\cite{cirac_ghz_w}. 

Further, there might be $4$-partite entanglement in the state of e.g. $A$, $B$, $C$ and $D$, which is defined in an analogous way as the tripartite entanglement. This can be continued up to the $n$-partite entanglement across all the subsystems. The criteria for multipartite separability are difficult to provide and are a topic of ongoing research. These are beyond the scope of this thesis and for more details the Reader is referred to~\cite{horodecki_entanglement}.


\section{Quantifying the amount of entanglement} \label{sec_neg}


Often one needs not only the knowledge of whether or not a state is entangled, but the amount of entanglement that it possesses.

A particularly useful measure of bipartite entanglement is the logarithmic negativity ${\cal E_N}$~\cite{log_neg_1,log_neg_2}. 
In a composite state $\rho_{AB}$, the logarithmic negativity between $A$ and $B$ is defined as:
\begin{equation} \label{qm_logneg}
{\cal E_N}(A|B) = \log\left[ \sum\limits_i (|\lambda_i| - \lambda_i) + 1 \right],
\end{equation}
where $\{\lambda_i\}$ is the set of all eigenvalues of the partial transpose of $\rho_{AB}$.

The logarithmic negativity is zero for all states separable according to the PPT criterion. It is also an \textit{entanglement monotone}, which means that it is non-increasing under a class of operations called LOCC {\it(local operations and classical communication)}. In general, these may involve a tensor product of operations applied to single subsystems, as well as communication in the classical sense.



\chapter{Elements of Gaussian quantum mechanics} \label{chgqm}

In this chapter we focus on a class of states called Gaussian states, and formulate the contents of Chapter \ref{chqm} in terms of the formalism of Gaussian quantum mechanics, which will be employed in some of the later parts of the thesis. First, in Sec. \ref{sec_gqm_gaussian} Gassian states are defined, and the formalism is introduced. In Sec. \ref{sec_gqm_comp_sys} composite systems are discussed. The chapter concludes with the treatment of quantum entanglement within this formalism in Sec. \ref{sec_gqm_ent}.

From here on we restrict our attention to spin-zero systems, unless otherwise stated. We thus work in the Fock space, with the usual definitions of the vacuum state, the creation and annihilation operators~\cite{peskin}. For a system with a countable number of degrees of freedom, labelled by indices $j$ and $l$, the quadrature operators are defined as:
\begin{align}   \label{gqm_quad}
\hat{q_j}&\defn \frac{1}{\sqrt{2}} (\hat{a_j}\d + \hat{a_j}),
\\
\hat{p_j}&\defn \frac{i}{\sqrt{2}} (\hat{a_j}\d - \hat{a_j}),
\end{align}
and they satisfy the canonical commutation relations:
\begin{equation} 
\begin{gathered} \label{gqm_comm}
[\hat{q}_j,\hat{q}_l] = [\hat{p}_j,\hat{p}_l] = 0,
\\
[\hat{q}_j,\hat{p}_l]=i \delta_{jl}.
\end{gathered}
\end{equation}
For systems with continuous degrees of freedom one may straightforwardly write analogous definitions and relations.


\section{Gaussian states}           \label{sec_gqm_gaussian}

Let us consider a state $\rho$, of a composite system consisting of $n$ subsystems, each one with corresponding quadrature operators $\hat{q}_j$ and $\hat{p}_j$. We define the vector of quadrature operators as:
\begin{equation}
{\bf \hat{x}} \defn (\hat{q}_1, \hat{p}_1, ...\, , \hat{q}_n, \hat{p}_n)^T,
\end{equation}
with $2n$ entries.

One defines first moments of the state as $\avg{\hat{x}_j}$, where the average is evaluated on the state $\rho$. Thus, the vector of first moments is:
\begin{equation}   \label{gqm_def_x}
\vec{X} \defn \avg{{\bf \hat{x}}}.
\end{equation}
The quadrature operators are Hermitian, thus \eqref{gqm_def_x} is a real vector. The second moments are defined as $\avg{\hat{x}_j\hat{x}_l}$. Out of them, the covariance matrix is constructed: 
\begin{equation} \label{gqm_def_cov}
\sigma_{jl} \defn \avg{ \hat{x}_j \hat{x}_l + \hat{x}_l \hat{x}_j } - 2 \avg{\hat{x}_j} \avg{\hat{x}_l}.
\end{equation}
It is a real, symmetric matrix with dimensions $2n \times 2n$.

Higher moments may also be defined analogously, but they will not be needed in this thesis. A state is \textit{Gaussian} iff its third and higher moments are expressed in terms of the first and second moments. It is thus fully characterized by its vector of first moments, and the covariance matrix.


Let us now consider a partition of the same composite state $\rho$ into another set of subsystems. It is known that if a state is Gaussian when expressed using the former family of subsystems, it remains Gaussian for all sets of subsystems that can be related to the former set of subsystems by a unitary transformation~\cite{schumaker_gaussian}.

Examples of Gaussian states include coherent, squeezed and thermal states, which are of particular importance in quantum information. For the vacuum state $\vec{X}=\vec{0}$ and $\boldsymbol\sigma=\I$~\cite{schumaker_gaussian}. Upon the action of a displacement operator we obtain a general coherent state with a non-zero vector of first moments, but still with $\boldsymbol\sigma=\I$. If we do not act with a displacement operator, we may keep $\vec{X}=\vec{0}$ and consider e.g. squeezed and thermal states centered around the origin in phase space. Fock states in turn, are not Gaussian. 

The canonical commutation relations \eqref{gqm_comm} may be compactified as follows~\cite{nick_gaussian}:
\begin{equation}
[\hat{x}_i , \hat{x}_j] = i \Omega_{ij},
\end{equation}
where $\Omega_{ij}$ is called the symplectic form, and defined as:
\begin{align}
\Omega \defn 
\bigoplus_{i=1}^n
	\begin{pmatrix}
		0 & 1 \\
		-1 & 0
	\end{pmatrix}.
\end{align}
A matrix $S$ is \textit{symplectic} iff it satisfies:
\begin{equation}
S^T \Omega S = S \Omega S^T = \Omega.
\end{equation}
If a state $\rho$ evolves unitarily according to eq. \eqref{qm_evo}, the corresponding evolution of the vector of first moments and the covariance matrix is expressed with symplectic matrices~\cite{eric_dets_for_probing}:
\begin{align}
\vec{X}(t) &= S(t) \,\vec{X}(0),
\\         \label{gqm_sympl_ev}
\boldsymbol\sigma(t) &= S(t) \,\boldsymbol\sigma(0) \,S(t)^T.
\end{align}         

\section{Composite systems}        \label{sec_gqm_comp_sys}

Consider a composite system with three subsystems $A$, $B$ and $C$. The vector of the first moments of the composite system has the following form:
\begin{equation}
\vec{X} = (\avg{\hat{q}_A}, \avg{\hat{p}_A}, \avg{\hat{q}_B}, \avg{\hat{p}_B}, \avg{\hat{q}_C}, \avg{\hat{p}_C})^T.
\end{equation}
It consists of three 2-element vectors of first moments corresponding to the reduced states of $A$, $B$ and $C$ consecutively. 

The covariance matrix of the composite system is $6\times 6$ and, as can be seen from the definition \eqref{gqm_def_cov}, it can be analogously divided into $2\times 2$ blocks:
\begin{equation}       \label{gqm_sigma_abc}
\boldsymbol{\sigma}_{ABC}=\left( \begin{array}{ccc}
              \boldsymbol\sigma_A & \boldsymbol\gamma_{AB} & \boldsymbol\gamma_{AC} \\
              \boldsymbol\gamma_{AB}^T & \boldsymbol\sigma_B & \boldsymbol\gamma_{BC} \\
              \boldsymbol\gamma_{AC}^T & \boldsymbol\gamma_{BC}^T & \boldsymbol\sigma_C \end{array} \right),
\end{equation}
where $\boldsymbol\sigma_A$ is the covariance matrix of the subsystem $A$ and similarly for the other subsystems. Furthermore the definition of $\boldsymbol\gamma_{AB}$ and the other off-diagonal block matrices, derives from \eqref{gqm_def_cov}. The off-diagonal block matrices carry information about the correlations between the respective subsystems. When they are zero, the subsystems are not entangled. However, it is not true that when they are non-zero, the subsystems are entangled. This will be discussed in the next section.

In order to construct a reduced state of a Gaussian state on a certain subsystem, one merely needs to cut out an appropriate subvector of the composite vector of first moments, and an appropriate submatrix of the composite covariance matrix. This operation is the equivalent of the partial trace \eqref{qm_pTr} in this formalism. Furthermore, the partial trace preserves Gaussianity. This means that if the composite state is Gaussian, so will be the reduced states of all of the subsystems~\cite{holevo_werner_channel}.

\section{Quantum entanglement}     \label{sec_gqm_ent}

We finish the chapter by providing a formula for the logarithmic negatity corresponding to \eqref{qm_logneg} within the covariance matrix formalism. For a composite state of $A$ and $B$, the covariance matrix is of the form:
\begin{equation}
\boldsymbol\sigma_{AB}=\left( \begin{array}{cc}
              \boldsymbol\sigma_A & \boldsymbol\gamma_{AB} \\
              \boldsymbol\gamma_{AB}^T & \boldsymbol\sigma_B 
              \end{array} \right),
\end{equation}
where the block submatrices are defined similarly as in eq. \eqref{gqm_sigma_abc}.

The logarithmic negativity corresponding to the bipartite entanglement between the subsystems $A$ and $B$ is~\cite{adesso_log_neg}:
\begin{equation}            \label{gqm_neg}
{\cal E_N}(A|B) = \textrm{max}(0,-\textrm{log}_2\tilde{\nu}_{-}),
\end{equation}
where $\tilde{\nu}_{-}$ is calculated from:
\begin{equation}
2\tilde{\nu}^2_{-}=\tilde{\Delta}-\sqrt{\tilde{\Delta}^2-4\,\textrm{det}\boldsymbol\sigma_{AB}},
\end{equation}
where $\tilde{\Delta}=\textrm{det}\boldsymbol\sigma_A + \textrm{det}\boldsymbol\sigma_B
-2\textrm{det}\boldsymbol\gamma_{AB}$.

It is more complicated to compute the logarithmic negativity across a bipartion of a composite system into two parts whose Hilbert spaces have different dimension. E.g. in Part \ref{part4} we will need to compute the negativity between one harmonic oscillator and two harmonic oscillators. We will discuss our approach in Sec. \ref{sec_tri_tri}.


\chapter{Minkowski and Rindler coordinates} \label{chgr}

Let us now introduce the coordinate systems, that are used in this thesis. We work in an inertial or a uniformly accelerated frame of reference. In Sec. \ref{chgr_general} general definitions of a line element and the proper time along a spacetime trajectory are given. Further, sections \ref{sec_gr_Minkowski} and \ref{sec_gr_rind} are devoted to Minkowski and Rindler coordinates respectively.

The metric signature is chosen to be $(+,-, ...\, ,-)$.

\section{General notions}   \label{chgr_general}

Let us consider spacetime with $n_D+1$ dimensions, described with a metric $g_{\mu\nu}$. Throughout this thesis $n_D$ is chosen to be $1$ or $3$, and the spacetime is flat. A line element equals:
\begin{equation}     \label{gr_line}
\x{d}s = g_{\mu\nu} \x{d}x^\mu \x{d}x^\nu,
\end{equation}
where $x^\mu$ are the coordinates and the Einstein summation convention is used. 

Integrating \eqref{gr_line} along a timelike path, one obtains the proper time elapsed along it~\cite{hartle}:
\begin{equation}   \label{gr_clock_general}
\tau = \int_\x{path} \sqrt{g_{\mu\nu} \x{d}x^\mu \x{d}x^\nu}.
\end{equation}
The \textit{clock postulate} states that the proper time is the time, that an {\it ideal clock} would measure, while travelling along a given trajectory. The {\it coordinate time}, $x^0$ in turn, generally does not correspond to any physically measurable quantity. Further discussion on the implications and limitations of the clock postulate are the topic of the investigation in Part \ref{part2}.

A surface described by $x^0=\x{const.}$ is called a \textit{Cauchy surface}.


\section{Minkowski coordinates}      \label{sec_gr_Minkowski}

The Minkowski coordinates describe an inertial frame in a flat spacetime. The Minkowski metric is:
\begin{equation}
g_{\mu\nu}=\x{diag}(1,-1,...\, ,-1).
\end{equation}
From this and \eqref{gr_clock_general} it follows that the proper time along a timelike path becomes:
\begin{equation}   \label{gr_clock_m}
\tau = \int_\x{path} \sqrt{1-v^2(t)} \,\,\x{d}t,
\end{equation}
where $v(t)$ is the speed of the body moving along this trajectory, at time $t$.

Throughout this thesis let us denote the Minkowski coordinates as:
\begin{equation}
(x^0,x^1,x^2,...\,,x^{n_D})=(t,x,{\bf x_\pe}),
\end{equation}
where ${\bf x_\pe} \defn (x^2, ...\, , x^{n_D})$.

\section{Rindler coordinates}    \label{sec_gr_rind}

The Rindler coordinates describe the flat spacetime from a uniformly accelerated frame of reference. The Minkowski and Rindler coordinates are related to each other via a Rindler transformation~\cite{rindler, rindler_relativity}. Let us direct the acceleration along the $x$ axis of the Minkowski coordinates. The coordinates perpendicular to the direction of acceleration remain the same in both frames of reference. 

\begin{figure}
\centering
\includegraphics[width=0.6\linewidth]{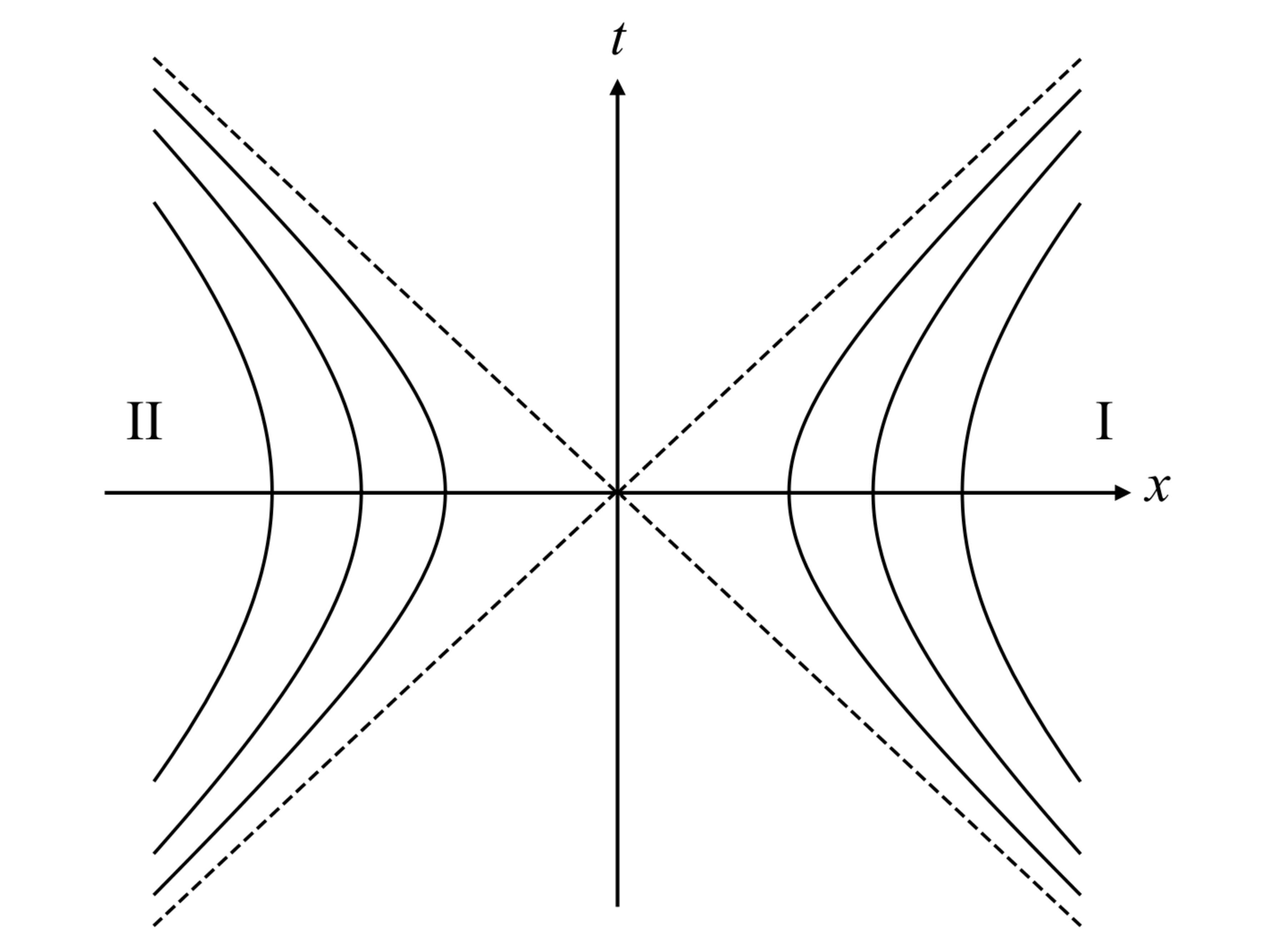}
\put(-150,146){\small{F}}
\put(-150,16){\small{P}}
\caption{\small{Four regions of the Rindler coordinate chart, drawn in the $xt$ plane of the Minkowski coordinates. The solid curves are the uniformly accelerated observers' trajectories \eqref{gr_rind_hyp} and the dashed lines are the future and the past event horizon.}}\label{fig_rindler}
\end{figure}

The Rindler transformation divides the Minkowski spacetime into 4 regions (wedges), as shown in Fig. \ref{fig_rindler}. The boundary between the regions I and F and between the regions II and F, called the \textit{future event horizon}, is described by $t=|x|$. Similarly, the boundary between the regions I and P and between the regions II and P, called the \textit{past event horizon}, is described by $t=-|x|$. One may note that no information can travel between regions I and II. Furthermore, while information can get inside the region F from any other wedge, once it is inside, it cannot get back outside. Information from region P in turn, can access any other region, but no information can get inside it from the outside. In this thesis, we are not interested in trajectories inside regions F and P.

The transformation between the Minkowski coordinates $(t,x,{\bf x_\pe})$ and the Rindler coordinates $(\eta,\chi,{\bf x'_\pe})$
is of the form:
\begin{equation} \label{gr_rindler1d}
\begin{aligned}
&t = \pm\chi\sinh a\eta,
\\
&x = \pm\chi\cosh a\eta,
\\
&{\bf x_\pe} = {\bf x'_\pe},
\end{aligned}
\end{equation}
where $a$ is a positive parameter. The upper signs refer to region I and the lower signs refer to region II. The corresponding Rindler metric yields:
\begin{equation} \label{gr_rmetric}
g_{\mu\nu}=\x{diag}(a^2\chi^2,-1,...\,,-1).
\end{equation}
The inverse transformation of \eqref{gr_rindler1d} is: 
\begin{equation}
\begin{aligned}
&\eta = \frac{1}{a}\x{atanh} \frac{t}{x},
\\
&\chi = \pm\sqrt{x^2-t^2},
\\
&{\bf x'_\pe} = {\bf x_\pe},
\end{aligned}
\end{equation}
where $\pm$ applies as in \eqref{gr_rindler1d}. 

In Minkowski coordinates, trajectories corresponding to uniformly accelerated observers moving parallel to the $x$ axis, are the arms of hyperbolae of the form:
\begin{equation} \label{gr_rind_hyp}
\frac{1}{\A}=\pm\sqrt{x^2-t^2},
\end{equation}
where $\pm$ applies as in \eqref{gr_rindler1d} and ${\bf x_\pe}$ is kept constant. Also, at $t=0$ the observers are located at $x=\pm\tfrac{1}{\A}$. Several hyperbolae of this form are shown in Fig. \ref{fig_rindler}. Here, $\A$ is the modulus of the proper acceleration of a given uniformly accelerated trajectory~\cite{rindler_relativity}. All the hyperbolae share a common center, at the origin of the Minkowski coordinates, and are asymptotic to the future and the past event horizon. The wedge I contains all of the uniformly accelerated trajectories described with \eqref{gr_rind_hyp}, accelerating to the right, and the wedge II contains all of the uniformly accelererated trajectories described with \eqref{gr_rind_hyp}, accelerating to the left. 

In Rindler coordinates, rays with initial points located at the origin of the Minkowski coordinates constitute Cauchy surfaces $\eta=\x{const}$. The proper time of a given hyperbola in Rindler coordinates, between the Cauchy surface corresponding to the coordinate time equal to $0$, and the Cauchy surface corresponding to the coordinate time equal to $\eta$ is:
\begin{equation}     \label{gr_tau_eta}
\tau = \frac{a}{\A}\,\eta.
\end{equation}
Therefore, there exists one special hyperbola, called the \textit{reference hyperbola}, whose proper time equals the coordinate time. It is the one with proper acceleration ${\A}= a$. Otherwise, the parameter $a$ has no physical significance.

The $\chi$ axis may be relabelled as follows:
\begin{equation}  \label{gr_rc}
\xi=\frac{1}{a}\log(\pm a\chi),
\end{equation}
where $+$ refers to wedge I and $-$ refers to wedge II. One should bear in mind that the coordinates $\chi$ and $\xi$ have different ranges. E.g. in wedge I we have $\chi\in[0,+\infty]$ whereas $\xi\in[-\infty,+\infty]$. The coordinates $(\xi,\eta, {\bf x_\pe})$ are sometimes called {\it Rindler conformal coordinates}.


\chapter{Introduction to quantum field theory in an inertial and a uniformly accelerated frame} \label{chqft}

In this chapter we formulate the basics of the description of a quantum scalar field in a relativistic inertial and a uniformly accelerated frame. We operate on Fock space and focus on a spin-zero field. In Sec. \ref{sec_qft_m} a flat spacetime described with Minkowski coordinates is discussed, and in Sec.~\ref{sec_qft_r} we move on to the case of the Rindler coordinates. Sec. \ref{sec_qft_conf} discusses the conformal invariance of the massless field in a Rindler spacetime with $1+1$~dimensions. The chapter concludes with Sec. \ref{sec_rqi_bogo} devoted to the Bogolyubov transformation.

\section{Quantum field theory in Minkowski coordinates}   \label{sec_qft_m}

The equation governing the evolution of a spin-zero quantum field $\hat{\Phi}$, is the Klein-Gordon equation~\cite{peskin, birrell_davies}. We work in a Minkowski spacetime with $n_D+1$ dimensions. In the Minkowski coordinates the equation takes the form:
\begin{equation} \label{qft_KG}
\left(
\partial^2_t - \partial^2_x - \nabla^2_{\bf x_\pe} + m^2
\right)
\hat{\Phi}
=0,
\end{equation}
where $m$ is the mass of the field, and the Laplacian is understood to be taken over the coordinates perpendicular to $x$.

There is a scalar product associated with the Klein-Gordon equation, of the form:
\begin{equation} \label{qft_scalar}
(\phi_1,\phi_2) = i \int_\Sigma \x{d}^{n_D} {\bf\Sigma}
\left(
\phi_1\s \partial_t \phi_2 -
\phi_2 \partial_t \phi_1\s
\right),
\end{equation}
where it is taken along a spacelike Cauchy surface $\Sigma$. Unless stated otherwise, the chosen surface will be $t=0$. The infinitesimal vector $\x{d}^{n_D} {\bf\Sigma}$ is a Cauchy surface element, directed normally to the surface. From the form of \eqref{qft_scalar} it follows that the scalar product satisfies the properties:
\begin{equation} \label{qft_scalar_prop}
(\phi_1,\phi_2) = (\phi_2,\phi_1)\s = -(\phi_2\s , \phi_1\s).
\end{equation}

\subsection{Free field}

The Klein-Gordon equation may be solved for normal mode solutions. This yields a family of plane waves:
\begin{equation} \label{qft_uk}
u_k = \frac{1}{\sqrt{2\omega_k \left(2\pi\right)^{n_D}}} \,
e^{i{\bf k}\cdot{\bf x} - i\omega_k t},
\end{equation}
with $\omega_k=\sqrt{k^2+m^2}$ being the frequency of the mode, and $\bf k$ being an ${n_D\text{-dimensional}}$ wave vector. Plane waves written in such a way, are normalized to a delta function in the sense of the following orthonormality relations:
\begin{equation}
\begin{aligned}
(u_{\bf k},u_{\bf k'}) &= \delta^{n_D}({\bf k}-{\bf k'}),
\\
(u_{\bf k}\s,u_{\bf k'}\s) &= -\delta^{n_D}({\bf k}-{\bf k'}),
\\
(u_{\bf k},u_{\bf k'}\s) &= 0.
\end{aligned}
\end{equation}
Using these modes, the quantum field can be decomposed as follows:
\begin{equation}  \label{qft_dec_a}
\hat{\Phi} = 
\int \x{d}^{n_D}{\bf k} \left(
u_{\bf k} \hat{a}_{\bf k} + u_{\bf k}\s \hat{a}_{\bf k}\d
\right),
\end{equation}
where $\hat{a}_{\bf k}$ and $\hat{a}_{\bf k}\d$ are the creation and annihilation operators associated with the respective modes, satisfying the commutation relations:
\begin{equation}
\begin{aligned}
\left[\hat{a}_{\bf k} , \hat{a}_{\bf k'}\right] & = 0,
\\
[\hat{a}_{\bf k} , \hat{a}_{\bf k'}\d] &= \delta^{n_D}({\bf k-k'}).
\end{aligned}
\end{equation}
The set of modes $\{u_{\bf k}\}$ is often referred to as {\it positive frequency modes} and $\{u_{\bf k}\s\}$ as {\it negative frequency modes}. Physically it means that upon the action of the free Hamiltonian on the state of $u_{\bf k}$, one obtains a positive energy, and from the state of $u_{\bf k}\s$ one gets a negative energy.

\subsection{Cavity field}

The plane wave solutions extend over the whole space, and form a continuous basis. Such is the case, when the field is free. Let us now study a different scenario, whereby the field is confined in a finite box (a cavity), and subject to certain boundary conditions. Let the cavity extend in space between $x^j=\sigma^j_-$ and $x^j=\sigma^j_+$, where $j$ is an integer ranging from $1$ to $n_D$. In the direction $x^j$ the cavity has length $L_j=\sigma^j_+ -\sigma^j_-$. The mode solutions are obtained by applying certain conditions to \eqref{qft_uk}. The resultant set of modes is infinite and discrete. We shall focus on two types of boundary conditions. 

The first type is Dirichlet boundary conditions with the modes, and hence the field vanishing at the walls of the cavity. The mode solutions then take the form:
\begin{equation}  \label{qft_u_cav_m}
U_{\bf n} = \frac{1}{\pi}\sqrt{\frac{2^{n_D-1}}{\omega_{\bf n}}}\,
\prod\limits_{j=1}^{n_D} \frac{1}{\sqrt{L_j}}
\,\sin\left[ \frac{n_j\,\pi}{L_j} (x^j -\sigma^j_-) \right]e^{-i\omega_{\bf n} t},
\end{equation}
where ${\bf n}=(n_1,...\,,n_{n_D})$ is a vector of positive integers labelling the modes. Furthermore, the frequency of a given mode is: 
\begin{equation}
\omega_{\bf n}=\sqrt{\sum\limits_{j=1}^{n_D}
\left(\frac{n_j\,\pi}{L_j} \right)^2+m^2}.
\end{equation}

The second type is periodic boundary conditions, meaning that the value of a mode 
needs to be equal at the opposite walls of the cavity. In this thesis we limit our attention only to the modes with $n_D=1$. The cavity is assumed to extend between $x=\sigma_-$ and $x=\sigma_+$ and to have length $L$, the form of the modes is as follows:
\begin{equation}   \label{qft_u_cav_m_per}
U_n = \frac{1}{\sqrt{4 \pi |n|}}\,e^{i\frac{2n\pi}{L}(x-\sigma_-) - i\omega_n t},
\end{equation}
with the frequency $\omega_n=\sqrt{\left( \frac{2n\pi}{L} \right)^2+m^2}$ and $n\in \mathbb{Z}\setminus \{0\}$ (we omit the zero mode, i.e. $n=0$).

It should be understood that the modes \eqref{qft_u_cav_m} and \eqref{qft_u_cav_m_per} only exist inside the cavity, so their value outside is zero. The scalar product \eqref{qft_scalar} can be thought of being taken along the entire space with the regions outside the cavity giving zero contribution, or equivalently it can be taken over the cavity region only.

The mode orthonormality relations take the following form:
\begin{equation}
\begin{aligned}
(U_{\bf n},U_{\bf n'}) &= \delta_{{\bf n}{\bf n'}},
\\
(U_{\bf n}\s,U_{\bf n'}\s) &= -\delta_{{\bf n}{\bf n'}},
\\
(U_{\bf n},U_{\bf n'}\s) &= 0.
\end{aligned}
\end{equation}
The decomposition of the field operator inside the cavity is:
\begin{equation}   \label{qft_dec_A}
\hat{\Phi} = 
\sum\limits_\mathbf{n} \left(
U_\mathbf{n} \hat{A}_\mathbf{n} + U_\mathbf{n}\s \hat{A}_\mathbf{n}\d
\right),
\end{equation}
where the corresponding creation and annihilation operators satisfy the commutation relations:
\begin{equation}
\begin{aligned}
\left[\hat{A}_\mathbf{n} , \hat{A}_\mathbf{n'}\right] & = 0,
\\
\left[\hat{A}_\mathbf{n} , \hat{A}_\mathbf{n'}\d\right] &= \delta_\mathbf{nn'}.
\end{aligned}
\end{equation}

\section{Quantum field theory in Rindler cooridnates}   \label{sec_qft_r}

The Klein-Gordon equation \eqref{qft_KG} in a flat spacetime covered with Rindler coordinates, can be obtained using the Rindler transformation \eqref{gr_rindler1d}. The outcome is~\cite{takagi}:
\begin{equation} \label{qft_KG_r}
\left[\frac{1}{a^2\chi^2}\frac{\partial^2}{\partial\eta^2}-\frac{\partial^2}{\partial\chi^2}-\frac{1}{\chi}\frac{\partial}{\partial\chi}-\nabla^2_{\bf x'_\pe}+m^2\right]\hat{\Phi} = 0.
\end{equation}
In the Rindler conformal coordinates, the equation \eqref{qft_KG_r} takes the form:
\begin{equation} \label{qft_KG_rc}
\left[\,\frac{\partial^2}{\partial\eta^2}-\frac{\partial^2}{\partial\xi^2}+
\left( - \nabla^2_{\bf x'_\pe}+m^2\right)e^{2a\xi}\,\,\right]\hat{\Phi}=0.
\end{equation}
The scalar product assciated with \eqref{qft_KG_r} is:
\begin{equation} \label{qft_scalar_r}
(\phi_1,\phi_2) = i \int_\Sigma \x{d}^{n_D-1}{\bf x'_\perp} \int \frac{\x{d}\chi}{a\chi}
\left(
\phi_1\s \partial_\eta \phi_2 -
\phi_2 \partial_\eta \phi_1\s
\right).
\end{equation}
Similarly, for \eqref{qft_KG_rc} one gets:
\begin{equation} \label{qft_scalar_rc}
(\phi_1,\phi_2) = i \int_\Sigma \x{d}^{n_D-1}{\bf x'_\perp} \int \x{d}\xi
\left(
\phi_1\s \partial_\eta \phi_2 -
\phi_2 \partial_\eta \phi_1\s
\right),
\end{equation}
bearing in mind that $\xi$ extends only over one wedge. Here, both scalar products are taken along the Cauchy surface $\eta=0$.

\subsection{Free field}

The modes in Rindler coordinates are~\cite{crispino_higuchi}:
\begin{align}      \label{qft_wi_m_r}
\wiOk=&
\sqrt{\frac{\sinh\left(\frac{\pi\Omega}{a}\right)}{\pi^2a\,(2\pi)^{n_D-1}}}\,
K_{\frac{i\Omega}{a}}\left(\sqrt{k_\perp^2+m^2}\,\,\chi\right)
e^{i{\bf k_\perp}\cdot{\bf x'_\perp}-i\Omega\eta}
\quad \text{in I},
\\
\wiiOk=&         \label{qft_wii_m_r}
\sqrt{\frac{\sinh\left(\frac{\pi\Omega}{a}\right)}{\pi^2a\,(2\pi)^{n_D-1}}}\,
K_{\frac{i\Omega}{a}}\left(-\sqrt{k_\perp^2+m^2}\,\,\chi\right)
e^{i{\bf k_\perp}\cdot{\bf x'_\perp}+i\Omega\eta}
\quad \text{in II},
\end{align}
where ${\bf k_\perp}=(k^2, ...\, , k^{n_D})$ is the component of the wave vector perpendicular to the direction of acceleration, $\Omega$ is a nonnegative parameter, and $K$ is the modified Bessel function of the second kind~\cite{nist}. Also, it should be understood that each mode is defined as above in its respective region, and zero elsewhere.

When one uses the Rindler conformal coordinates, one obtains:
\begin{align} \label{qft_wi_m_rc}
\wiO &= \sqrt{\frac{\sinh\left(\frac{\pi\Omega}{a}\right)}{\pi^2a\,(2\pi)^{n_D-1}}}\, 
K_{i\frac{\Omega}{a}}\left( \frac{\sqrt{k_\perp^2+m^2}}{a}\, e^{a\xi} \right)
e^{i{\bf k_\perp}\cdot{\bf x'_\perp}-i\Omega\eta}  
\quad \text{in I},
\\             \label{qft_wii_m_rc}
w_{\text{II}\Omega}&= \sqrt{\frac{\sinh\left(\frac{\pi\Omega}{a}\right)}{\pi^2a\,(2\pi)^{n_D-1}}}\, 
K_{i\frac{\Omega}{a}}\left( \frac{\sqrt{k_\perp^2+m^2}}{a}\, e^{a\xi} \right)
e^{i{\bf k_\perp}\cdot{\bf x'_\perp}+i\Omega\eta} 
\quad \text{in II},
\end{align}
but one should keep in mind that all $\xi\in\mathbb{R}$ in $\wiO$ correspond to $\chi>0$ and all $\xi\in\mathbb{R}$ in $\wiiO$ correspond to $\chi<0$.

Using \cite{passian}, one may verify that the modes \eqref{qft_wi_m_r}-\eqref{qft_wii_m_r} and \eqref{qft_wi_m_rc}-\eqref{qft_wii_m_rc} are normalized with respect to the scalar products \eqref{qft_scalar_r} and \eqref{qft_scalar_rc} respectively, as:
\begin{equation}
\begin{aligned}
(\wiOk,\wiXl) &= \delta(\Omega-\Xi)\,\,\delta^{n_D-1}({\bf k_\perp}-{\bf l_\perp}),
\\
(\wiOk\s , \wiXl\s) &= -\delta(\Omega-\Xi)\,\,\delta^{n_D-1}({\bf k_\perp}-{\bf l_\perp}),\\
(\wiOk , \wiXl\s) &= 0.
\end{aligned}
\end{equation}
The same relations hold for $\wiiOk$. Because wedges I and II are disjoint, the overlap of modes from region I with the modes from region II is zero.

The decomposition of the field operators into Rindler modes can be written as:
\begin{equation} \label{qft_dec_b_3d}
\begin{aligned}
\hat{\Phi}=\int_0^{\infty}\text{d}\Omega\int\text{d}^{n_D-1}{\bf k_\perp}
\big(
\wiOk &\biOk + \wiOk\s \biOk\d 
\\
+ \,&\wiiOk \biiOk + \wiiOk\s \biiOk\d
\big),
\end{aligned}
\end{equation}
where $\biOk$ and $\biiOk$ are the creation and annihilation operators associated with the corresponding modes. They satisfy the algebra:
\begin{equation}
\begin{aligned}
\left[\biOk , \biXl\right] & = 0,
\\
\left[\biOk , \biXl\d\right] &= \delta(\Omega-\Xi) \,\, \delta^{n_D-1}({\bf k_\perp}-{\bf l_\perp}).
\end{aligned}
\end{equation}
The same relations hold for $\biiOk$. Because wedges I and II are disjoint, all the commutators of ladder operators from region I with the ladder operators from region II vanish.

\subsection{Cavity field}

We shall now find also the modes in a cavity extending from $\chi=\sigma^1_-$ to ${\chi=\sigma^1_+}$ within region I only or region II only, and between $x^j=\sigma^j_-$ and $x^j=\sigma^j_+$, where the integer $j$ here ranges from $2$ to $n_D$. In the direction $\chi$ the cavity has proper length $L_\pa$, whereas in the direction $x^j$ the cavity has length $L_j=\sigma^j_+ -\sigma^j_-$. The cavity is assummed to be stationary in the Rindler frame, and to accelerate uniformly as seen by an inertial observer. More precisely, its walls at ${\chi=\sigma^1_\pm}$ accelerate uniformly, with proper accelerations $\A_\pm$ respectively, such that the cavity keeps a fixed proper length $L_\pa$ in the Rindler frame. For this to be true, we need $L_\pa=\left|\frac{1}{\A_-}-\frac{1}{\A_+}\right|$. We focus on Dirichlet boundary conditions, with the field vanishing at the walls.

The mode solutions are of the form~\cite{andrzej_mody_besseli}:
\begin{equation}  \label{qft_w_cav_m}
\begin{aligned}
W_{\bf n} &\propto
\prod\limits_{j=2}^{n_D}
\sin\left[ \frac{n_j\,\pi}{L_j} (x^j -\sigma^j_-) \right]
\\
&\times\Im\left\{
I_{-i\omega_{\bf n}}(m_{\text{eff}}|\sigma^1_-|)
I_{i\omega_{\bf n}}(m_{\text{eff}}|\chi|)
\right\}
e^{\mp i\omega_{\bf n}t},
\end{aligned}
\end{equation}
where $-$ refers to wedge I and $+$ refers to wedge II. Also, ${\bf n}=(n_1,...\,,n_{n_D})$ is a vector of positive integers labelling the modes, $I$ is the modified Bessel function of the first kind~\cite{nist}, and ${m_{\x{eff}}=\sqrt{\sum\limits_{j=2}^{n_D}\left(\frac{n_j\,\pi}{L_j}\right)^2+m^2}}$ can be interpreted as an effective mass. Furthermore, $\omega_{\bf n}$ is a set of values for which the boundary conditions are satisfied, that can be evaluated only numerically. These values are also labelled by the set of positive integers ${\bf n}$. 

The normalization constant may only be computed numerically, to satisfy the usual orthonormality relations:
\begin{equation}
\begin{aligned}
(W_{\bf n},W_{\bf n'}) &= \delta_{\bf nn'},
\\
(W_{\bf n}\s,W_{\bf n'}\s) &= -\delta_{\bf nn'},
\\
(W_{\bf n},W_{\bf n'}\s) &= 0,
\end{aligned}
\end{equation}
with respect to the scalar product \eqref{qft_scalar_r}.

The decomposition of the field operator inside the cavity is:
\begin{equation}   \label{qft_dec_B}
\hat{\Phi} = 
\sum\limits_{\bf n} \left(
W_{\bf n} \hat{B}_{\bf n} + W_{\bf n}\s \hat{B}_{\bf n}\d
\right),
\end{equation}
where the corresponding creation and annihilation operators satisfy the commutation relations:
\begin{equation}
\begin{aligned}
\left[\hat{B}_{\bf n} , \hat{B}_{\bf n'}\right] & = 0,
\\
\left[\hat{B}_{\bf n} , \hat{B}_{\bf n'}\d\right] &= \delta_{\bf nn'}.
\end{aligned}
\end{equation}

\section{Conformal invariance}             \label{sec_qft_conf}

Let us now study a special case of \eqref{qft_KG_r} with $m=0$ and $n_D=1$. In this case the modes following from \eqref{qft_KG_r} take the form~\cite{birrell_davies}
\begin{align}    \label{qft_wi1d}
\wiO &= \frac{1}{\sqrt{4\pi\Omega}}\,
e^{\pm i \frac{\Omega}{a}\log\chi - i\Omega\eta}
\qquad \text{in I},
\\               \label{qft_wii1d}
\wiiO &= \frac{1}{\sqrt{4\pi\Omega}}\,
e^{\mp i \frac{\Omega}{a}\log\chi + i\Omega\eta}
\qquad \text{in II},
\end{align}
where the solutions with the upper sign are referred to as the {\it right-moving} solutions and the solutions with the lower sign are referred to as the {\it left-moving} solutions. Here, $\Omega$ is a nonnegative parameter. 

One may note that for $m=0$ and $n_D=1$ the equations \eqref{qft_KG} and \eqref{qft_KG_rc} have the same form. This phenomenon is known as the {\it conformal invariance}. It manifests itself in the fact that the above solutions have the form of plane waves, when expressed in the Rindler conformal coordinates with \eqref{gr_rc}:
\begin{align} \label{qft_wi_m0_rc}
\wiO &= \frac{1}{\sqrt{4\pi\Omega}}\,
e^{\pm i \Omega\xi - i\Omega\eta}
\qquad \text{in I},
\\            \label{qft_wii_m0_rc}
\wiiO &= \frac{1}{\sqrt{4\pi\Omega}}\,
e^{\mp i \Omega\xi + i\Omega\eta}
\qquad \text{in II}.
\end{align}
It should be understood that $\wiO$ vanishes outside region I and similarly $\wiiO$ vanishes outside region II. 

There is an important subtlety here. Mathematically the equation \eqref{qft_KG_r} has two orthogonal families of modes - the Bessel $K$ and Bessel $L$ functions. The latter are less commonly used, but the Reader may find the definition in~\cite{dunster}. In the limit of $\xi\rightarrow \infty$ one may use the following approximations~\cite{dunster}:
\begin{align}
K_{i\frac{\Omega}{a}}\left( \frac{m}{a} e^{a\xi} \right) &\approx
-\sqrt{\frac{a\pi}{\Omega\sinh\left( \frac{\pi\Omega}{a} \right)}}
\sin\left(
\Omega\xi + \frac{\Omega}{a}\log\left( \frac{m}{2a} \right) - \nu_\frac{\Omega}{a}
\right),
\\
L_{i\frac{\Omega}{a}}\left( \frac{m}{a} e^{a\xi} \right) &\approx
+\sqrt{\frac{a\pi}{\Omega\sinh\left( \frac{\pi\Omega}{a} \right)}}
\cos\left(
\Omega\xi + \frac{\Omega}{a}\log\left( \frac{m}{2a} \right) - \nu_\frac{\Omega}{a}
\right),
\end{align}
where $k_\pe$ has been put to zero and $\nu_\frac{\Omega}{a}$ is the following phase of the Gamma function:
\begin{equation}
\nu_\frac{\Omega}{a} = \arg\{ \Gamma\left( 1+i\frac{\Omega}{a} \right) \}.
\end{equation}
Substituting these approximations in \eqref{qft_wi_m_rc} and \eqref{qft_wii_m_rc} does yield a family of modes equivalent to \eqref{qft_wi_m0_rc} and \eqref{qft_wii_m0_rc}. However, the Bessel $L$ functions diverge at $\xi\rightarrow\infty$, and are not normalizable and hence have to be discarded. Thus, we end up only with one family of mode solutions in the case when $m\neq 0$. For this reason the massive and the massless cases need to be investigated separately in Rindler coordinates at $n_D=1$. In other words, the solutions \eqref{qft_wi_m_r}-\eqref{qft_wii_m_r} do not apply when $m=0$ and $n_D=1$.

For a scalar field in a cavity extending between $x=\sigma_-$ and $x=\sigma_+$, at $m=0$ and $n_D=1$, from applying Dirichlet boundary conditions to the mode \eqref{qft_wi_m0_rc} or \eqref{qft_wii_m0_rc}, one obtains:
\begin{equation} \label{qft_Wn_cav_r_m0}
W_n = \frac{1}{\sqrt{\pi n}}\sin\left[ \omega_n \left(\xi-\frac{1}{a}\log(a\sigma_-)\right) \right]e^{\mp i\omega_n \eta},
\end{equation}
where $-$ refers to wedge I and $+$ refers to wedge II. The frequency of the mode is $\omega_n=\frac{a n\pi}{\log\frac{\sigma_+}{\sigma_-}}$ and $n\in \mathbb{N}_+$.

\section{Bogolyubov transformation}    \label{sec_rqi_bogo}

In general, quantum states get altered under a change of the reference frame. If a quantum state has a certain form in one frame of reference, in order to express it in a different frame of reference, the Bogolyubov transformation is performed~\cite{birrell_davies}. 

In quantum field theory all states can be written in terms of creation and annihilation operators acting on the \textit{vacuum state}. In a given reference frame the vacuum state $\ket{0}$ is defined as follows:
\begin{equation}   \label{rqi_vac_general}
\hat{a}\ket{0} \defn 0,
\end{equation}
where $\hat{a}$ is any annihilation operator in this reference frame. It thus suffices to express the creation and annihilation operators and the vacuum state in one reference frame in terms of the creation and annihilation operators and the vacuum state in the other reference frame, in order to obtain the expression for a transformed state. This is how the Bogolyubov transformation operates.

The contents of this section apply to a basis labelled by a single, continuous index. However, they can be straightforwardly generalized to a basis labelled by a discrete variable or to a basis labelled with multiple variables.

Suppose a scalar field operator has two decompositions, which may correspond to any two frames of reference:
\begin{equation}   \label{rqi_bogo_decs}
\hat{\Phi} =
\int\x{d}k \left(u_k \hat{a}_k + u_k\s \hat{a}_k\d \right)
=
\int\x{d}l \left(w_l \hat{b}_l + w_l\s \hat{b}_l\d \right).
\end{equation}
We assume that the modes are orthonormal, and the equations below hold:
\begin{equation}
\begin{aligned}
&(u_k, u_p) = \delta(k-p),
\\
&(u_k\s , u_p\s) = -\delta(k-p),
\\
&(u_k, u_p\s) = 0,
\end{aligned}
\end{equation}

\vspace{0.1cm}

\begin{equation}
\begin{aligned}
&(w_l, w_s) = \delta(l-s),
\\
&(w_l\s , w_s\s) = -\delta(l-s),
\\
&(w_l, w_s\s) = 0,
\end{aligned}
\end{equation}

\vspace{0.1cm}

\begin{equation}  \label{rqi_bogo_a_comm}
\begin{aligned}
&\left[\hat{a}_k , \hat{a}_p\right] = 0,
\\
&[\hat{a}_k , \hat{a}_p\d] = \delta(k-p),
\end{aligned}
\end{equation}

\vspace{0.1cm}

\begin{equation}  \label{rqi_bogo_b_comm}
\begin{aligned}
&[\hat{b}_l , \hat{b}_s] = 0,
\\
&[\hat{b}_l , \hat{b}_s\d] = \delta(l-s),
\end{aligned}
\end{equation}
where we do not assume any particular form of the scalar product. We only assume that it satisfies the properties \eqref{qft_scalar_prop}.

Taking the scalar product of both sides of \eqref{rqi_bogo_decs} with $w_s$, i.e. $(w_s, ...)$, and renaming the indices yields:
\begin{equation}  \label{rqi_bogo_b}
\hat{b}_l
=
\int\x{d}k \left(\alpha\s_{lk} \hat{a}_k - \beta\s_{lk} \hat{a}_k\d \right),
\end{equation}
where we have defined the {\it Bogolyubov coefficients} as:
\begin{equation}
\begin{aligned}
\alpha_{lk} &\defn (u_k, w_l),
\\
\beta_{lk} &\defn -(u_k\s, w_l),
\end{aligned}
\end{equation}
and used the properties \eqref{qft_scalar_prop}.
Complex conjugating \eqref{rqi_bogo_b}, one obtains an expression for $\hat{b}_l\d$. Substituting this into \eqref{rqi_bogo_b_comm}, one gets the {\it Bogolyubov identities}:
\begin{equation}
\begin{aligned}
&\int\x{d}k\left(
\alpha_{lk} \alpha\s_{sk} - \beta_{lk} \beta\s_{sk} \right) = \delta(l-s),
\\
&\int\x{d}k\left(
\alpha_{lk} \beta_{sk} - \beta_{lk} \alpha_{sk} \right) = 0.
\end{aligned}
\end{equation}

One may also take the scalar product of both sides of \eqref{rqi_bogo_decs} with $u_p$, i.e. $(u_p, ...)$. This way we obtain:
\begin{equation}  \label{rqi_bogo_a}
\hat{a}_k
=
\int\x{d}l \left(\alpha_{lk} \hat{b}_l + \beta\s_{lk} \hat{b}_l\d \right),
\end{equation}
which is the inverse transformation of \eqref{rqi_bogo_b}.

Upon the substitution of \eqref{rqi_bogo_b} into \eqref{rqi_bogo_decs} and using \eqref{rqi_bogo_a_comm}, one obtains the transformation of a mode, corresponding to a change of a reference frame:
\begin{equation}
u_k = \int\x{d}l \left( \alpha_{lk}\s w_l - \beta_{lk} w_l\s \right).
\end{equation}
The inverse transformation of a mode may be derived similary, and takes the form:
\begin{equation}
w_l = \int\x{d}k \left( \alpha_{lk} u_k + \beta_{lk}\s u_k\s \right).
\end{equation}



\chapter{Unruh effect} \label{chrqi}

In this chapter we discuss the Unruh effect~\cite{unruh_effect} for a massive scalar field in $1+1$ dimensions. The transformation of the Minkowski vacuum state, upon a change of the reference frame to a uniformly accelerated one, is the topic of Sec. \ref{chrqi_sec_unruh}. Further, Sec. \ref{sec_vac_ent} concerns the quantum entanglement contained in this state.

\section{Bogolyubov transformation of the Minkowski vacuum state}    \label{chrqi_sec_unruh}

Let us now derive the form of the vacuum state of the inertial frame of reference $\ket{0}_\x{M}$ (the {\it Minkowski vacuum state}), as seen from a relativistic uniformly accelerated reference frame. We assume that the field under consideration is a massive scalar field with mass $m$, in $1+1$~dimensions. The results can be analogously obtained for different fields and dimensionalities.

We begin by calculating the Bogolyubov coefficients of the transformation from the inertial frame to a uniformly accelerated frame, wherein the parameter of the Rindler transformation is $a$. Computing the scalar product \eqref{qft_scalar} of the Minkowski modes \eqref{qft_uk} with the Rindler modes \eqref{qft_wi1d} and \eqref{qft_wii1d}, one obtains~\cite{crispino_higuchi}:
\begin{equation}   \label{rqi_u_bogo1d}
\begin{aligned}   
\alpha_{\Omega k}^{(\text{I})} &\defn (u_k, \wiO) 
=\frac{1}{\sqrt{4\pi\omega_k a \sinh\left(\frac{\pi\Omega}{a}\right)}}
\left(\frac{\omega_k+k}{\omega_k-k}\right)^{-i\frac{\Omega}{2a}}e^{\frac{\pi\Omega}{2a}},
\\
\alpha_{\Omega k}^{(\text{II})} &\defn (u_k, \wiiO) 
= {\alpha_{\Omega k}^{(\text{I})}}^\star,
\\
\beta_{\Omega k}^{(\text{I})} &\defn -(u_k\s, \wiO)
= -e^{-\frac{\pi\Omega}{a}}\alpha_{\Omega k}^{(\text{I})},
\\
\beta_{\Omega k}^{(\text{II})} &\defn -(u_k\s, \wiiO)
=-e^{-\frac{\pi\Omega}{a}} {\alpha_{\Omega k}^{(\text{I})}}^\star,
\end{aligned}
\end{equation}
with $\omega_k=\sqrt{k^2+m^2}$.

The vacuum state in the inertial frame $\ket{0}_\x{M}$, is defined via:
\begin{equation}   \label{rqi_u_Mvac}
\hat{a}_k \ket{0}_\x{M}=0,
\end{equation}
for all $k$. Let us suppose that in the uniformly accelerated frame this state can be written as:
\begin{equation}        \label{rqi_ansatz}
\ket{0}_\x{M}
=\bigotimes_\Omega \sum\limits_n
C_{\Omega n} \ket{n}_{\x{I}\Omega} \otimes \ket{n}_{\x{II}\Omega},
\end{equation}
where $n$ is a nonnegative integer, and $C_{\Omega n}$ are constants to be determined.

Expressing $\hat{a}_k$ in terms of $\biO$ and $\biiO$ and substituting it into \eqref{rqi_u_Mvac} together with \eqref{rqi_ansatz} one can calculate $C_{\Omega n}$ and consequently one gets:
\begin{equation}   \label{rqi_tms}
\ket{0}_{\x{M}}
=
\bigotimes_\Omega \frac{1}{\cosh r_\Omega} \sum\limits_n
\tanh^n r_\Omega \ket{n}_{\x{I}\Omega} \otimes \ket{n}_{\x{II}\Omega},
\end{equation}
where the factor $r_\Omega$ equals:
\begin{equation}
r_\Omega=\x{atanh} \,\, e^{-\frac{\pi \Omega}{a}}.
\end{equation}
The expression \eqref{rqi_tms} is a tensor product of two-mode squeezed states, where the two modes correspond to regions I and II. 
Let us consider the form of this state, as seen by an observer accelerating to the right, along the right arm of the reference hyperbola. Tracing out the degrees of freedom associated with wedge II, one obtains the reduced state:
\begin{equation}
\begin{aligned}   \label{rqi_thermal}
\x{Tr}_{\x{II}} \, \ket{0}_{\x{M}}\bra{0}
=&
\bigotimes_\Omega \frac{1}{\cosh^2 r_\Omega} 
\sum\limits_n \tanh^{2n} r_\Omega
\ket{n}_{\x{I\Omega}}\bra{n}
\\=&
\bigotimes_\Omega \frac{1}{\sum\limits_n e^{-\frac{2\pi n\Omega}{a}}}
\sum\limits_n e^{-\frac{2\pi n\Omega}{a}}
\ket{n}_{\x{I\Omega}}\bra{n}.
\end{aligned}
\end{equation}
This is a thermal state with a temperature (in natural units):
\begin{equation}   \label{rqi_utemp_a}
T=\frac{a}{2\pi k_\x{B}},
\end{equation}
This is the termperature perceived by an observer travelling along the reference hyperbola. In order to consider other hyperbolae, one needs to take into account the Tolman-Ehrenfest effect~\cite{tolman1,tolman2}. It provides a thermal equilibrium condition for different locations in a given spacetime:
\begin{equation}
\sqrt{g_{00}}\,\,T=\x{const.}
\end{equation}
Recalling \eqref{gr_rmetric} and applying this to the reference hyperbola with $\chi=\frac{1}{a}$ and to another hyperbola at $\chi=\frac{1}{\A}$, one finds that the temperature for a uniformly accelerated observer, accelerating with proper acceleration $\A$ is:
\begin{equation}
T=\frac{\A}{2\pi k_\x{B}},
\end{equation}
which is known as the {\it Unruh temperature}.
An accelerated observer therefore perceives the Minkowski vacuum as a thermal bath at the above temperature, proportional to his or her proper acceleration. This is known as the {\it Unruh effect}. For everyday accelerations of the order of the gravitational acceleration at the Earth's surface, the Unruh temperature is $\sim 10^{-19} \x{K}$ which is less than negligible for all practical purposes as of today. The effect is so weak that it has not been detected yet even in the high-acceleration experiments. An empirical observation of the Unruh effect is a topic of ongoing research~\cite{ObsCirac,ObsRad,unruh_brasil}.

Let us now derive the Unruh effect using a different reasoning. Following~\cite{gerry_knight} the expression \eqref{rqi_tms} can be brought to the form:
\begin{equation}   \label{rqi_s12}
\ket{0}_{\x{M}}
=
\bigotimes_\Omega \,
e^{\tanh r_\Omega (\biO\d\biiO\d - \biO\biiO)}
\ket{0}_\x{R}
\defn
\hat{S}_{\x{I,II}}
\ket{0}_\x{R},
\end{equation}
where $\ket{0}_\x{R}$ is the vacuum state of the uniformly accelerated frame (the {\it Rindler vacuum state}). We have defined here the squeezing operator of the two regions I and II. Using its form, the following identities may be obtained:
\begin{equation}  \label{rqi_sbs}
\begin{aligned}
\hat{S}_{\x{I,II}}\d \biO \hat{S}_{\x{I,II}}
&= \biO \cosh r_\Omega + \biiO\d \sinh r_\Omega,
\\
\hat{S}_{\x{I,II}}\d \biiO \hat{S}_{\x{I,II}}
&= \biiO \cosh r_\Omega + \biO\d \sinh r_\Omega.
\end{aligned}
\end{equation}
Let us now calculate the mean particle number operator corresponding to a given $\Omega$, on this state:
\begin{equation}         \label{rqi_bdb}
\begin{aligned}
{}_\x{M}\bra{0} \biO\d \biX \ket{0}_\x{M}
&={}_\x{R}\bra{0} 
\hat{S}_{\x{I,II}}\d \biO\d \hat{S}_{\x{I,II}} \,
\hat{S}_{\x{I,II}}\d \biX \hat{S}_{\x{I,II}} 
\ket{0}_\x{R}
\\&=
\sinh r_\Omega \sinh r_\Xi \,\,\,{}_\x{R}\bra{0} \biiO \biiX\d \ket{0}_\x{R}
\\&=
\frac{\delta(\Omega-\Xi)}{e^\frac{2\pi\Omega}{a}-1},
\end{aligned}
\end{equation}
where the identities \eqref{rqi_sbs} have been used. Interpreting $\Omega$ as the energy of the mode (in natural units), one may see that this result is a Bose-Einstein distribution with the temperature \eqref{rqi_utemp_a}.

\section{Vacuum entanglement}       \label{sec_vac_ent}

The Minkowski vacuum state, as seen from the uniformly accelerated frame, \eqref{rqi_tms} is pure, but the reduced state of the wedge I \eqref{rqi_thermal} is mixed. This implies that the composite state of the wedges I and II, \eqref{rqi_tms} is entangled. The entanglement exists between the modes from wedge I and II with the same $\Omega$ . Thus, the Minkowski vacuum contains a certain amount of quantum entanglement, and it can be seen when one starts accelerating.

On the other hand, individual Rindler observers cannot detect this entanglement. No Rindler observer from region I can communicate with one from region II and vice versa, and individually each of them registers a thermal state, such as the one shown in \eqref{rqi_thermal}. Only considering the global state \eqref{rqi_tms}, one may note that the thermal noise is quantum-correlated.

The existence of correlations between the corresponding Rindler modes in wedge I and II is reflected in the non-vanishing expectation value:
\begin{equation}
\begin{aligned}
{}_\x{M}\bra{0} \biO \biiX \ket{0}_\x{M}
=
\frac{\delta(\Omega-\Xi)}{2\sinh\frac{\pi\Omega}{a}}.
\end{aligned}
\end{equation}

It has been proposed, that one could extract this entanglement, by swapping it onto a state of two particle detectors~\cite{reznik1,reznik2}. It could be done in order to investigate the structure and the properties of the vacuum entanglement, as well us in order to use it for further purposes, e.g. in quantum information protocols. This topic shall be discussed in detail in Part \ref{part4}.



\pusta
\part{Uniformly accelerating clocks} \label{part2}

\chapter{Ideal clocks - a convenient fiction} \label{chclock}

In this chapter we apply the theory introduced in Part \ref{part1} to show an exemplary consequence of the Unruh effect, i.e. how it affects the rate of ticking of a quantum mechanical clock. In Sec. \ref{ch_clk_motivn} the motivation for the work is provided. Further, in Sec. \ref{ch_clk_model} we discuss the model of the clock and compare the case when it is stationary (Sec. \ref{ch_clk_stat}) and when it is uniformly accelerating (Sec. \ref{ch_clk_acc}). The chapter is concluded in Sec. \ref{ch_clk_conc}. 


\section{Motivation}   \label{ch_clk_motivn}

In Sec. \ref{chgr_general} an immediate consequence of special relativity was stated, that a clock  
moving with a certain velocity, experiences time dilation given by a general formula \eqref{gr_clock_general}. In particular, when observed from an inertial frame, the proper time measured by it amounts to \eqref{gr_clock_m}~\cite{hartle}. 

One property of the formula \eqref{gr_clock_m} is that the time measured by a moving clock depends only on its instantaneous velocity. Let us perform a thought experiment. Consider an ideal clock oscillating along a sinusoidal path: $x(t) = A\sin\omega t$, where $A$ is the amplitude and $\omega$ is the frequency. Then the clock's velocity and acceleration vary according to: $\dot{x}(t) = A\omega\cos\omega t$, $\ddot{x}(t) = -A\omega^2\sin \omega t$. Let us consider a limit of small amplitudes and high frequencies such that $A\to 0$, $A\omega\to 0$ and $A\omega^2 \to \infty$. This is allowed since these parameters are independent, and may approach infinity at ``just the right rates'', in order for the above limits to hold. In such a case the clock would remain at rest, subject to an infinite, rapidly changing acceleration. From the clock postulate discussed in Sec. \ref{chgr_general}, it follows that the rate of the ideal clock should not be affected by this, and it should behave just as if it had no acceleration. Is this prediction realistic?

There exist various types of physical clocks and each type is affected by acceleration in a different way. If we accelerate a pendulum clock, the pendulum might start swinging differently or may even stop swinging if the clock is in free fall. An electric clock might still work in free fall, but most probably will break, once it undergoes a high acceleration when it hits the floor. Further, there also exist atomic clocks, or ones based on light reflecting between two mirrors, such as those analyzed in Einstein's thought experiments. Each of them can in principle get accefected by acceleration in a different way.

Hence, the clock postulate is introduced. A hypothetical device called the ideal clock is proposed, which always measures the proper time, regardless of the trajectory it travels along. The realistic clocks are understood to be approximations of the ideal clock. The latter one would be realised, if one managed to completely remove the influence of any acceleration on the clock. 

In this chapter the Author questions the feasibility of this construct, claiming that it is fundamentally impossible to realise an ideal clock, because of the Unruh effect, which inevitably affects any clock in Nature. The true clocks measure the proper time only approximately, when accelerations they are subject to, are sufficiently low. This also raises a question, whether the proper time (or time in general) should have such a fundamental role in theoretical physics, as it does now, if it cannot be measured by a physical device.

The previous work on the topic includes \cite{clk_andrzej}, where it has been shown that a finite-size clock based on the interference effect of motion along two different paths shows deviations from the ideal clock formula \eqref{gr_clock_m}. In this chapter we study a clock based on a decay time of an unstable particle. The existence of the effect of accelaration on such a decaying particle, is supported by models that have been studied previously~\cite{clk_muller}. Experimentally, no deviation from the ideal clock formula has been found so far for such clocks~\cite{BaileyN,Bailey,Mainwaring,Botermann}. Yet, increasingly higher accelerations are achievable in particle accelerators and it is possible that the Unruh effect may start affecting the results of the experiments performed in the near future. Another work has been developed recently, in which the quantitative size of the effect discussed in this chapter, is calculated, for a realistic scenario of an accelerated muon decay~\cite{clk_roberto}.

Let us now move on to the next section, whereby we set up the model of the clock under investigation.

\section{The model of the clock}   \label{ch_clk_model}

We intend to consider the most fundamental clock possible and from the way it functions, infer general properties of clocks. Since in the most fundamental theory of Nature as of today, i.e. in quantum field theory, all processes are ultimately described in terms of particle interactions, we choose a clock based on the decay time of an unstable particle. If such decay times do get affected by the presence of the Unruh thermal bath, ultimately durations of all known processes should also get affected, and in particular the readings of other types of clocks functioning according to quantum field theory. 

In this work the spacetime is assumed to be $1+1$-dimensional. In quantum field theory particles are considered to be excitations of localized wavepacket modes. In order to simplify the calculations, instead of wavepacket modes we take the ground ($n=1$) mode \eqref{qft_u_cav_m} of a cavity. The cavity is assumed to have proper length $L$ and to extend from $x=\sigma_-$ to $x=\sigma_+$ at $t=0$, where $\sigma_->0$. The quantum field $\hat{\Phi}_\x{cav}$ inside the cavity is taken to be the simplest field possible, i.e. a massless scalar field. We intend to model a decay of a single particle, thus the cavity mode initially is assumed to contain one boson.

In order to facilitate the decay, we also introduce another field $\hat{\Phi}_\x{ext}$, extending over the entire space inside and outside the cavity. The walls of the cavity are assumed to be transparent to it. The field $\hat{\Phi}_\x{ext}$ is chosen to be massive, with mass $m$\footnote{The goal of the work was to demonstrate the effect of the simplest possible case. We found that considering a massless external field leads to infrared divergences, which could have been dealt with using renormalization. However, we believe that these would only obscure the simplicity and the generality of our result, which is independent of technical details such as e.g. choosing a renormalization scheme.}.
It is initially in the vacuum state. 

The system of the two fields is assumed to be subject to the simplest possible coupling\footnote{Using other types of fields and couplings does not change qualitative conclusions presented here. One can also study a semi-classical model of coupling via Unruh-DeWitt Hamiltonian and similar conclusions can be drawn using such an approach.}, described by the following interaction Hamiltonian:
\begin{equation} \label{clk_HInt}
\hat{H}_\x{int}=\lambda\int\text{d}x\,\hat{\Phi}_\x{cav}\,\hat{\Phi}_\x{ext},
\end{equation}
where $\lambda$ is a small coupling strength. As a result of this interaction the state of the cavity field may decay to zero bosons, while the external field may get excited.

We calculate the probability of the decay of the particle in the cavity, to the vacuum state. The degrees of freedom associated with the external field will be traced out to take into account any possible final state of this field. The calculation is performed in the first order perturbation theory for two cases. First, a stationary cavity is considered, and then a uniformly accelerating one. The results are then compared and confronted in the light of the clock postulate. 

\section{Stationary clock}  \label{ch_clk_stat}

We start with the case whereby the cavity (or the clock) is stationary in the inertial frame. In this case we can write the decompositions of our quantum fields according to the formulae \eqref{qft_dec_a} and \eqref{qft_dec_A}:
\begin{equation}   \label{clk_decs}
\begin{aligned}  
\hat{\Phi}_\x{cav} &= 
\sum\limits_{n=1}^\infty \left(
U_n \hat{A}_n + U_n\s \hat{A}_n\d
\right),
\\
\hat{\Phi}_\x{ext} &= 
\int \x{d}k \left(
u_k \hat{a}_k + u_k\s \hat{a}_k\d
\right).
\end{aligned}
\end{equation}

We assume the initial state of the system to be such that the cavity field has one boson in the mode $U_1$ and no excitations in the other modes. Denoting the vacuum state of the stationary cavity field as $\ket{0}_\x{s.c.}$, we may write the initial state of the cavity field as $\hat{A}\d_1 \ket{0}_\x{s.c.}$. The external field is initially in the Minkowski vacuum state $\ket{0}_\x{M}$. 


We are interested in the situation, whereby the final state of the cavity field is the vacuum state $\ket{0}_\x{s.c.}$, while the final state of the external field is any Fock state, i.e.:
\begin{equation}    \label{clk_fock}
\ket{\{\zeta_j\}} \defn \prod\limits_j \frac{1}{\sqrt{\zeta_j!}} \,\hat{a}_j^{\dagger\zeta_j} \ket{0}_\x{M},\\
\end{equation}
where $\{\zeta_j\}$ is a set of nonnegative integers $\zeta_j$ labelled by an index $j$. The states \eqref{clk_fock} form a basis of the space of all possible final states of $\hat{\Phi}_\x{ext}$.

Let us first calculate the amplitude of the decay of interest. In the first order perturbation theory it is:
\begin{equation}
\mathscr{A}_\downarrow=
-i\,\int\limits_0^t\text{d}t' 
\,\,{}_\x{s.c.}\bra{0} \otimes \bra{\{\zeta_j\}} 
\,\, \hat{H}_\x{int} \, \hat{A}\d_1 \,\, 
\ket{0}_\x{s.c.} \otimes \ket{0}_\x{M},
\end{equation}
where $t$ is the interaction time in the inertial frame.
Writing out the Hamiltonian and substituting the field decompositions \eqref{clk_decs} one obtains:
\begin{equation}
\begin{aligned}
\mathscr{A}_\downarrow=&
-i\lambda\,\int\limits_0^t\text{d}t'  \int\text{d}x
\sum\limits_{n=1}^\infty \int \x{d}k
\Big[
\,\,{}_\x{s.c.} \bra{0}\big(U_n \hat{A}_n + U_n\s \hat{A}_n\d \big)\hat{A}\d_1 \ket{0}_\x{s.c.}
\\ &\,\,\qquad\qquad\qquad\qquad\qquad\,\,\,\, \times\,\,
\bra{\{\zeta_j\}}(u_k \hat{a}_k + u_k\s \hat{a}_k\d \big)\ket{0}_\x{M}
\Big].
\end{aligned}
\end{equation}
Further, we trace out the external field degrees of freedom and calculate the probability of the decay:
\begin{equation}
P_\downarrow=
\sum\limits_{\{\zeta_j\}}|\mathscr{A}_\downarrow|^2
=
\lambda^2 \int\x{d}k \left|\gamma_{k1}\right|^2,
\end{equation}
where the following time-integrated overlap is defined:
\begin{equation}
\gamma_{kn} \defn \int\limits_0^t\text{d}t'
\int\text{d}x\,\,
u_k\s U_n.
\end{equation}
Finally, let us substitute in the form of the cavity modes \eqref{qft_u_cav_m} with zero mass and the free field modes \eqref{qft_uk}. The resultant expression is:
\begin{align}   \label{clk_pstat_general}
P_\downarrow=
\frac{4\lambda^2}{L^2}
{\int\limits_{-\infty}^{+\infty}}\text{d}k\,
\frac{\cos^2\left(\frac{kL}{2}\right)
\sin^2\left((\omega_k-\frac{\pi}{L})\frac{t}{2}\right)}
{(\omega_k-\frac{\pi}{L})^2(k^2-\frac{\pi^2}{L^2})^2\,\omega_k},
\end{align}
where $\omega_k=\sqrt{k^2+m^2}$. In the long time limit $t\rightarrow\infty$ the expression $\sin^2 (\eta t)/\eta^2t$ becomes proportional to the Dirac delta function $\delta(\eta)$ (the proof is given in Appendix \ref{app_sec_delta}) and the integration can be approximated by:
\begin{equation} \label{clk_pstat_long}
P_\downarrow=
\frac{4\lambda^2\pi t\cos^2\left(\sqrt{\frac{\pi^2}{L^2}-m^2}\, \frac{l}{2}\right)}
{L^2m^4\sqrt{\frac{\pi^2}{L^2}-m^2}},
\end{equation}
for $\frac{\pi}{L}>m$, and zero otherwise. This has the following physical interpretation: in the limit of $t\rightarrow\infty$ energy in the decay is conserved therefore the transition must be resonant. For the cavity field, $\frac{\pi}{L}$ corresponds to the energy of the mode (in natural units). Moreover, $m$ is the smallest energy of a mode of the external field. If this energy is bigger than the energy of the cavity mode, the transition cannot occur because for infinite interaction times the energy has to be conserved.

Let us now perform an analogous calculation for an accelerated clock.

\section{Uniformly accelerated clock}  \label{ch_clk_acc}

Now we turn to the case whereby the cavity (or the clock) is uniformly accelerating in the inertial frame. 
We assume that the cavity is a rigid body, hence its walls must have different accelerations. Let us specify what is meant by the acceleration of the cavity. We assume that the walls of the cavity follow hyperbolic trajectories within region I, intersecting the $x$-axis at $x=\sigma_-$ and $x=\sigma_+$. The reference hyperbola of the Rindler coordinates is chosen to go through the center of the cavity, such that $a=\frac{2}{\sigma_+ +\sigma_-}$ refers to the average proper acceleration of the cavity. If $L\ll \frac{1}{a}$ then the accelerations corresponding to different positions within the cavity are similar, and one can talk about a single ``acceleration of the cavity'' to a good approximation.

We compute the decay rate of the particle in the cavity in the uniformly accelerated frame, and the calculations are performed in the Rindler conformal coordinates. In this case the decompositions of our quantum fields, according to the formulae \eqref{qft_dec_b_3d} and \eqref{qft_dec_B} are:
\begin{equation}  \label{clk_decs_acc}
\begin{aligned} 
\hat{\Phi}_\x{cav} &= 
\sum\limits_{n=1}^\infty \left(
W_n \hat{B}_n + W_n\s \hat{B}_n\d
\right),
\\
\hat{\Phi}_\x{ext} 
&=\int\limits_0^\infty \x{d}\Omega\,\left( \wiO\biO + \wiO\s\biO\d + \wiiO\biiO + \wiiO\s\biiO\d \right).
\end{aligned}
\end{equation}

Similarly to the case of a stationary cavity, the initial state of the system is assumed to be such that the cavity field has one boson in the mode $W_1$ and no excitations in the other modes. It can be written as $\hat{B}\d_1\ket{0}_\x{a.c.}$, where $\ket{0}_\x{a.c.}$ is the vacuum state of the accelerated cavity. The external field is in the Minkowski vacuum state, which means that it is in the state \eqref{rqi_s12} as seen from the uniformly accelerated frame.

The final state of the system is taken to be vacuum of the cavity field~$\ket{0}_\x{a.c.}$. The final state of the external field may be written as:
\begin{equation}           \label{clk_fock_acc}
\ket{\{\zeta_{\Lambda j}\}} \defn \hat{S}_\x{I,II} \,
\prod\limits_j \frac{1}{\sqrt{\zeta_{\x{I}j}! \zeta_{\x{II}j}!}} \,\hat{b}_{\x{I}j}^{\dagger\zeta_{\x{I}j}} \hat{b}_{\x{II}j}^{\dagger\zeta_{\x{II}j}} \ket{0}_\x{R},
\end{equation}
where $\{\zeta_{\Lambda j}\}$ is a set of nonnegative integers $\zeta_{\Lambda j}$ labelled by indices $j$ and $\Lambda\in\{\x{I},\x{II}\}$. The states \eqref{clk_fock_acc} form a basis of the space of all possible final states of $\hat{\Phi}_\x{ext}$.

Writing down the decay amplitude in the first order perturbation theory, we get:
\begin{equation}
\mathscr{A}_\downarrow=
-i\,\int\limits_0^\tau\text{d}\tau' 
\,\,{}_\x{a.c.}\bra{0} \otimes \bra{\{\zeta_{\Lambda j}\}} 
\,\, \hat{S}_\x{I,II}\d \, \hat{H}_\x{int} \, \hat{S}_\x{I,II} \, \hat{B}\d_1 \,\,
\ket{0}_\x{a.c.} \otimes \ket{0}_\x{R},
\end{equation}
where $\tau$ is the proper time of the interaction in the Rindler coordinates. The description is made in the accelerated frame, therefore we use the Hamiltonian $\hat{H}_\x{int}$ in a constant $\tau$ foliation.

Writing out the Hamiltonian, substituting the field decompositions \eqref{clk_decs_acc}, performing the sum over the cavity modes and making use of the identities \eqref{rqi_sbs} for the external field terms, one obtains:
\begin{align}\nonumber
\mathscr{A}_\downarrow=
-i\lambda \int\text{d}\tau' \int\text{d}\xi \int &\x{d}\Omega\,\,
\Big\{ \bra{\{\zeta_{\Lambda j}\}}
\big[(\cosh r_\Omega \,\biO + \sinh r_\Omega \,\biiO\d) \wiO \\
+\,&(\cosh r_\Omega \,\biO\d + \sinh r_\Omega \,\biiO) \wiO\s \\ \nonumber
+\,&(\cosh r_\Omega \,\biiO + \sinh r_\Omega \,\biO\d) \wiiO \\ \nonumber
+\,&(\cosh r_\Omega \,\biiO\d +\sinh r_\Omega \,\biO) \wiiO\s \big]
\ket{0}_\x{R} W_1\Big\}.
\end{align}
Tracing out the degrees of freedom of the external field, we get:
\begin{equation}
P_\downarrow 
=\sum\limits_{\{\zeta_{\Lambda j}\}}
|\mathscr{A}_\downarrow|^2
=
\lambda^2 \int\x{d}\Omega
\left[\left|\gamma_{\Omega 1}\right|^2 + \sinh^2r_{\Omega}
\left(\left|\gamma_{\Omega 1}\right|^2 + \left|\bar{\gamma}_{\Omega 1}\right|^2\right)
\right],
\end{equation}
with the following definitions:
\begin{equation}
\begin{aligned}
\gamma_{\Omega n}&\defn
\int\limits_0^\tau \text{d}\tau' \int\text{d}\xi \, \wiO\s W_n,
\\
\bar{\gamma}_{\Omega n}&\defn
\int\limits_0^\tau \text{d}\tau' \int\text{d}\xi \, \wiO W_n.
\end{aligned}
\end{equation}
We now evaluate the probability of the decay for the cavity modes \eqref{qft_Wn_cav_r_m0} and the external field modes \eqref{qft_wi_m_rc}-\eqref{qft_wii_m_rc}. This yields the following expression:
\begin{align}  \label{clk_pacc_general}
P_\downarrow=&
\frac{4\lambda^2}{a\pi^3}
{\int\limits_{0}^{+\infty}}\text{d}\Omega\,\,
\sinh\left(\frac{\pi\Omega}{a}\right)
\left|\,
{\int\limits_{\xi_-}^{\xi_+}}\text{d}\xi\,
K_{\frac{i\Omega}{a}}\left(\frac{m}{a}e^{a\xi}\right)
\sin\left(\omega_1\left(\xi-\xi_-\right)\right)
\right|^2 
\\\times\,& \nonumber
\Bigg[
\frac{\sin^2\left[\left(\Omega-\omega_1\right)\frac{\tau}{2}\right]}
{\left(\Omega-\omega_1\right)^2}
+
\frac{1}{e^{\frac{2\pi\Omega}{a}}-1}
\Bigg(
\frac{\sin^2\left[\left(\Omega-\omega_1\right)\frac{\tau}{2}\right]}
{\left(\Omega-\omega_1\right)^2}
+
\frac{\sin^2\left[\left(\Omega+\omega_1\right)\frac{\tau}{2}\right]}
{\left(\Omega+\omega_1\right)^2}
\Bigg)\Bigg],
\end{align}
where $\omega_n$ is defined below the Eq. \eqref{qft_Wn_cav_r_m0} and $\xi_\pm\defn\frac{1}{a}\log a\sigma_\pm$ are the Rindler conformal positions of the cavity walls. Let us now compute the long time limit $\tau\rightarrow\infty$ of \eqref{clk_pacc_general}:
\begin{equation}   \label{clk_pacc_long}
P_\downarrow=\frac{\lambda^2\tau e^{\frac{\pi\omega_1}{a}}}{a \pi^2}
\left|\,\,
{\int\limits_{\xi_-}^{\xi_+}}\text{d}\xi\,
K_{\frac{i\omega_1}{a}}\left(\frac{m}{a}e^{a\xi}\right)
\sin\left(\omega_1\left(\xi-\xi_-\right)\right)
\right|^2.
\end{equation}

If the clock under our consideration was an ideal clock, the result would be the same as \eqref{clk_pstat_long} with $t$ replaced with $\tau$. It is not the case. 

Let us now ask the question if we retrieve the form \eqref{clk_pstat_long} in the low acceleration limit. Taking $a\rightarrow 0$, we obtain the form \eqref{clk_pstat_long} with superimposed rapid oscillations with respect to $a$. We average the result over acceleration around the chosen value $a$, which corresponds to a finite uncertainty about the value of acceleration. We perform this averaging for $\frac{\pi}{l}>M$ \cite{dunster, nist} in the limit of small $a$, keeping the cavity size $L$ fixed, to obtain: 
\begin{equation} \label{clk_pacc_long_asmall}
P_\downarrow=
\frac{4\lambda^2\pi\tau\cos^2\left(\sqrt{\frac{\pi^2}{L^2}-m^2}\frac{L}{2}\right)}
{L^2m^4\sqrt{\frac{\pi^2}{L^2}-m^2}}.
\end{equation}
This is of the same form as \eqref{clk_pstat_long}, which shows that the clock that we investigated is an ideal clock only for small accelerations.

\section{Conclusions}  \label{ch_clk_conc}

We have shown that the ticking rate of a clock based on a particle decay, is altered in a non-trivial way by the Unruh effect. It seems to be impossible to build a clock that could be completely isolated from the Unruh thermal bath, and hence no clock would correctly measure the proper time along arbitary high-acceleration trajectories. An ideal clock remains a fiction, an idealization that is convenient to use in thought experiments, but not achievable physically. The proper time therefore loses its operational meaning for arbitrarily high accelerations. 

An objection might be raised that perhaps one could calculate the correction due to the Unruh effect, and subtract it from the clock's readout to obtain the proper time for any trajectory. While it is in principle true, a knowledge of the full trajectory would be required, but that would also allow for calculating the proper time itself. This is not how one expects a clock to function. A clock should measure time without using any information about the trajectory that it moves along.

Furthermore, we have only considered a uniformly accelerated trajectory and a particular interaction Hamiltonian. This work is a proof of principle. We have not proven that the effect occurs for every accelerated trajectory and for every interaction Hamiltonian. Yet, if the effect occurs for the simple case studied here, there is no reason to expect that it would not occur in general.

In this work the description of the decay of a physical particle was highly simplified. We have considered a $1+1$-dimensional scenario and the fields $\hat{\Phi}_\x{cav}$ and $\hat{\Phi}_\x{ext}$ were taken to be real and scalar. It is impossible to draw from our results any prediction about the magnitude of the acceleration of a physical decaying particle, for which the effect might be expected to be measurable. Further research has been performed recently, that aims at analyzing a more realistic process and producing a quantitative prediction~\cite{clk_roberto}. In the aforementioned work a decay is investigated, of a muon accelerated along a circle by a constant magnetic field. The calculation is performed in $3+1$ dimensions and the particles are treated as scalar fields for simplicity. The results show a deviation from the ideal clock formula. However, it becomes visible only at accelerations of the order of $10^{28}g$. Let us now move on to the next part of the thesis.



\part{Effect of uniform acceleration on localized two-mode Gaussian quantum states} 
\label{part3}

This part of thesis concerns the question how quantum states given in an inertial frame are seen by uniformly accelerated observers. Sec. \ref{chrqi_sec_unruh} discusses the transformation of the Minkowski vacuum state under such a change of frame of reference. Studying the alteration of a general state under the Bogolyubov transformation shown in Sec. \ref{chrqi_sec_unruh} is a complicated task. 

An alternative approach is investigating how particle detectors get excited, while travelling along various trajectories. A {\it particle detector}~\cite{blhu_dets} is typically a point-like system with a certain number of energy levels. It interacts with a quantum field and can get excited by this interaction. It is a model of a physical detector, whose state is altered by the interaction with the field. One may subsequently measure this state and gain information about the state of the field. A particle detector may move along various trajectories, and it only interacts with the quantum field at its position. The most frequently used model of a particle detector is the Unruh-DeWitt detector~\cite{birrell_davies,unruh_effect,dewitt}. The advantage of its usage, is clearly the fact that one does not need to perform calculations in a non-inertial frame of reference. On the other hand, the results may depend on a chosen type of coupling between the field and the detector. Particle detectors will be discussed in more detail in Sec. \ref{sec_tri_mot}.

The approach employed in this work is inspired by the localized projective measurement formalism presented in~\cite{andrzej_loc_proj}. We restrict our attention to Gaussian states, thus it suffices to transform the vector of first moments and the covariance matrix of a given state, to a uniformly accelerated frame of reference. The creation and annihilation operators in the inertial frame of reference are expressed as linear combinations of the creation and annihilation operators in a uniformly accelerated frame of reference via a Bogolyubov transformation. Then one may calculate the expectation values necessary to write down the resultant vector of first moments and the covariance matrix. Hence, one may obtain equations of a Gaussian quantum channel, which for any input Gaussian state in the inertial frame, may produce an output state, as perceived by a uniformly accelerated observer. This approach allows to obtain a result without resorting to perturbation theory. Also, the results do not depend on any particular field-detector coupling model. However, the properties of a physical detector are encoded in the chosen observer modes, which are assumed to be sufficiently localized in both the position and the frequency space. One expects such features from realistic detectors, because they have finite size and are sensitive to a finite window of frequencies. 

In this part of the thesis we further limit our attention to two-mode states. This allows us to consider the effects of acceleration on the input state, and in particular on the entanglement between the two accelerating modes. Previously, the case of one accelerated and one inertial mode was investigated in~\cite{channel_andrzej}, but only an input squeezed state was analyzed. Here we extend this work to allow for any desired input Gaussian state. We further apply the general framework to the Minkowski vacuum state, in order to draw conclusions about the entanglement present in the quantum vacuum, and its dependence on parameters characterizing the modes, including their proper accelerations and their relative distance. It should be noted that the approach can be straightforwardly generalized to an arbitrary number of modes~\cite{channel_kacper}.

Let us now outline the cases of the framework that are considered in this part of the thesis. We begin with the real scalar field. In Chapter \ref{chb1d} the $1+1$-dimensional case is discussed, with the possibility of a relative shift of the Rindler wedges. In Chapter \ref{chb3d} we present the $3+1$-dimensional case, including a proposal to consider the dependence of the vacuum entanglement on the relative angle between the observers' paths. Finally, in Chapter~\ref{chf1d} we also discuss the work on a Dirac spinor field in $1+1$-dimensional spacetime.


\chapter{Real scalar field in 1+1-dimensional spacetime} \label{chb1d} 

In this chapter we develop the framework for calculating how a given Gaussian state of two inertial modes of a real scalar field, is perceived by two accelerated observers, who have access to another set of two localized modes that uniformly accelerate. The $1+1$-dimensional case of the problem is studied. The framework is finally applied to investigate the dependence of the bipartite entanglement of the vacuum on parameters characterizing the oberservers' modes. In Sec. \ref{sec_opus_b1_mod_rind}, we introduce the coordinate system that allows for treating the distance between the observers and their accelerations independently. In Sec. \ref{sec_opus_b1_setup} all the assumptions about the problem are outlined and the quantum channel transforming the state of the inertial modes into the state of the observers' modes, is defined. Further, in Sec. \ref{sec_opus_M} and~\ref{sec_opus_b1_N} the two matrices that characterize the channel, are derived. In Sec. \ref{sec_opus_a} we show that the channel does not depend on $a$ - the unphysical parameter of the Rindler transformation. In Sec. \ref{opus_modes} we choose the modes suitable to study the entanglement of the vacuum, and finally in Sec. \ref{sec_opus_b1_results} the results of this study are discussed.


\section{Modified Rindler coordinates}          \label{sec_opus_b1_mod_rind}

In this work we consider four localized wavepacket modes of a massive, real scalar field. $\phi_\x{I}$ and $\phi_\x{II}$ are assumed to be inertial and are observed at ${t=0}$ (the time $t$ always refers to the Minkowski frame) by two accelerating observers, having access to modes $\psi_\x{I}$ and $\psi_\x{II}$ respectively. Let us begin with introducing the coordinate system which they are described in. At the time of the observation $t=0$, the modes $\phi_\x{I}$ and $\psi_\x{I}$ are taken to be localized within the region I of the Rindler coordinates, and $\phi_\x{II}$ and $\psi_\x{II}$ are localized within region II. In this section we however generalize slightly what is meant by region I and II.
 
\begin{figure}
\centering
\includegraphics[width=0.6\linewidth]{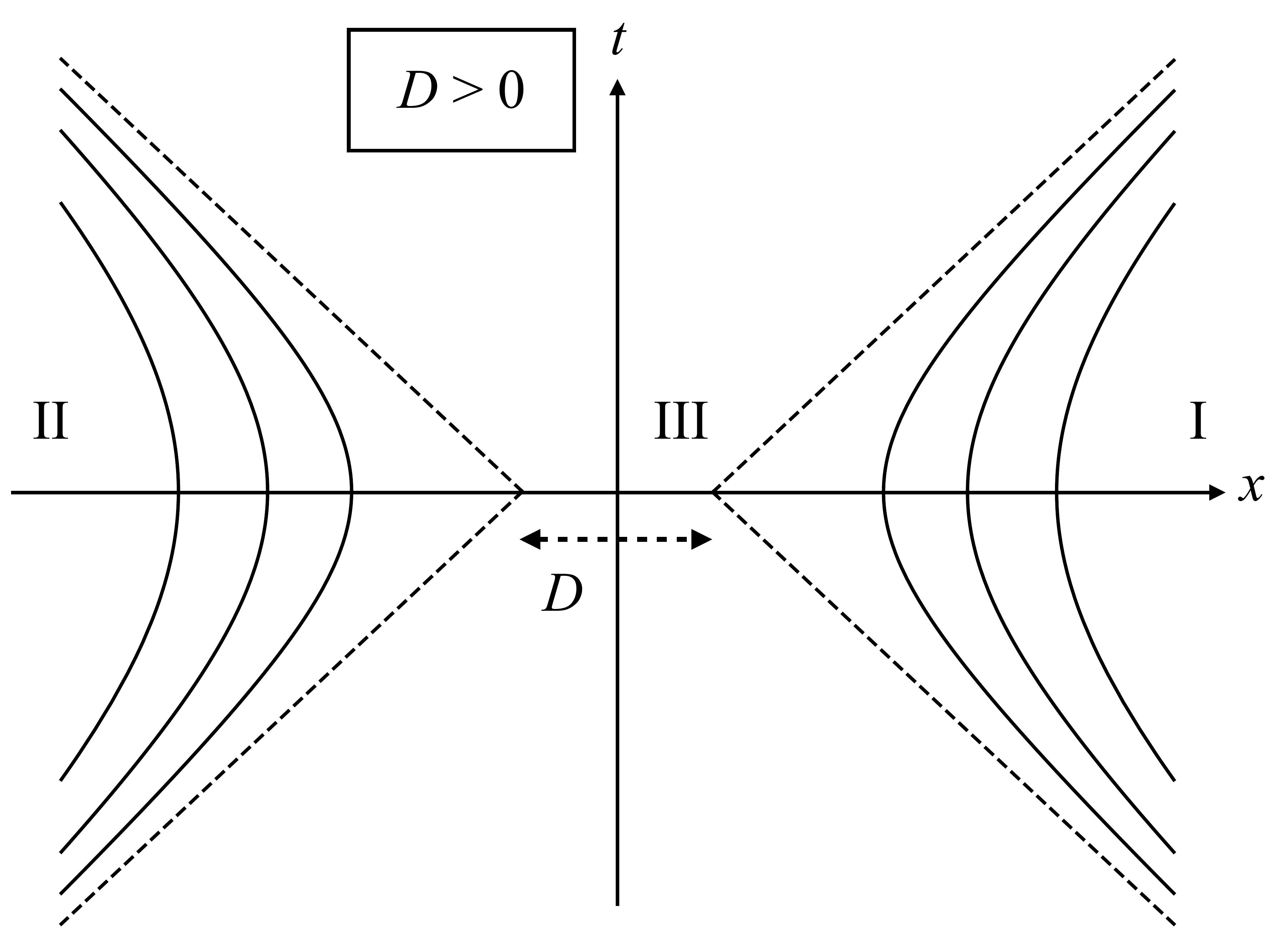}
\caption{\small{The modified Rindler coordinate system with $D>0$.}}\label{fig_opus1d_Dp}
\end{figure}

\begin{figure}
\centering
\includegraphics[width=0.6\linewidth]{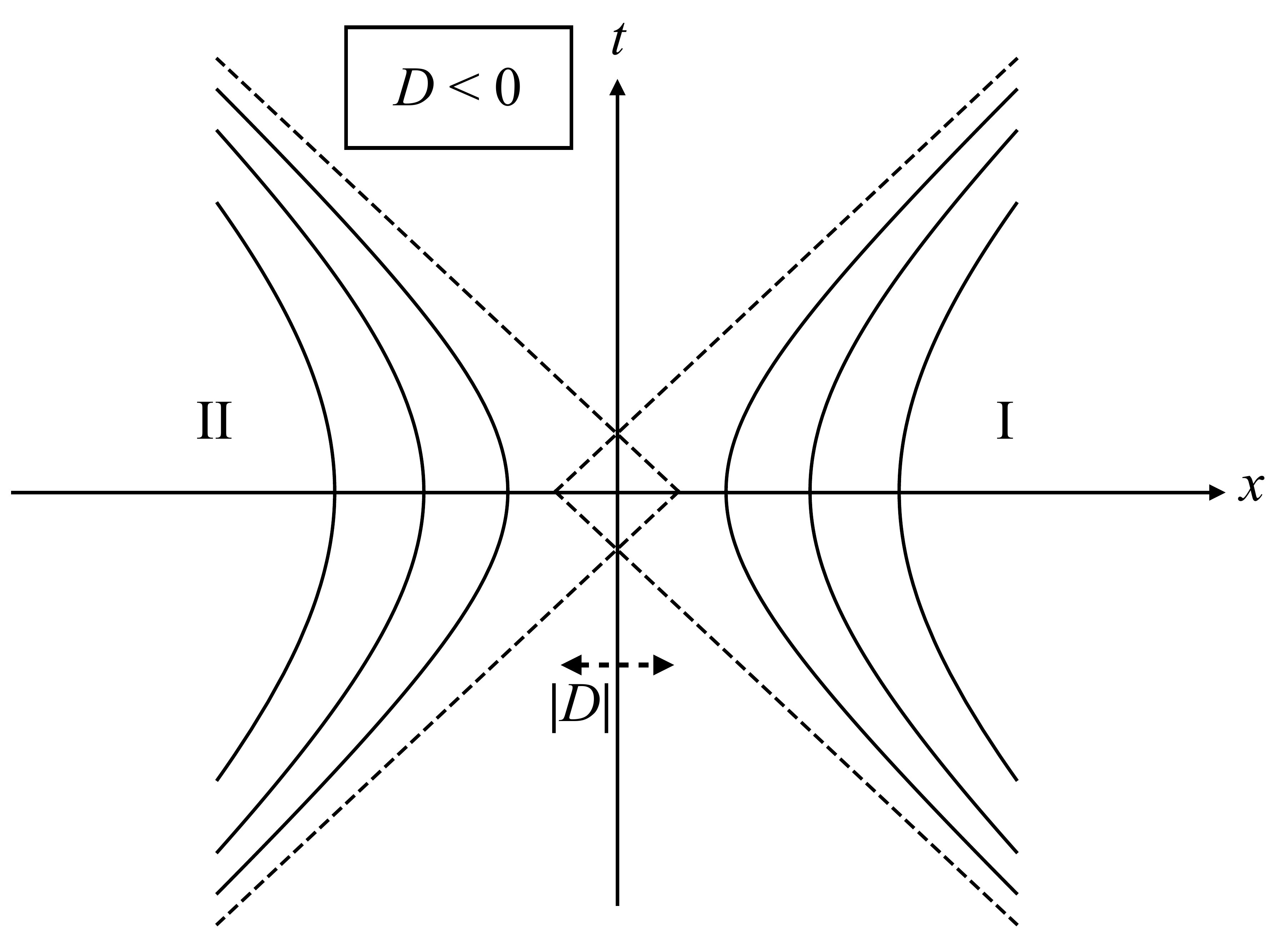}
\caption{\small{The modified Rindler coordinate system with $D<0$.}}
\label{fig_opus1d_Dn}
\end{figure}

When we use the standard Rindler coordinate chart introduced in Sec. \ref{sec_gr_rind}, we are allowed only to consider observers described with the family of hyperbolae \eqref{gr_rind_hyp}. We may only vary the proper acceleration $\A$, and the distance between the arms of a given hyperbola is determined by the value of $\A$. In order to allow for a general respective displacement of the accelerated observers~\cite{dnon0_nick}, we move apart the wedges I and II of the $1+1$-dimensional Rindler chart by distance $D$, where $D>0$. The region between the wedges is named III. The resultant {\it modified Rindler coordinate} chart is shown in Fig. \ref{fig_opus1d_Dp}. If we let $D<0$, region III does not exist and we obtain the chart presented in Fig. \ref{fig_opus1d_Dn}.

The corresponding modified Rindler transformation from the Minkowski coordinates to the coordinates covering wedges I and II or the modified Rindler coordinate system, takes the following form:
\begin{equation} \label{opus_b1_mod_Rindler}
\begin{aligned}
t &= \pm\chi\sinh a\eta,
\\
x &= \pm\chi\cosh a\eta \pm \frac{D}{2},
\end{aligned}
\end{equation}
where $+$ refers to the coordinates covering region I, and $-$ refers to the coordinates covering region II, and $D$ may be positive or negative. We are not interested in this work in coordinates covering region III for $D>0$.

When the Rindler coordinates are modified, the mode decomposition \eqref{qft_dec_b_3d} is also altered. For any $D$ let us write:
\begin{align} \label{opus_dec_b_1d}
\hat{\Phi} 
=\int\limits_0^\infty \x{d}\Omega\,\left( \wiO\biO + \wiO\s\biO\d + \wiiO\biiO + \wiiO\s\biiO\d \right) + \hat{\Phi}_\x{III}(D).
\end{align}
When $D>0$, the region III arises, and it has to be covered with certain modes for the basis to be complete. The operator $\hat{\Phi}_\x{III}(D)$ corresponds to these modes. When $D<0$ it compensates for the fact that the wedges overlap and the basis becomes overcomplete. The explicit form of $\hat{\Phi}_\x{III}(D)$ is not relevant in our framework, except for the case when $D=0$, in which $\hat{\Phi}_\x{III}(D)=0$. However, it should still be kept in mind, that when $D\neq 0$, both wedges are shifted, thus the Rindler modes \eqref{qft_wi_m_r}-\eqref{qft_wii_m_r} and creation and annihilation operators apearing in the formula \eqref{opus_dec_b_1d} are modified.

\begin{figure}
\centering
\includegraphics[width=0.6\linewidth]{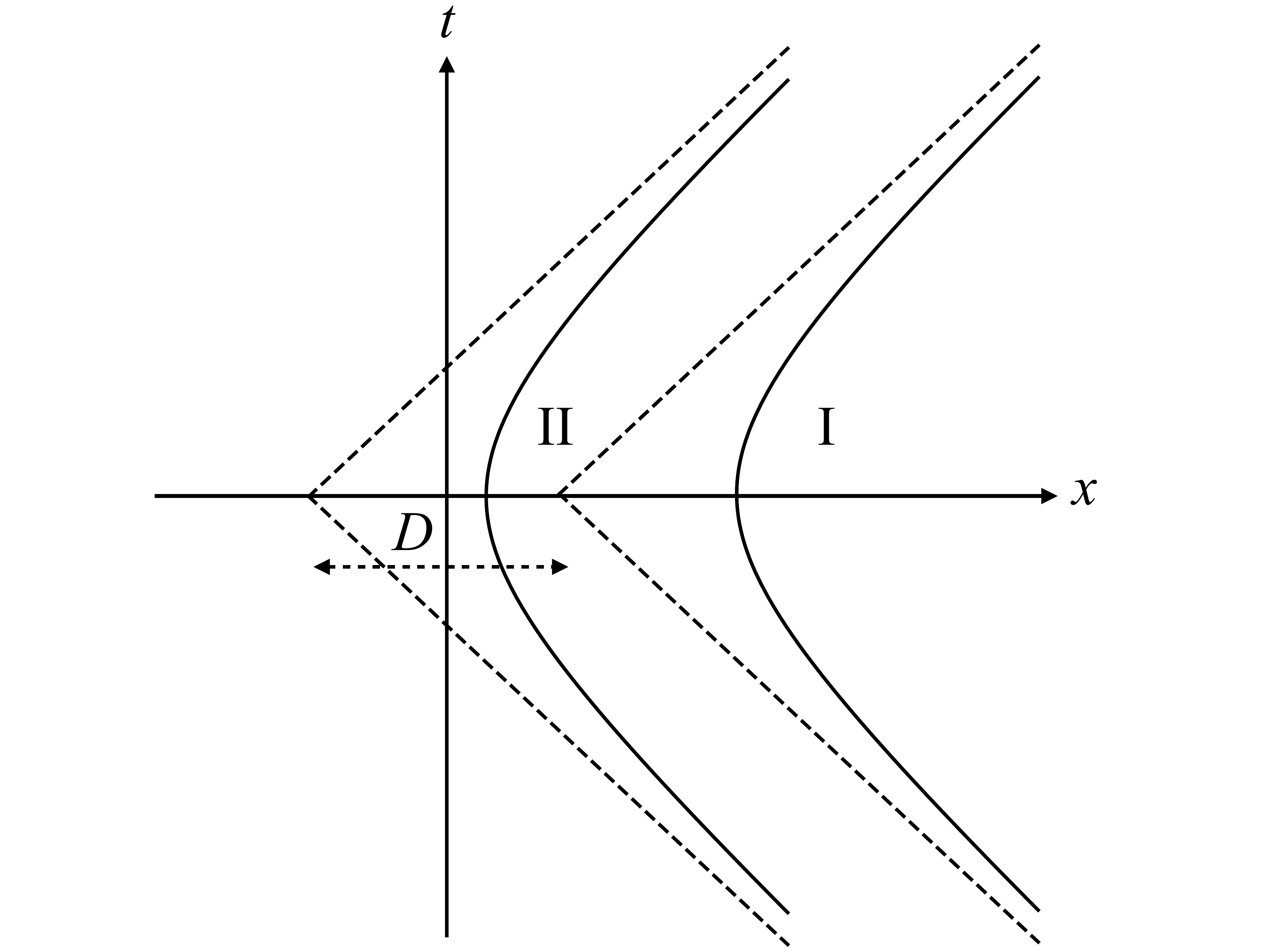}
\caption{\small{The modified Rindler coordinate system with $D>0$, whereby the proper accelerations of the observers are in the same direction.}}\label{fig_opus1d_Para}
\end{figure}

Having the freedom of varying $D$ and $\A$ independently, allows for considering two observers at an arbitrary distance and with arbitrary accelerations in opposite directions. We may further modify the Rindler coordinate system to make it possible to study a case, whereby the accelerations of the observers are parallel. This situation, shown in Fig. \ref{fig_opus1d_Para}, involves a mirror reflection of the wedge~II.

The corresponding modified Rindler coordinate transformation, takes the following form:
\begin{equation} \label{opus_b1_mod_Rindler_par}
\begin{aligned}
t &= \chi\sinh a\eta,
\\
x &= \chi\cosh a\eta \pm \frac{D}{2},
\end{aligned}
\end{equation}
where $+$ refers to the coordinates covering region I, and $-$ refers to the coordinates covering region II, and $D>0$ without loss of generality. In this case we may decompose the field into the Rindler modes associated with wedge I, wedge II and a basis of modes covering the region $x<-\frac{D}{2}$ which is arbirary and later is traced over. Because the wedges I and II overlap, this basis is overcomplete.

We may now move on to the discussion of the setup under consideration.

\section{The setup and the channel}        \label{sec_opus_b1_setup}

We work with a $1+1$D massive Klein-Gordon field $\hat{\Phi}$, with mass $m$\footnote{The entire calculation may be repeated for a massless field, but it leads to an infrared divergence, which would have to be removed by renormalization. We chose to avoid such considerations, not to obscure the main purpose of this work.}. The mass~$m$ is assumed to be strictly positive for the reasons explained in Sec.~\ref{sec_qft_conf}. We begin with a certain Gaussian state of two localized inertial modes $\phi_\x{I}$ and~$\phi_\x{II}$, with associated annihilation operators $\hat{f}_\x{I}$ and $\hat{f}_\x{II}$. The mode $\phi_\x{I}$ at $t=0$ is localized within region I, and $\phi_\x{II}$ at $t=0$ is localized within region II. We assume that the modes are orthogonal\footnote{We exclude the situation whereby the wavepackets e.g. occupy the same position within the overlapping part of the wedges for $D<0$.}, hence:
\begin{align}         \label{opus_b1_ass1}
(\phi_\x{I},\phi_\x{II}^{(\star)}) &= 0,
\\
[\hat{f}_\x{I} , \hat{f}_\x{II}^{(\dagger)}] &= 0,
\end{align}
where ${}^{(\star)}$ denotes the presence or absence of complex conjugation, and ${}^{(\dagger)}$~denotes the presence or absence of Hermitian conjugation. 

Futhermore, we also consider two localized accelerating modes $\psi_\x{I}$ and $\psi_\x{II}$, with associated annihilation operators $\hat{d}_\x{I}$ and $\hat{d}_\x{II}$. The mode $\psi_\x{I}$ is at all times localized within region I, and $\psi_\x{II}$ is at all times localized within region II. They are also assumed to be orthogonal, thus:
\begin{align}
(\psi_\x{I},\psi_\x{II}^{(\star)}) &= 0,
\\           \label{opus_b1_comm_dd}
[\hat{d}_\x{I} , \hat{d}_\x{II}^{(\dagger)}] &= 0.
\end{align}
Physically, these might model two localized detector modes, which observe the two inertial modes localized at different positions in space. The state of the accelerating modes is under investigation in this work. The trajectories of the accelerating modes may be arbitrary arms of hyperbolae of the modified Rindler coordinate chart, but we are interested in a situation, whereby at $t=0$ the mode $\psi_\x{I}$ occupies the same location in space as $\phi_\x{I}$ and $\psi_\x{II}$ occupies the same location in space as $\phi_\x{II}$. We also demand:
\begin{align}            \label{opus_b1_noov12}
(\psi_\x{I},\phi_\x{II}^{(\star)}) = (\psi_\x{II}, \phi_\x{I}^{(\star)}) &= 0,
\\                       \label{opus_b1_ass6}
[\hat{f}_\x{I} , \hat{d}_\x{II}^{(\dagger)}] = [ \hat{f}_\x{II} , \hat{d}_\x{I}^{(\dagger)}] &= 0.
\end{align}

We can decompose the field $\hat{\Phi}$ in the orthonormal bases corresponding to the wavepacket modes, as follows:
\begin{equation}       \label{opus_dec_wavepackets}
\hat{\Phi} =
\sum\limits_n \phi_n \hat{f}_n + \phi_n\s \hat{f}_n\d =
\sum\limits_n \psi_n \hat{d}_n + \psi_n\s \hat{d}_n\d,
\end{equation}
where $n$ may be equal to I, II or may label the remaining modes of the orthonormal basis, which in principle may be constructed. Another two decompositions of the field, are with respect to the Minkowski modes \eqref{qft_dec_a} and with respect to the modified Rindler modes \eqref{opus_dec_b_1d}.

Computing Bogolyubov transformations between the modes of the decompositions \eqref{qft_dec_a}, \eqref{opus_dec_b_1d} and \eqref{opus_dec_wavepackets}, yields the following expressions for the annihilation operators:
\begin{align}      \label{opus_b1_fina_n}
\hat{f}_\Lambda &= \int\x{d}k
\left[
(\phi_\Lambda , u_k) \,\hat{a}_k +
(\phi_\Lambda , u_k\s) \,\hat{a}_k\d
\right]
,
\\                 \label{opus_b1_dinb_n}
\hat{d}_\Lambda &= \int\x{d}\Omega
\left[
(\psi_\Lambda , w_{\Lambda\Omega}) \,\bLO +
(\psi_\Lambda , w_{\Lambda\Omega}\s) \,\bLO\d
\right],
\end{align}
where throughout this part of the thesis $\Lambda\in\{\x{I},\x{II}\}$.
This however would imply a possible detection of particles in the vacuum of the rest frame of a given detector. Consider the mean particle number operator corresponding to~$\phi_\Lambda$ on the Minkowski vacuum state, and the mean particle number operator corresponding to $\psi_\Lambda$ on the Rindler vacuum state:
\begin{align}
{}_\x{M}\bra{0} \hat{f}_\Lambda\d \hat{f}_\Lambda \ket{0}_\x{M}
=&
\int\x{d}k\,
\left|
(\phi_\Lambda , u_k\s)
\right|^2,
\\
{}_\x{R}\bra{0} \hat{d}_\Lambda\d \hat{d}_\Lambda \ket{0}_\x{R}
=&
\int\x{d}\Omega\,
\left|
(\psi_\Lambda , w_{\Lambda\Omega}\s)
\right|^2.
\end{align}
One would expect these mean particle number operators to vanish, because physically an observer should not detect particles in the vacuum of his or her rest frame. We thus demand that the wavepackets have no overlap with negative frequency modes associated with their rest frames, i.e.:
\begin{gather}
(\phi_\Lambda , u_k\s) = 0,
\\
(\psi_\Lambda , w_{\Lambda\Omega}\s) = 0.
\end{gather}
Taking this into account in \eqref{opus_b1_fina_n}-\eqref{opus_b1_dinb_n} yields the following expressions for the annihilation operators:
\begin{align}     \label{opus_b1_fina}
\hat{f}_\Lambda &= \int\x{d}k\,
(\phi_\Lambda , u_k) \,\hat{a}_k,
\\            \label{opus_b1_dinb}
\hat{d}_\Lambda &= \int\x{d}\Omega\,
(\psi_\Lambda , w_{\Lambda\Omega}) \,\bLO.
\end{align}

From \eqref{opus_dec_wavepackets} one also obtains the Bogolyubov transformation between the creation and annihilation operators corresponding to the wavepackets:
\begin{equation}         \label{opus_bogo_fd}
\hat{d}_\Lambda=
\sum_n\left[ (\psi_\Lambda , \phi_n) \hat{f}_n +
(\psi_\Lambda , \phi_n\s) \hat{f}_n\d \right].
\end{equation}  
In order to construct the vector of first moments and the covariance matrix of the composite state of the two observers' modes, we need to perform a Bogolyubov transformation from the inertial to the uniformly accelerated frame, and tracing out the modes with $n\notin\{\x{I},\x{II}\}$. These operations preserve Gaussianity, i.e. if the input state of the transformation was Gaussian, so is the output state. A general form of a quantum channel describing such a transformation, which is a Gaussian quantum channel, is of the form~\cite{holevo_werner_channel}:
\begin{align}           \label{opus_b1_channel_x}
\vec{X}^{(d)} &=  M \vec{X}^{(f)}, 
\\                      \label{opus_b1_channel_s}
\boldsymbol\sigma^{(d)} &=  M\boldsymbol\sigma^{(f)} M^T+N,
\end{align}
where $\vec{X}^{(f)}$ and $\boldsymbol\sigma^{(f)}$ are the vector of first moments and the covariance matrix of the inertial wavepackets' state respectively, and similarly $\vec{X}^{(d)}$ and $\boldsymbol\sigma^{(d)}$ are the vector of first moments and the covariance matrix of the observers' state respectively. 
Since we work with two-mode states, the matrices $M$, $N$, $\boldsymbol\sigma^{(f)}$ and $\boldsymbol\sigma^{(d)}$ have dimensions $4\times 4$. The covariance matrices have to be real and symmetric, thus $M$ and $N$ are real and $N=N^T$. In general, the transformed vector of first moments could include a constant displacement vector, but it vanishes in our case, because a Bogolyubov transformation due to motion is always homogeneous. 

The matrix $N$ is the noise matrix describing the noise due to the fact that the accelerating modes with $n\notin\{\x{I},\x{II}\}$, which do mix with the inertial modes with $n\in\{\x{I},\x{II}\}$ via the Bogolyubov transformation \eqref{opus_bogo_fd}, are traced out. The interpretation of the matrix $M$ is discussed in Sec. \ref{sec_opus_M}.

If we derive expressions for $M$ and $N$ matrices, in principle we could calculate an output (observed) state, for any input Gaussian state. Let us proceed with this derivation.

\section{Derivation of the $M$ matrix}       \label{sec_opus_M}

Let us start by deriving the $M$ matrix using Eq.~\eqref{opus_b1_channel_x}. The left hand side vector of first moments, from \eqref{gqm_quad}, can be written as follows:
\begin{equation}         \label{opus_b1_MXd}
\vec{X}^{(d)} 
=
\begin{pmatrix}
1 & 1 & 0 & 0  \\
-i & i & 0 & 0  \\
0 & 0 & 1 & 1 \\
0 & 0 & -i & i 
\end{pmatrix}
\begin{pmatrix}
\langle\hat{d}_\text{I}\rangle \\
\langle\hat{d}_\text{I}\d\rangle \\
\langle\hat{d}_\text{II}\rangle \\
\langle\hat{d}_\text{II}\d\rangle 
\end{pmatrix},
\end{equation}
where it is understood that the expectation values of the ladder operators are taken in the state of the inertial wavepackets. Then, taking into account the assumption \eqref{opus_b1_noov12}, the Bogolyubov transformation can be written in the matrix form as:
\begin{equation}        \label{opus_b1_Mbogo}
\begin{pmatrix}
\langle\hat{d}_\text{I}\rangle \\
\langle\hat{d}_\text{I}\d\rangle \\
\langle\hat{d}_\text{II}\rangle \\
\langle\hat{d}_\text{II}\d\rangle 
\end{pmatrix}
=
\begin{pmatrix}
\alpha_{\text{I}} & -\bei & 0 & 0 \\
-\bei\s & \ali\s & 0 & 0 \\
0 & 0 & \alii & -\bei  \\
0 & 0 & -\bei\s & \alii\s  \\
\end{pmatrix}
\begin{pmatrix}
\langle\hat{f}_\text{I}\rangle \\
\langle\hat{f}_\text{I}\d\rangle \\
\langle\hat{f}_\text{II}\rangle \\
\langle\hat{f}_\text{II}\d\rangle 
\end{pmatrix},
\end{equation}
where we define:
\begin{equation}        \label{opus_Mbogo_defn}
\begin{aligned}
\ali& \defn (\psi_\x{I} , \phi_\x{I}),  \\
\bei& \defn -(\psi_\x{I} , \phi_\x{I}\s),\\
\alii& \defn (\psi_\x{II} , \phi_\x{II}), \\
\beii& \defn -(\psi_\x{II} , \phi_\x{II}\s). 
\end{aligned}
\end{equation}
To express the right hand side vector in \eqref{opus_b1_Mbogo} in terms of $\vec{X}^{(f)}$, one needs to apply an inverse transformation of the one performed in \eqref{opus_b1_MXd}. Combining this, \eqref{opus_b1_MXd} and \eqref{opus_b1_Mbogo}, we infer the form of the $M$ matrix:
\begin{equation}         \label{opus_M}
M=
\left(\begin{matrix}
\text{Re}(\ali-\bei) & -\text{Im}(\ali+\bei) & 0 & 0 \\
\text{Im}(\ali-\bei) & \text{Re}(\ali+\bei) & 0 & 0 \\
0 & 0 & \text{Re}(\alii-\beii) & -\text{Im}(\alii+\beii) \\
0 & 0 & \text{Im}(\alii-\beii) & \text{Re}(\alii+\beii) 
\end{matrix}\right).
\end{equation}
The $M$ matrix has no dependence on $D$, because the transformation acts on the two wedges independently. Thus, this derivation holds for an arbitrary~$D$. 

Physically, the matrix $M$ describes the mismatch between the modes $\phi_\x{I}$ and $\psi_\x{I}$ and the mismatch between the modes $\phi_\x{II}$ and $\psi_\x{II}$, meaning that the corresponding modes do not perfectly overlap in any physical system whereby one of them is stationary, and the corresponding one is acclerating.
Let us now calculate the $N$ matrix.

\section{Derivation of the noise matrix}         \label{sec_opus_b1_N}

To compute the noise matrix, without loss of generality we substitute the vacuum initial state $\boldsymbol\sigma^{(f)} = \I$ into \eqref{opus_b1_channel_s}. Let us call the resultant output matrix $\sdvac$. Thus:
\begin{equation}
N = \sdvac -  M M^T.
\end{equation}
Since $M$ is already derived, the remaining task is to compute $\sdvac$.

From the definition of the covariance matrix \eqref{gqm_def_cov} and from \eqref{opus_b1_comm_dd} the following expressions for covariance matrix elements follow:
\begin{align}          \label{opus_b1_s11}
\sdvel_{11} &= 1 + 2\,\Re\,\,{}_M\bra{0} \hat{d}_\x{I} \hat{d}_\x{I} + \hat{d}_\x{I}\d \hat{d}_\x{I} \ket{0}_M,
\\
\sdvel_{12} &= 2\,\Im\,\,{}_M\bra{0} \hat{d}_\x{I} \hat{d}_\x{I} \ket{0}_M,
\\
\sdvel_{22} &= 1 + 2\,\Re\,\,{}_M\bra{0} -\hat{d}_\x{I} \hat{d}_\x{I} + \hat{d}_\x{I}\d \hat{d}_\x{I} \ket{0}_M,
\\
\sdvel_{13} &= 2\,\Re\,\,{}_M\bra{0} \hat{d}_\x{I} \hat{d}_\x{II} + \hat{d}_\x{I} \hat{d}_\x{II}\d \ket{0}_M,
\\
\sdvel_{23} &= 2\,\Im\,\,{}_M\bra{0} \hat{d}_\x{I} \hat{d}_\x{II} + \hat{d}_\x{I} \hat{d}_\x{II}\d \ket{0}_M,
\\
\sdvel_{14} &= 2\,\Im\,\,{}_M\bra{0} \hat{d}_\x{I} \hat{d}_\x{II} - \hat{d}_\x{I} \hat{d}_\x{II}\d \ket{0}_M,
\\                        \label{opus_b1_s24}
\sdvel_{24} &= -2\,\Re\,\,{}_M\bra{0} \hat{d}_\x{I} \hat{d}_\x{II} - \hat{d}_\x{I} \hat{d}_\x{II}\d \ket{0}_M,
\end{align}
where $\sdvel_{33}$, $\sdvel_{34}$ and $\sdvel_{44}$ are obtained from $\sdvel_{11}$, $\sdvel_{12}$ and $\sdvel_{22}$ respectively by replacing subscripts I with~II. The remaining covariance matrix elements follow from the fact that the matrix is symmetric. In this computation we also use the fact that ${}_M\bra{0} \hat{d}_\Lambda \ket{0}_M = 0$, which follows from combining \eqref{rqi_bogo_b} with \eqref{opus_b1_dinb}.

In the expressions for the covariance matrix elements, six independent expectation values arise, which are expressed below in terms of the modified Rindler creation and annihilation operators via \eqref{opus_b1_dinb}:
\begin{align}             \label{opus_b1_d1d1}
{}_M\bra{0} \hat{d}_\x{I} \hat{d}_\x{I} \ket{0}_M &= 
\iint\x{d}\Omega\x{d}\Xi \,\, (\psi_\x{I} , \wiO) (\psi_\x{I} , \wiX)
\,\,{}_M\bra{0} \biO \biX \ket{0}_M,
\\                        \label{opus_b1_d1d1d}
{}_M\bra{0} \hat{d}_\x{I}\d \hat{d}_\x{I} \ket{0}_M &= 
\iint\x{d}\Omega\x{d}\Xi \,\, (\psi_\x{I} , \wiO)\s (\psi_\x{I} , \wiX)
\,\,{}_M\bra{0} \biO\d \biX \ket{0}_M,
\\                       \label{opus_b1_d1d2}
{}_M\bra{0} \hat{d}_\x{I} \hat{d}_\x{II} \ket{0}_M &= 
\iint\x{d}\Omega\x{d}\Xi \,\, (\psi_\x{I} , \wiO) (\psi_\x{II} , \wiiX)
\,\,{}_M\bra{0} \biO \biiX \ket{0}_M,
\\                       \label{opus_b1_d1d2d}
{}_M\bra{0} \hat{d}_\x{I} \hat{d}_\x{II}\d \ket{0}_M &= 
\iint\x{d}\Omega\x{d}\Xi \,\, (\psi_\x{I} , \wiO) (\psi_\x{I} , \wiX)\s
\,\,{}_M\bra{0} \biO \biiX\d \ket{0}_M.
\end{align}
The remaining two expectation values follow from replacing subscripts I with~II in \eqref{opus_b1_d1d1} and \eqref{opus_b1_d1d1d}. The Reader may note, that only the expectation values \eqref{opus_b1_d1d2} and \eqref{opus_b1_d1d2d} depend on $D$, as they involve operators that are shifted relative to each other. Thus, we may calculate the remaining ones for $D=0$ without loss of generality. 

Let us now compute the necessary expectation values of products of modified Rindler ladder operators. One of them has already been given in Eq.~\eqref{rqi_bdb}. The expectation values present in \eqref{opus_b1_d1d1} and \eqref{opus_b1_d1d1d} can be calculated similarly. The resultant outcomes are:
\begin{align}            \label{opus_b1_b1b1}
{}_\x{M}\bra{0} \hat{b}_{\Lambda\Omega} \hat{b}_{\Lambda\Xi} \ket{0}_\x{M}
&=
0,
\\                       \label{opus_b1_b1b1d}
{}_\x{M}\bra{0} \hat{b}\d_{\Lambda\Omega} \hat{b}_{\Lambda\Xi} \ket{0}_\x{M}
&=
\frac{\delta(\Omega-\Xi)}{e^\frac{2\pi\Omega}{a}-1}.
\end{align}
The expectation values that appear in \eqref{opus_b1_d1d2} and \eqref{opus_b1_d1d2d} are substantially more complicated and their calculation is split up between the cases of antiparallel and parallel accelerations of the observers. Also, the case of $D=0$ is shown, obtained in the above way. 

Let us sum up the results calculated so far. Combining Eq.'s~\eqref{opus_b1_s11}-\eqref{opus_b1_b1b1d} we get the following structure of $\sdvac$:
\begin{equation}       \label{opus_sdvac}
\sdvac=
\left(\begin{matrix}
1+N_{\text{I}}
&
0
&
\Re \,N^+_{\text{I,II}}
&
\Im\,N^-_{\text{I,II}}
\\
0
&
1+N_{\text{I}}
&
\Im\,N^+_{\text{I,II}}
&
-\Re\,N^-_{\text{I,II}}
\\
\Re\,N^+_{\text{I,II}}
&
\Im\,N^+_{\text{I,II}}
&
1+N_{\text{II}}
&
0
\\
\Im\,N^-_{\text{I,II}}
&
-\Re\,N^-_{\text{I,II}}
&
0
&
1+N_{\text{II}}
\end{matrix}\right),
\end{equation}
with the following definitions:
\begin{align}        \label{opus_b1_Nlambda}
N_\Lambda &\defn 2\int\x{d}\Omega \,\,
\frac{\left| (\psi_\Lambda , w_{\Lambda\Omega}) \right|^2}{e^\frac{2\pi\Omega}{a}-1},
\\                   \label{opus_N12general}
N_\x{I,II}^\pm &\defn 2
{}_M\bra{0} \hat{d}_\x{I} \hat{d}_\x{II} \pm \hat{d}_\x{I} \hat{d}_\x{II}\d \ket{0}_M.
\end{align}

\subsection{Antiparallel accelerations, $D=0$}

This case is rather straightforward, as we compute the remaining expectation values of products of the modified Rindler ladder operators, using the same method as in \eqref{rqi_bdb}. The outcome is:
\begin{align}            \label{opus_b1_b1b2_D0}
{}_\x{M}\bra{0} \hat{b}_{\x{I}\Omega} \hat{b}_{\x{II}\Xi} \ket{0}_\x{M}
&=
\frac{\delta(\Omega-\Xi)}{2\sinh\left( \frac{\pi\Omega}{a} \right)},
\\                       \label{opus_b1_b1b2d_D0}
{}_\x{M}\bra{0} \hat{b}_{\x{I}\Omega} \hat{b}_{\x{II}\Xi}\d \ket{0}_\x{M}
&=0
.
\end{align}
Combining this with \eqref{opus_b1_d1d2}, \eqref{opus_b1_d1d2d} and \eqref{opus_N12general}, we obtain:
\begin{equation}         \label{opus_b1_N12_D0}
N_\x{I,II}^\pm = \int\x{d}\Omega \,\,
\frac{(\psi_\x{I} , \wiO) (\psi_\x{II} , \wiiO) }{\sinh\left( \frac{\pi\Omega}{a} \right)}.
\end{equation}
This completes the derivation of the noise matrix for $D=0$. Now we shall proceed with a derivation for a general $D$.

\subsection{Antiparallel accelerations, $D\neq 0$}

Let us investigate how the Bogolyubov coefficients \eqref{rqi_u_bogo1d} are modified when $D\neq 0$. As depicted in Figs.~\ref{fig_opus1d_Dp} and \ref{fig_opus1d_Dn}, region~I is shifted by $\frac{D}{2}$, therefore to modify the Bogolyubov coefficient $\alpha_{\Omega k}^{(\text{I})}$ we can equivalently shift the Minkowski mode function in the opposite direction: $u_k(x,t)$ is replaced by $\tilde{u}_k(x,t)\equiv u_k(x+\frac{D}{2},t)=e^{i\frac{D}{2}k}u_k(x,t)$. Using the antilinearity of the scalar product we find the modified $\alpha_{\Omega k}^{\text{(I)}}$ coefficient to be: 
\begin{align}        \label{opus_b1_modbogo}
(\tilde{u}_k,w_{\text{I}\Omega})
&=e^{-i \frac{D}{2}k}(u_k,w_{\text{I}\Omega})=e^{-i \frac{D}{2}k}\alpha_{\Omega k}^{\text{(I)}}.
\end{align}
Similarly one can verify that the other Bogolyubov coefficients are modified as: 
\begin{equation}     \label{opus_b1_modbogos}
\begin{aligned}
\alpha_{\Omega k}^{\text{(I)}}& \to e^{-i \frac{D}{2}k}\alpha_{\Omega k}^{\text{(I)}}, \qquad
&\beta_{\Omega k}^{\text{(I)}} &\to e^{i \frac{D}{2}k}\beta_{\Omega k}^{\text{(I)}},
\\
\alpha_{\Omega k}^{\text{(II)}}& \to e^{i \frac{D}{2}k}\alpha_{\Omega k}^{\text{(II)}}, \qquad
&\beta_{\Omega k}^{\text{(II)}}& \to e^{-i \frac{D}{2}k}\beta_{\Omega k}^{\text{(II)}}.
\end{aligned}
\end{equation}
Using the modified Bogolyubov coefficients and \eqref{rqi_bogo_b}, we compute the expectation values of products of the modified Rindler creation and annihilation operators arising in the off-diagonal blocks of the noise matrix. This yields:
\begin{align}            \label{opus_b1_b1b2_Dn0}
I_- \defn
{}_\x{M}\bra{0} \hat{b}_{\x{I}\Omega} \hat{b}_{\x{II}\Xi} \ket{0}_\x{M}
&=
-\int\x{d}k\,\, {\alpha_{\Omega k}^{\x{(I)}}}\s {\beta_{\Xi k}^{\x{(II)}}}\s e^{iDk}
=
\int\x{d}k\,\, {\alpha_{\Omega k}^{\x{(I)}}}\s \alpha_{\Xi k}^{\x{(I)}}\, e^{iDk},
\\                       \label{opus_b1_b1b2d_Dn0}
I_+ \defn
{}_\x{M}\bra{0} \hat{b}_{\x{I}\Omega} \hat{b}_{\x{II}\Xi}\d \ket{0}_\x{M}
&=
\int\x{d}k\,\, {\alpha_{\Omega k}^{\x{(I)}}}\s \alpha_{\Xi k}^{\x{(II)}} e^{iDk} 
=
\int\x{d}k\,\, {\alpha_{\Omega k}^{\x{(I)}}}\s {\alpha_{\Xi k}^{\x{(I)}}}\s e^{iDk}
,
\end{align}
where the last equalitites have been obtained using relations \eqref{rqi_u_bogo1d}.
We rewrite the resultant integrals using the explicit form of the coefficient $\alpha^\text{(I)}_{\Omega k}$, as given in \eqref{rqi_u_bogo1d}, to obtain:
\begin{align}
I_\pm&=\frac{e^{\frac{\pi(\Omega+\Xi)}{2a}}}{4\pi a\sqrt{\sinh\left(\frac{\pi\Omega}{a}\right)\sinh\left(\frac{\pi\Xi}{a}\right)}}\int\frac{\mathrm{d}k}{\omega_k}\left(\frac{\omega_k+k}{\omega_k-k}\right)^{i\frac{\Omega\pm\Xi}{2a}}e^{iDk}.
\end{align}
Let us introduce $\theta_\pm\equiv\frac{\Omega\pm\Xi}{a}$ and $\Delta\equiv mD$. We continue by changing the integration variable to $x = \text{asinh} \frac{k}{m}$, and treating the integral as a distribution, to get:
\begin{equation}      
\begin{aligned}
I_\pm &\frac{4\pi a\sqrt{\sinh\left(\frac{\pi\Omega}{a}\right)\sinh\left(\frac{\pi\Xi}{a}\right)}}{e^{\frac{\pi(\Omega+\Xi)}{2a}}}=\int_{-\infty}^{\infty}\mathrm{d}x\,e^{i\left(\Delta\sinh x+\theta_\pm x\right)}
\\
&=\int_{-\infty}^{\infty}\mathrm{d}x\,\cos(\Delta\sinh x + \theta_\pm x)
\\
&=2\int_{0}^{\infty}\mathrm{d}x\,\cos\left(|\Delta|\sinh x\right)\cos(\theta_\pm x)
\\
&-2 \frac{\Delta}{|\Delta|}\int_{0}^{\infty}\mathrm{d}x\, \sin\left(|\Delta|\sinh x\right)\sin (\theta_\pm x).
\end{aligned}
\end{equation}
The above integrals are of the form that is proportional to the following integral representation of the modified Bessel function~\cite{nist}:
\begin{equation}
\begin{aligned}
K_{i\nu}(\delta)&=\frac{1}{\cosh\frac{\pi\nu}{2}}\int_0^\infty \mathrm{d}x \cos(\delta\sinh x)\cos(\nu x)
\\
&=\frac{1}{\sinh\frac{\pi\nu}{2}}\int_0^\infty \mathrm{d}x \sin(\delta\sinh x)\sin(\nu x),
\end{aligned}
\end{equation}
valid for $\delta >0$. Implementing the above identities into $I_\pm$ gives us:
\begin{align}      \label{opus_b1_Ipm}
I_\pm&=\frac{ e^{\frac{\pi(\Omega+\Xi)}{2a}}}{2\pi a\sqrt{\sinh\left(\frac{\pi\Omega}{a}\right)\sinh\left(\frac{\pi\Xi}{a}\right)}} e^{-\frac{D}{|D|}\frac{\pi\theta_\pm}{2}}K_{i\theta_\pm}(|\Delta|).
\end{align}
This result can be combined with \eqref{opus_b1_d1d2}, \eqref{opus_b1_d1d2d} and \eqref{opus_N12general} to yield:
\begin{equation}    \label{opus_b1_N12_Dn0}
\begin{aligned}   
N_\x{I,II}^\pm
=
&\frac{1}{\pi a}
\!\int\!\!\!\!\int\mathrm{d}\Omega\mathrm{d}\Xi
\frac{(\psi_\text{I},\wiO)}{\sqrt{
\sinh\left(\frac{\pi\Omega}{a}\right)
\sinh\left(\frac{\pi\Xi}{a}\right)}}
\\
&\times\big[
e^{\frac{\pi(\Omega-\Xi)}{2a}(1-\frac{D}{|D|})}
(\psi_\text{II},\wiiX) K_{i(\frac{\Omega-\Xi}{a})}(m|D|) 
\\
&\,\,\,\pm
e^{\frac{\pi(\Omega+\Xi)}{2a}(1-\frac{D}{|D|})}
(\psi_\text{II},\wiiX)^\star K_{i(\frac{\Omega+\Xi}{a})}(m|D|)
\big].
\end{aligned} 
\end{equation}
This completes the derivation of the noise matrix for $D\neq 0$, and completes the case of counter-accelerating observers. 

Let us study the asymptotic behavior of $N_{\text{I,II}}$ given by Eq.~\eqref{opus_b1_N12_Dn0} as $D\to 0$. For this purpose we employ the following property of the modified Bessel function: $\lim_{\epsilon\to0^+} K_{i\nu}(\epsilon)=\pi\delta(\nu)$, proven in Appendix \ref{App_N12limits}. Therefore in the limit of $|D|\to 0$ the first term of the integrand of \eqref{opus_b1_N12_Dn0} becomes proportional to $\delta\left({\frac{\Omega-\Xi}{a}}\right)$, while the second one vanishes (the argument of the delta function is positive). This allows one to perform the integration over~$\Xi$ in \eqref{opus_N12general} and leads to:
\begin{align}          \label{opus_b1_N12_lim_D0}
\lim_{|D|\to 0}
N_\x{I,II}^\pm
&=
\int\x{d}\Omega\,
\frac{(\psi_\x{I},\wiO)(\psi_\x{II},\wiiO)}{\sinh\left(\frac{\pi \Omega}{a}\right)}.
\end{align}
Thus, the result \eqref{opus_b1_N12_D0} obtained for $D=0$ is correctly retrieved.

To show the asymptotic behavior of $N_{\text{I,II}}^\pm$ as $D\to\infty$ we use the asymptotic form of the modified Bessel function for the large argument $|x|$: $K_{i\nu}(|x|)\approx \sqrt{\frac{\pi}{2|x|}}e^{-|x|}$~\cite{nist}. Since in this limit the function becomes independent of the order $\nu$ and vanishes for large arguments, we can take the modified Bessel functions appearing in Eq.~\eqref{opus_b1_N12_Dn0} outside the integral. Then in the limit of $|D|\to\infty $ the entire integral $N_{\text{I,II}}^\pm$ vanishes:
\begin{align}
\lim_{|D|\to \infty}N_{\text{I,II}}^\pm &=0.
\end{align}

\subsection{Parallel accelerations}   \label{sec_opus_b1_par}

In this case the wedges not only get shifted by $\pm\frac{D}{2}$, but also the wedge II is reflected with respect to its apex, as shown in Fig. \ref{fig_opus1d_Para}. We perform a reasoning analogous to the one used in the case of counter-accelerating modes, to obtain the following modified Bogolyubov coefficients:
\begin{align}          \label{opus_b1_modbogos_par}
\alpha_{\Omega k}^{\text{(I)}}& \longrightarrow e^{-i \frac{D}{2}k}\alpha_{\Omega k}^{\text{(I)}},
\nonumber\\
\beta_{\Omega k}^{\text{(I)}}&  \longrightarrow e^{i \frac{D}{2}k}\beta_{\Omega k}^{\text{(I)}}
= -\,e^{i \frac{D}{2}k} e^{-\frac{\pi \Omega}{a}} \alpha_{\Omega k}^{\text{(I)}},
\nonumber\\
\alpha_{\Omega k}^{\text{(II)}}& \longrightarrow e^{i \frac{D}{2} k}\alpha_{\Omega k}^{\text{(I)}},
\nonumber\\
\beta_{\Omega k}^{\text{(II)}}&  \longrightarrow e^{-i \frac{D}{2}k}\beta_{\Omega k}^{\text{(I)}}
= -\,e^{-i \frac{D}{2}k} e^{-\frac{\pi \Omega}{a}} \alpha_{\Omega k}^{\text{(I)}}.
\end{align}  
Using the above coefficients, \eqref{rqi_bogo_b} and \eqref{rqi_u_bogo1d}, we compute the necessary expectation values of products of the modified Rindler ladder operators. The outcomes can be related to the integrals $I_\pm$, defined in \eqref{opus_b1_b1b2_Dn0} and \eqref{opus_b1_b1b2d_Dn0}:
\begin{align}            \label{opus_b1_b1b2_Dn0_par}
{}_\x{M}\bra{0} \hat{b}_{\x{I}\Omega} \hat{b}_{\x{II}\Xi} \ket{0}_\x{M}
&=
\int\x{d}k\,\, {\alpha_{\Omega k}^{\x{(I)}}}\s {\alpha_{\Xi k}^{\x{(I)}}}\s\,e^{-\frac{\pi\Xi}{a}} e^{iDk} = I_+ \,e^{-\frac{\pi\Xi}{a}},
\\                       \label{opus_b1_b1b2d_Dn0_par}
{}_\x{M}\bra{0} \hat{b}_{\x{I}\Omega} \hat{b}_{\x{II}\Xi}\d \ket{0}_\x{M}
&=
\int\x{d}k\,\, {\alpha_{\Omega k}^{\x{(I)}}}\s \alpha_{\Xi k}^{\x{(I)}} \,\,e^{iDk}
=I_-
.
\end{align}
This, together with \eqref{opus_b1_d1d2}, \eqref{opus_b1_d1d2d}, \eqref{opus_N12general} and \eqref{opus_b1_Ipm} leads to:
\begin{equation}    \label{opus_b1_N12_Dn0_par}
\begin{aligned}   
N_\x{I,II}^\pm
=
&\frac{1}{\pi a}
\!\int\!\!\!\!\int\mathrm{d}\Omega\mathrm{d}\Xi
\frac{(\psi_\text{I},\wiO)}{\sqrt{
\sinh\left(\frac{\pi\Omega}{a}\right)
\sinh\left(\frac{\pi\Xi}{a}\right)}}
\\
&\times\big[
e^{\frac{\pi}{2a}[(\Omega-\Xi)-(\Omega+\Xi)\frac{D}{|D|}]}
(\psi_\text{II},\wiiX) K_{i(\frac{\Omega-\Xi}{a})}(m|D|) 
\\
&\,\,\,\pm
e^{\frac{\pi}{2a}[(\Omega+\Xi)-(\Omega-\Xi)\frac{D}{|D|}]}
(\psi_\text{II},\wiiX)^\star K_{i(\frac{\Omega+\Xi}{a})}(m|D|)
\big].
\end{aligned} 
\end{equation}
This completes the derivation of the noise matrix for the case of parallel-accelerating observers.

Analogously to the previous scenario, we use the limit of the modified Bessel function: $\lim_{\epsilon\to0^+} K_{i\nu}(\epsilon)=\pi\delta(\nu)$, from which we find:
\begin{align}
\lim_{|D|\to 0}N_{\text{I,II}}^\pm &= \pm\int \mathrm{d}\Omega \,\frac{{(\psi_\x{I},\wiO)(\psi_\x{II},\wiiO)}\s}{\sinh(\frac{\pi\Omega}{a})}\,e^{\frac{\pi\Omega}{a}}.
\end{align}
Using the asymptotic form of the modified Bessel function for large arguments: $K_{i\nu}(|x|)\approx \sqrt{\frac{\pi}{2|x|}}e^{-|x|}$~\cite{nist}, we find that when the separation $|D|$ between the wedges becomes large, the noise correlations in the channel vanish:
\begin{align}
\lim_{|D|\to\infty}N_{\text{I,II}}^\pm=0.
\end{align}

Thus, the noise matrix has been derived in all cases. However, the expressions \eqref{opus_b1_Nlambda}, \eqref{opus_b1_N12_Dn0} and \eqref{opus_b1_N12_Dn0_par} may suggest that the noise matrix depends on the unphysical parameter $a$. Let us look in more detail at why this dependence in merely apparent.

\section{$a$-independence of the channel}    \label{sec_opus_a}

No measurable physical quantity computed based on the output covariance matrix $\sdvac$ should depend on the parameter $a$ of the Rindler transformation. Furthermore, $a$ appears in our expressions only because we have decided to expand the field in the (modified) Rindler modes \eqref{opus_dec_b_1d}. Had we made a different choice of the coordinate system, $a$ would not be present. In this section we show that the $N$ matrix in fact does not depend on $a$, as $a$ can be removed from our expressions via a change of variables.

The expressions \eqref{opus_b1_Nlambda}, \eqref{opus_b1_N12_Dn0} and \eqref{opus_b1_N12_Dn0_par} involve overlaps of an observer's wavepacket with a respective Rindler mode. The Rindler mode \eqref{qft_wi_m_r} (the same reasoning can be repeated for \eqref{qft_wii_m_r}) itself depends on $a$ as well as on the coordinate time $\eta$. Let us express the coordinate time in terms of the proper time $\tau$ along the trajectory of the corresponding observer, via \eqref{gr_tau_eta}. This yields:
\begin{equation}
\wiO = \sqrt{\frac{\sinh\left(\frac{\pi\Omega}{a}\right)}{\pi^2 a}} 
K_{i\frac{\Omega}{a}}\left( m\chi \right)e^{-i\frac{\Omega}{a} \tau \A_\x{I}}  
,
\end{equation}
where $\A_\x{I}$ is the proper acceleration of the observer in the wedge I. Thus, the Rindler mode can we written as follows:
\begin{equation}
\wiO = \frac{1}{\sqrt{a}} \,{F}_1 \left( \frac{\Omega}{a} \right),
\end{equation}
where $F_1$ is a certain function, whose form is uniportant in this section. Using~\eqref{qft_scalar_r} let us now write our overlap explicitly, expressing the coordinate time in terms of the proper time:
\begin{equation}
\begin{aligned}
(\psi_\x{I}, \wiO) 
= 
i \int_{t=0} \frac{\x{d}\chi}{a\chi}
&\left(
\psi_\x{I}\s \partial_\eta \wiO -
\wiO \partial_\eta \psi_\x{I}\s
\right)
\\
=
i \int_{t=0} \frac{\x{d}\chi}{\A_\x{I}\chi}
&\left(
\psi_\x{I}\s \partial_\tau \wiO -
\wiO \partial_\tau \psi_\x{I}\s
\right).
\end{aligned}
\end{equation}
The overlap depends on $\Omega$ and $a$ only via $\wiO$, thus we can also express it as:
\begin{equation}       \label{opus_b1_a-ind_overlap}
(\psi_\x{I}, \wiO) = \frac{1}{\sqrt{a}} \,{F}_2 \left( \frac{\Omega}{a} \right),
\end{equation}
where $F_2$ is a certain function.
Now we may note that the expression \eqref{opus_b1_Nlambda} may be written as:
\begin{equation}
N_\Lambda = \int\frac{\x{d}\Omega}{a} \,\, F_3 \left( \frac{\Omega}{a} \right),
\end{equation}
where $F_3$ is a certain function.
Therefore, changing the variable of integration to $\frac{\Omega}{a}$ makes the expression for $N_\x{\Lambda}$ explicitly $a$-independent.

Similarly \eqref{opus_b1_N12_Dn0} and \eqref{opus_b1_N12_Dn0_par}, explicitly depend on $a$ only via $\frac{\Omega}{a}$ and $\frac{\Xi}{a}$. Since each of them involves a product of two overlaps, we may express $N_\x{I,II}^\pm$~as follows:
\begin{equation}
N_\x{I,II}^\pm = \iint\frac{\x{d}\Omega}{a}\frac{\x{d}\Xi}{a} \,\, F_4 \left( \frac{\Omega}{a}, \frac{\Xi}{a} \right),
\end{equation}
where $F_4$ is a certain function.
Variables can be changed to $\frac{\Omega}{a}$ and $\frac{\Xi}{a}$, which makes $N_\x{I,II}$ explicitly $a$-independent for all cases.

Thus, we see that our channel does not depend on the unphysical parameter $a$. Let us now apply the channel to a specific set of modes, and obtain physical results.

\section{The choice of the modes}    \label{opus_modes}

So far we have been working with general inertial and uniformly accelerated modes. In the derivation we assumed, that at $t=0$, the modes $\phi_\x{I}$ and $\psi_\x{I}$ were localized within region I, whereas $\phi_\x{II}$ and $\psi_\x{II}$ were localized within region II. Furthermore, we demanded that their corresponding annihilation operators could be written as \eqref{opus_b1_fina} and \eqref{opus_b1_dinb}. In this section let us assume specific mode functions that are used onwards.

Since the channel is fully expressed in terms of scalar products of modes taken at $t=0$, it suffices to specify the form of the mode, as well as its time derivative, at that instant. Similarly to~\cite{channel_andrzej,andrzej_loc_proj,andrzej_mody_kolejne}, for $\phi_\Lambda$ we choose to work with modes that emulate the cavity modes \eqref{qft_u_cav_m}, but have a smooth envelope. The choice is:
\begin{equation}      \label{opus_b1_phi}
\begin{aligned}
\phi_\Lambda(x,0)&= \mathcal{N}_\phi \,\,
e^{-2\left( \frac{1}{{\A}_\Lambda L}\log({\A}_\Lambda x) \right)^2}
\sin\left[\sqrt{\Omega_0^2-m^2} \left(x \mp \frac{1}{{\A}_\Lambda}\right)\right],
\\
{\partial}_t \phi_\Lambda(x,0)&=-i  \Omega_0\phi_\Lambda(x,0),
\end{aligned}
\end{equation}
where the upper sign refers to $\Lambda=\x{I}$, the lower sign refers to $\Lambda=\x{II}$. Also, ${\A}_\Lambda$ are the proper accelerations of the accelerating observers (they appear here, so that the inertial wavepackets are centered where the accelerating observers have zero velocity), $L$ denotes the size of the wavepackets and $\mathcal{N}_\phi$ is a normalization constant. The wavepackets are centered around $\frac{1}{{\A}_\Lambda}$. Because they have finite size, we assume that they are sufficiently localized, such that a single proper acceleration can be attributed to each of them, i.e. $\frac{1}{{\A}_\Lambda} \gg L$. Furthermore, $\Omega_0$ is a frequency which the spectrum of the mode function is centered around. It has to be sufficiently large in order to prevent the negative frequency contributions, i.e., $\Omega_0 \gg \frac{1}{L}$. Additionally we impose an extra cut-off at zero frequency to completely eliminate the negative frequency contribution, in order for the assumptions \eqref{opus_b1_fina} and \eqref{opus_b1_dinb} to be satisfied. Such a cut-off modifies the spatial profile given by \eqref{opus_b1_phi} only negligibly. The parameters $L$ and $\Omega_0$ can be chosen independently for each region, i.e., $\Lambda\in \{\text{I}, \text{II}\}$. The exact form of the exponential envelope appearing in the definition \eqref{opus_b1_phi} is chosen for later computational convenience, but to a good approximation it can be treated as a Gaussian: $e^{-2\left( \frac{1}{{\A}_\Lambda L}\log({\A}_\Lambda x)\right)^2} \approx e^{-2\left(\frac{x-1/{\A}_\Lambda}{L}\right)^2}$ as long as $|x-\frac{1}{{\A}_\Lambda}| < L$. For the arguments $x$ that are further away from $\frac{1}{{\A}_\Lambda}$ the tails of the chosen function vanish faster than the Gaussian tails.

\begin{figure}
\centering
\includegraphics[width=0.6\linewidth]{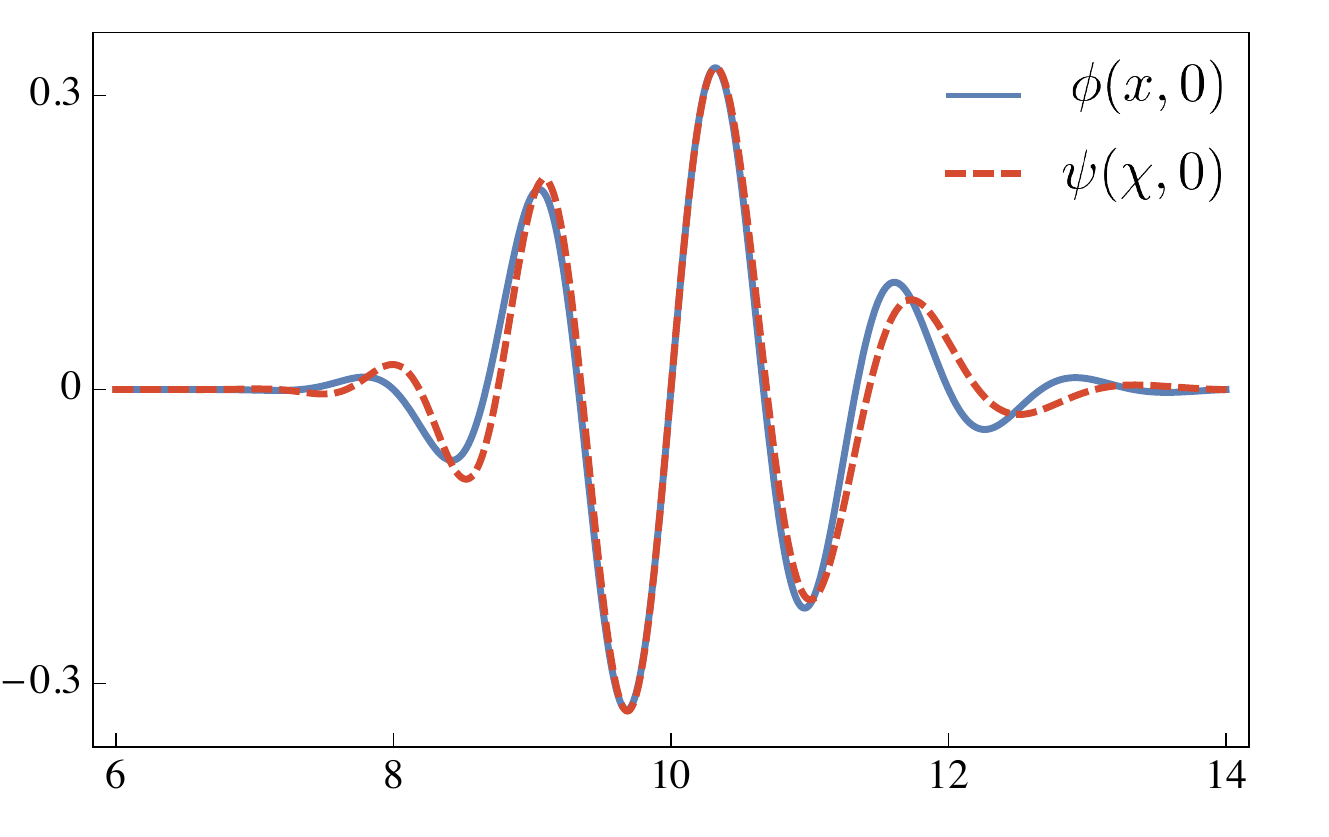}
\caption{\label{fig_opus1d_modescomparison}\small{Comparison between the spatial profiles of modes $\phi_\Lambda$ and $\psi_\Lambda$ for the following choice of parameters: ${\cal A}_\Lambda=0.15$, $L=2$, $\Omega_0\approx 5$, $m=0.1$.}}
\end{figure}

Turning to the accelerated modes $\psi_\Lambda$, we choose to take the accelerated cavity modes \eqref{qft_w_cav_m}, with a smooth envelope. The choice is:
\begin{equation}         \label{opus_b1_psi_pass}
\begin{aligned}
&\psi_\Lambda(\chi,0)=\mathcal{N}_\psi\,\,  
e^{-2\left( \frac{1}{{\A}_\Lambda L}\log({\A}_\Lambda x)\right)^2} 
\Im \left[
I_{-i\frac{\Omega_0}{{\A}_\Lambda}}\left(\frac{m}{{\A}_\Lambda}\right)
I_{i\frac{\Omega_0}{{\A}_\Lambda}}(m|\chi|)\right], 
\\
&{\partial}_\tau \psi_\Lambda(\chi,0)=\mp i  \Omega_0\psi_\Lambda(\chi,0),
\end{aligned}
\end{equation}
where the upper (lower) sign refers to $\Lambda=\x{I(II)}$ and $\mathcal{N}_\psi$ is a normalization constant. It should be understood that this mode is defined in the respective wedge and is zero elsewhere. The dependence of the modes on the proper accelerations is analogous to the way cavity eigenmodes depend on the proper acceleration of the cavity. The comparison of the spatial profiles of \eqref{opus_b1_phi} and \eqref{opus_b1_psi_pass} is shown in the Fig. \ref{fig_opus1d_modescomparison}. There exists an inevitable mismatch of the modes due to a prence of a relative acceleration. The higher the relative acceleration is, the more prominent the effect becomes.

\section{Results}        \label{sec_opus_b1_results}

In this section we apply the framework that we have developed, with the modes chosen in the previous section, to the Minkowski vacuum state as the input state of our channel. We compute the amount of entanglement in the vacuum, as seen by the uniformly accelerated observers, as a function of the parameters characterizing the accelerated modes.

Substituting the Minkowski vacuum state as the input state, we have $\vec{X}^{(f)}=\vec{0}$ and $\boldsymbol\sigma^{(f)} = \I$, therefore the output state reduces to $\vec{X}^{(d)}=\vec{0}$ and \eqref{opus_sdvac} with $N_{\text{I}}$ and $N_{\text{II}}$ given by \eqref{opus_b1_Nlambda}. $N_{\text{I,II}}^\pm$ appearing in the off-diagonal terms, in the counter-accelerating case are given by \eqref{opus_b1_N12_Dn0} and in the parallel-accelerating case by \eqref{opus_b1_N12_Dn0_par}. Let us note that the results presented in this section also hold for input coherent states, for which the only difference is that $\vec{X}^{(f)} \neq  0$.

To quantify the amount of entanglement, the logarithmic negativity discussed in Sec. \ref{sec_neg}, is chosen. It is given by the formula \eqref{gqm_neg} with the substitutions: $\boldsymbol\sigma_{AB} = \sdvac\,$ and $\tilde{\Delta} = (1+N_{\text{I}})^2 +  (1+N_{\text{II}})^2 + \left| N_{\text{I,II}} \right|^2$.

\begin{figure}
\centering
\includegraphics[width=0.6\linewidth]{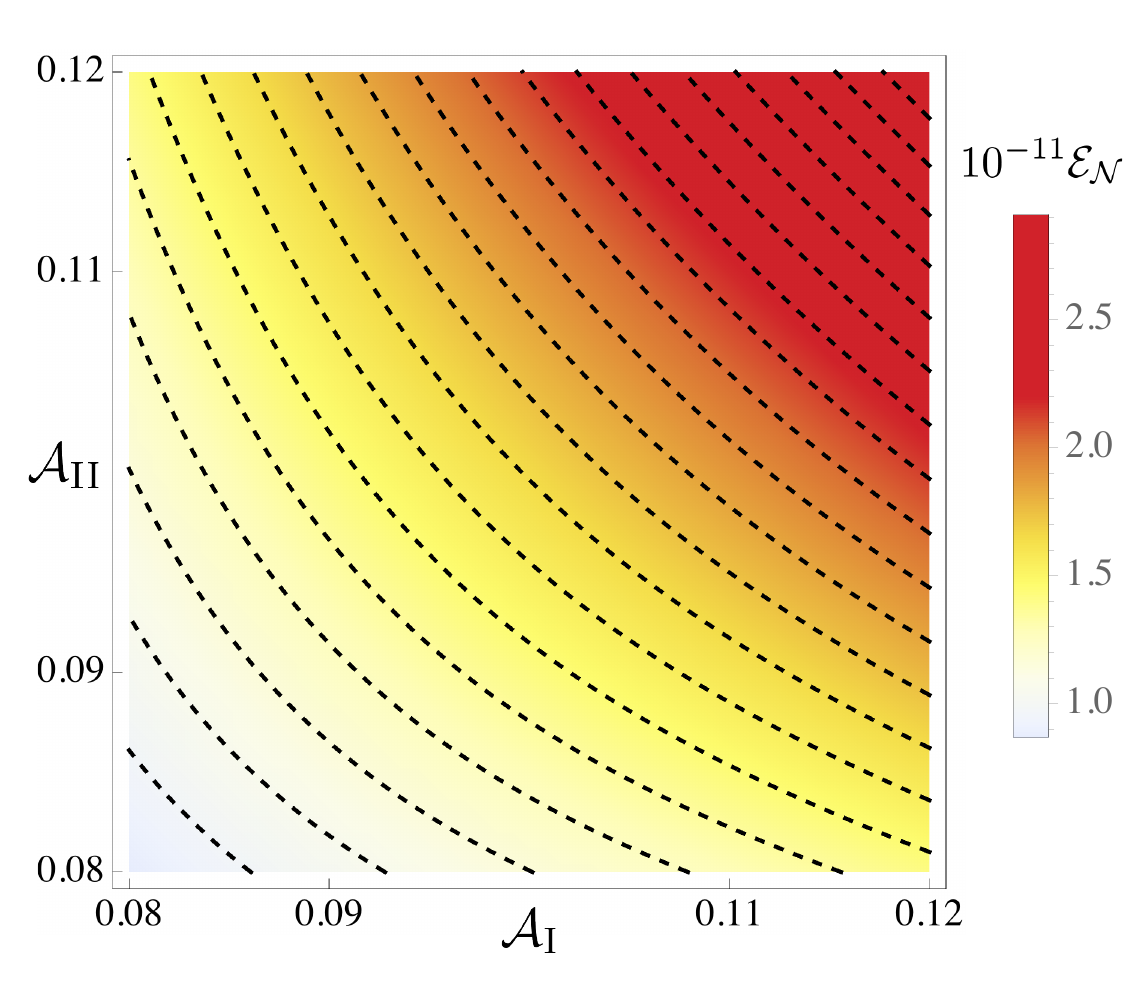}
\caption{\label{fig_opus1d_D0_vs_Accelerations}\small{Logarithmic negativity of the Minkowski vacuum for two counter-accelerated modes, as a function of their proper accelerations for $D=0$. We have chosen $L=2$, $m=0.1$ and $\Omega_0\approx 5$.}}
\end{figure}

\begin{figure}
\centering
\includegraphics[width=0.6\linewidth]{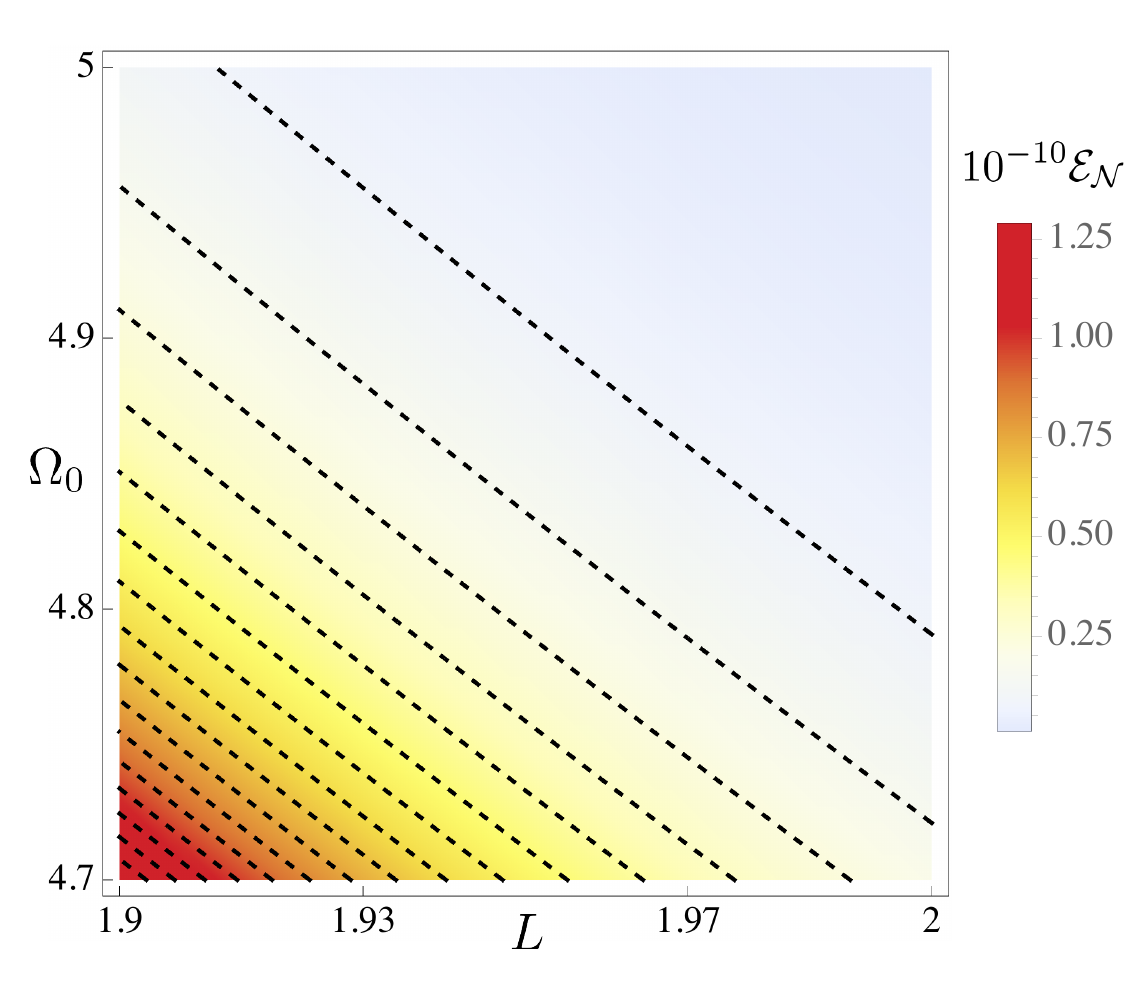}
\caption{\label{fig_opus1d_Negativity_vs_LOmega}\small{Logarithmic negativity of the Minkowski vacuum for two counter-accelerated modes characterized by the same proper acceleration ${\cal A}_\Lambda=0.1$ as a function of $L$ and $\Omega_0$. We have chosen $D=0$ and $m=0.1$.}}
\end{figure}

\begin{figure}
\centering
\includegraphics[width=0.7\linewidth]{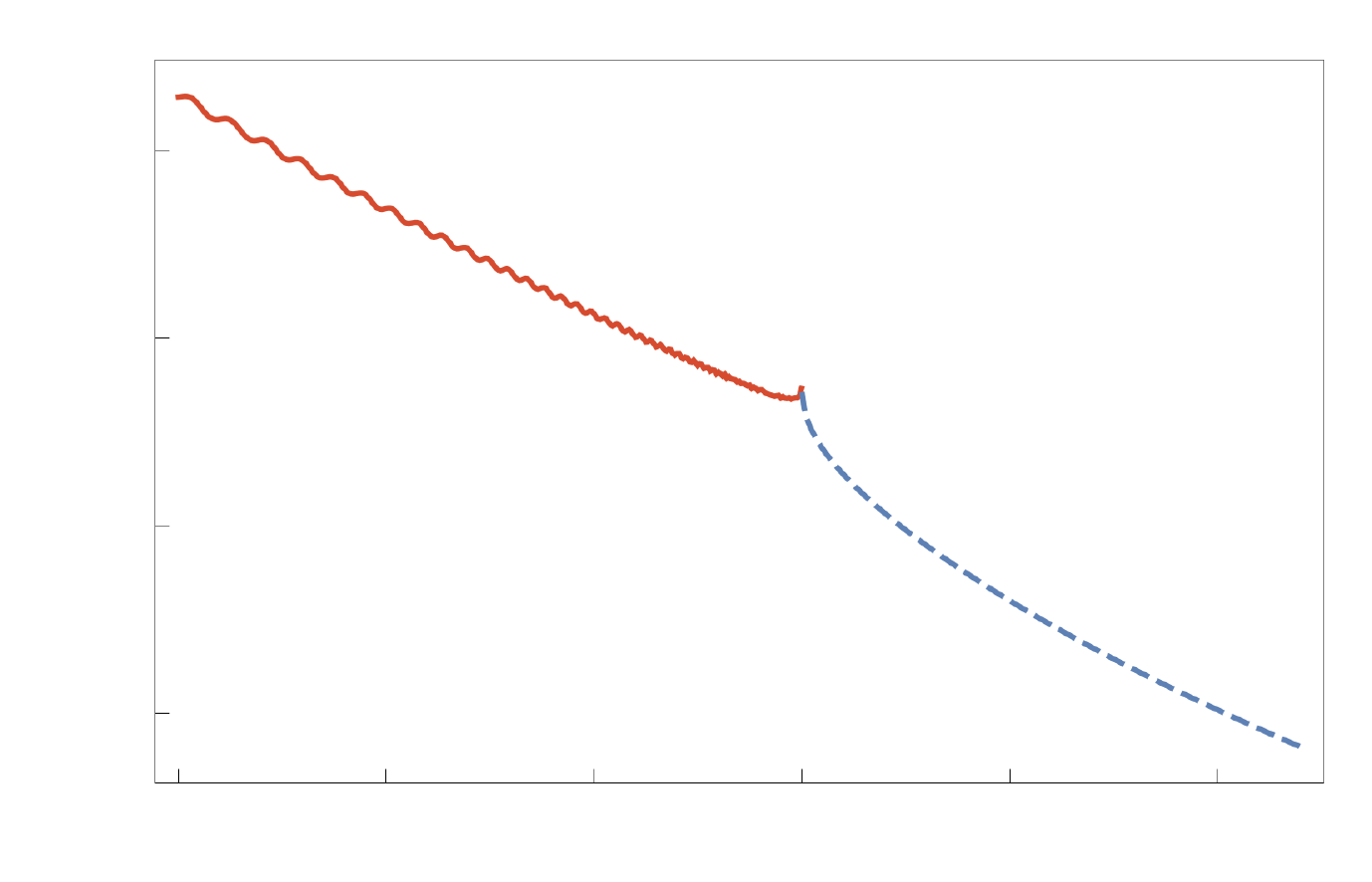}
\put(-130,-2){$D$}
\put(-249,10){\scriptsize{$-1.5$}}
\put(-203,10){\scriptsize{$-1$}}
\put(-166,10){\scriptsize{$-0.5$}}
\put(-115,10){\scriptsize{$0$}}
\put(-77,10){\scriptsize{$0.5$}}
\put(-32.5,10){\scriptsize{$1$}}
\put(-249,31){\scriptsize{$1$}}
\put(-256,68){\scriptsize{$1.4$}}
\put(-256,105){\scriptsize{$1.8$}}
\put(-256,143){\scriptsize{$2.2$}}
\put(-275,68){\rotatebox{90}{$10^{-11}{\cal E}_{{\cal N}}$}}
\vspace{30px}
\includegraphics[width=0.7\linewidth]{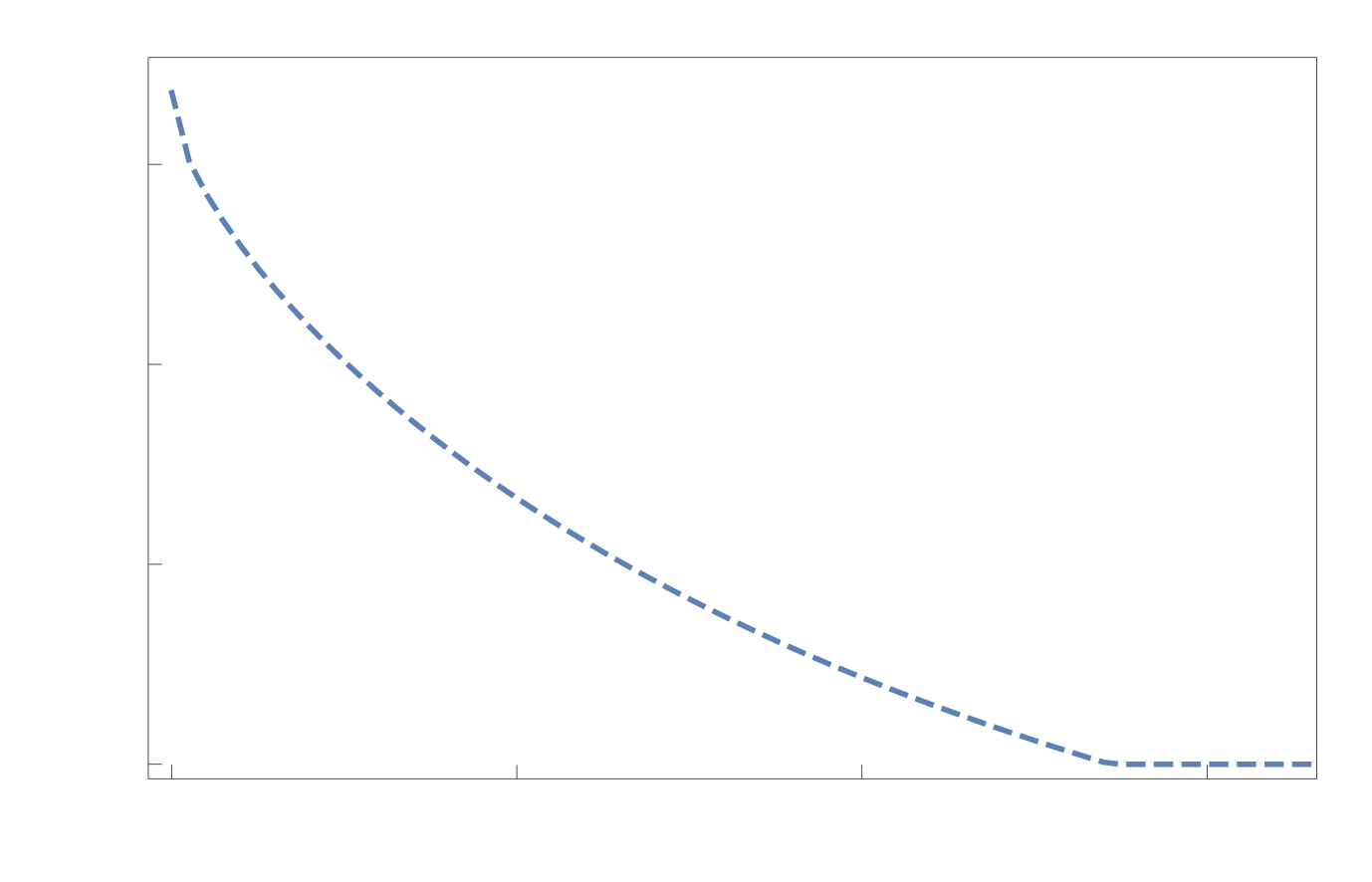}
\put(-130,-2){$D$}
\put(-241.5,11){\scriptsize{$0$}}
\put(-172,11){\scriptsize{$2$}}
\put(-104,11){\scriptsize{$4$}}
\put(-34,11){\scriptsize{$6$}}
\put(-251,21){\scriptsize{$0$}}
\put(-257,61){\scriptsize{$0.5$}}
\put(-251,100.5){\scriptsize{$1$}}
\put(-257,141){\scriptsize{$1.5$}}
\put(-275,68){\rotatebox{90}{$10^{-11}{\cal E}_{{\cal N}}$}}
\vspace{20px}
\caption{\label{fig_opus1d_Negativity_vs_D}\small{Logarithmic negativity of the Minkowski vacuum for two counter-accelerated modes, for fixed and equal proper accelerations ${\cal A}_{\text{I}}={\cal A}_{\text{II}} = 0.1$ as a function of the distance $D$. We have chosen $L=2$, $m=0.1$, and $\Omega_0\approx 5$. In the upper plot we focus on the neighborhood of $D=0$ and the lower plot shows the behavior of the negativity in a larger scale $D>0$. Solid lines correspond to $D<0$ and the dashed lines correspond to $D>0$.}}
\end{figure}

\begin{figure}
\centering
\includegraphics[width=0.7\linewidth]{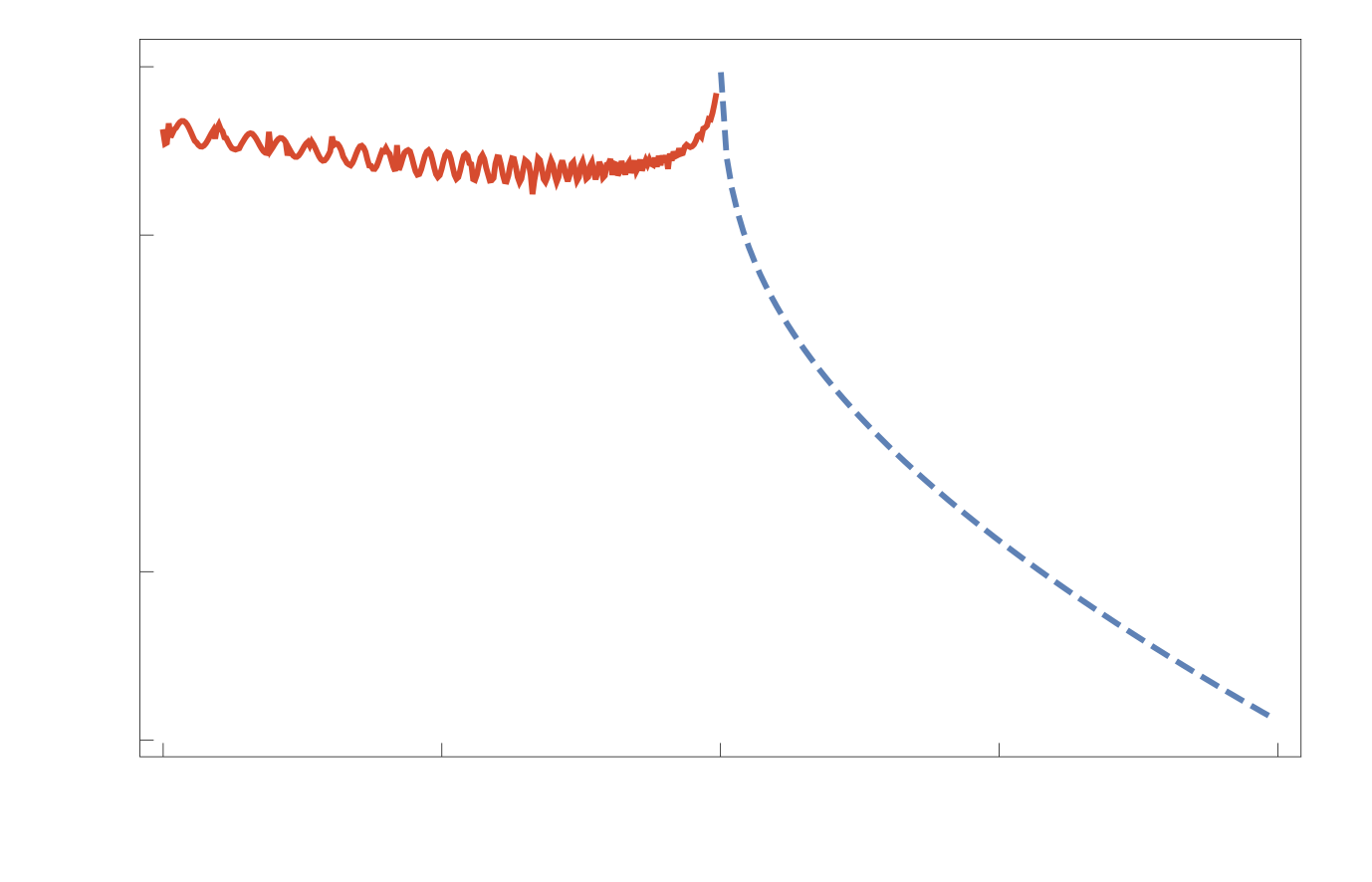}
\put(-130,-2){${\cal A}_\Lambda$}
\put(-250,15){\scriptsize{$0.099$}}
\put(-198,15){\scriptsize{$0.0995$}}
\put(-135,15){\scriptsize{$0.1$}}
\put(-87,15){\scriptsize{$0.1005$}}
\put(-29,15){\scriptsize{$0.101$}}
\put(-263,26){\scriptsize{$1.46$}}
\put(-263,58){\scriptsize{$1.52$}}
\put(-263,125){\scriptsize{$1.64$}}
\put(-259,159){\scriptsize{$1.7$}}
\put(-275,73){\rotatebox{90}{$10^{-11}{\cal E}_{{\cal N}}$}}
\vspace{30px}
\includegraphics[width=0.7\linewidth]{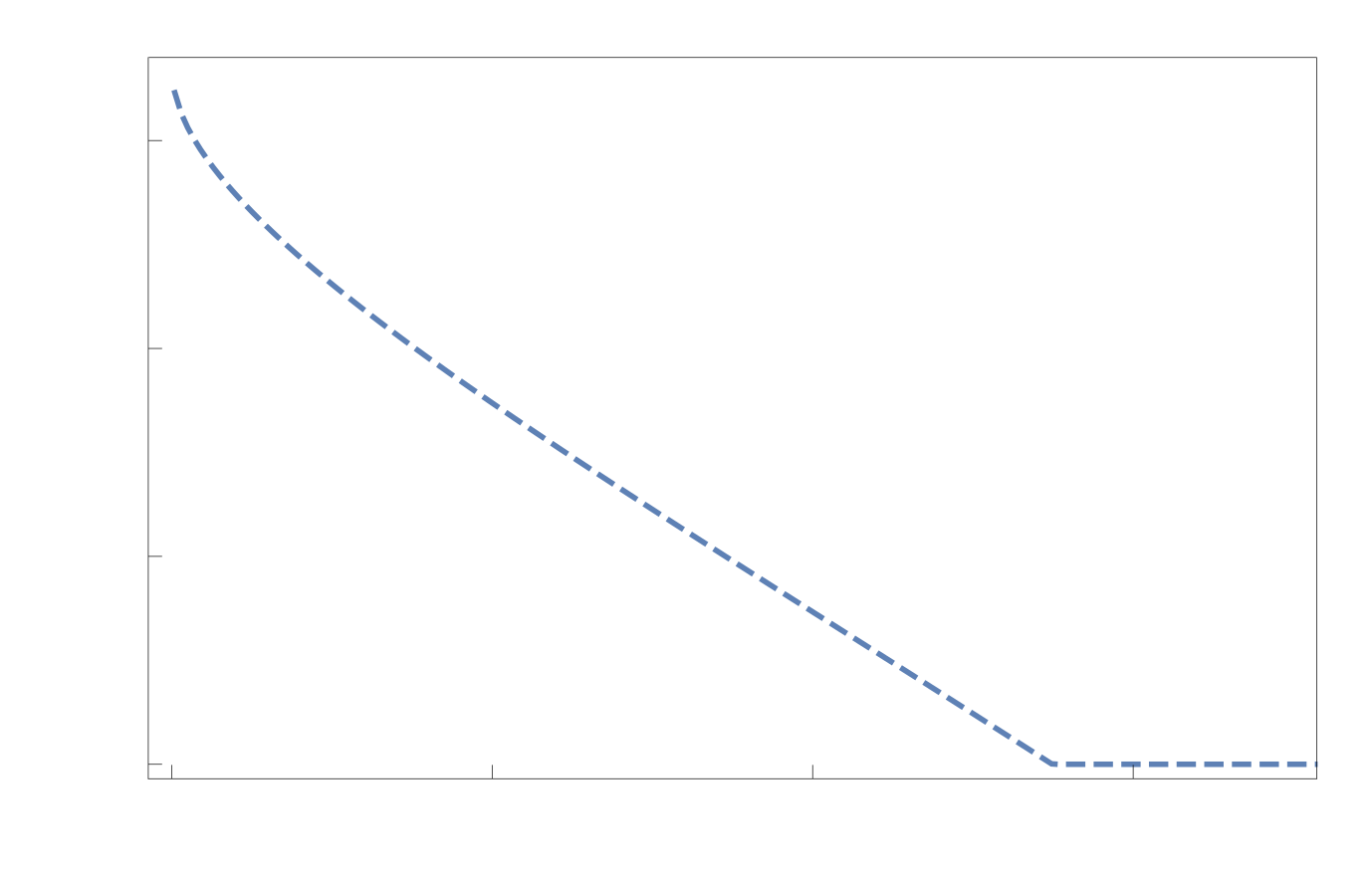}
\put(-130,-2){${\cal A}_\Lambda$}
\put(-244,11){\scriptsize{$0.1$}}
\put(-183,11){\scriptsize{$0.11$}}
\put(-119,11){\scriptsize{$0.12$}}
\put(-55,11){\scriptsize{$0.13$}}
\put(-251,21){\scriptsize{$0$}}
\put(-257,62){\scriptsize{$0.5$}}
\put(-251,103.5){\scriptsize{$1$}}
\put(-257,144.5){\scriptsize{$1.5$}}
\put(-275,73){\rotatebox{90}{$10^{-11}{\cal E}_{{\cal N}}$}}
\vspace{20px}
\caption{\label{fig_opus1d_Negativity_fixed_distance}\small{Logarithmic negativity of the Minkowski vacuum for two counter-accelerated modes, as a function of proper acceleration ${\cal A}_{\text{I}} = {\cal A}_{\text{II}}\equiv {\cal A}_\Lambda$ for $D = 20-\frac{2}{{\cal A}_\Lambda}$ such that the separation between modes is fixed and equal to $20$. We have chosen $L=2$, $m=0.1$, and $\Omega_0\approx 5$. In the upper plot we focus on the neighborhood of $D=0$ and the lower plot shows the behavior of the negativity in a larger scale, when $D>0$. Solid lines correspond to $D<0$ and the dashed lines correspond to $D>0$.}}
\end{figure}

Let us first discuss the results for the counter-accelerating observers. The integrals $\eqref{opus_b1_Nlambda}$ and $\eqref{opus_b1_N12_Dn0}$ are calculated numerically. The former one is simple, but the latter is very challenging. It includes a double integration of the overlap integral, which is also computed numerically. Producing each plot for the $D\neq 0$ case takes several days. For $D=0$ the simpler expression \eqref{opus_b1_N12_D0} may be used. The plots of the results are presented in the Figures \ref{fig_opus1d_D0_vs_Accelerations}-\ref{fig_opus1d_Negativity_fixed_distance}.

In Fig.~\ref{fig_opus1d_D0_vs_Accelerations}, we graph the logarithmic negativity for the state of the output modes characterized by the same width $L=0.1$ and central frequency $\Omega_0\approx 5$ as a function of the proper accelerations ${\cal A}_{\text{I}}$ and ${\cal A}_{\text{II}}$. Here, we show the results for the case of $D=0$. As discussed in Sec. \ref{opus_modes}, we only consider proper accelerations such that ${\cal A}_\Lambda L\ll 1$. Fig.~\ref{fig_opus1d_D0_vs_Accelerations} shows that the entanglement present in the vacuum state, as seen by the accelerating modes $\psi_\Lambda$ is an increasing function of both proper accelerations ${\cal A}_{\text{I}}$ and ${\cal A}_{\text{II}}$. This is consistent with the results known in the literature \cite{andrzej_loc_proj}. Next, in Fig.~\ref{fig_opus1d_Negativity_vs_LOmega} we investigate how the entanglement of the output state depends on the central frequency of the modes, $\Omega_0$, and their width, $L$. For Fig.~\ref{fig_opus1d_Negativity_vs_LOmega} we consider both modes to be identical and characterized by a fixed proper acceleration ${\cal A}_{\text{I}} = {\cal A}_{\text{II}} = 0.1$, and the separation $D=0$. It turns out that the extracted entanglement is larger in the infrared end of the spectrum and it also increases with decreasing $L$. These results are not very surprising and find confirmation in the literature on entanglement of the vacuum \cite{andrzej_loc_proj}.

A more challenging and surprising result is shown in Fig.~\ref{fig_opus1d_Negativity_vs_D}, where we compute the logarithmic negativity for a pair of identical, counter-accelerating output modes characterized by the same proper accelerations ${\cal A}_{\text{I}}={\cal A}_{\text{II}}=0.1$, $L=2$, and $\Omega_0\approx 5$, as a function of the distance $D$. Our framework allows us to fix proper accelerations and therefore study the effect of the spatial separation only. In the upper plot in Fig.~\ref{fig_opus1d_Negativity_vs_D} we consider the range of separations $D$ that are close to zero. When $D$ is positive, the entanglement slowly decreases as a function of $D$, but when $D$ changes sign, the behavior of the plot changes and we observe rapid oscillations on the increasing curve, as $D$ decreases. Note that in order to satisfy the condition $[\hat{d}_{\text{I}},\hat{d}_{\text{II}}^{(\dagger)}]=0$, we only consider negative values of $D$ such that $\frac{2}{{\cal A}_\Lambda} + D \gg L$. In \eqref{opus_b1_N12_lim_D0} we have analytically proven that the off-diagonal terms of the noise matrix, $N^\pm_{\text{I},\text{II}}$,  are continuous functions at $D=0$, and so is the logarithmic negativity. In the lower plot of Fig.~\ref{fig_opus1d_Negativity_vs_D} we show how ${\cal E_N}$ changes for large, positive values of $D$. As expected, the detected entanglement vanishes for increasing positive separations $D$. We discover a sudden death of entanglement that occurs for a finite distance $D$.

The results plotted in Fig.~\ref{fig_opus1d_Negativity_fixed_distance} are complementary to the results shown in Fig.~\ref{fig_opus1d_Negativity_vs_D}. The setting is also symmetric, i.e., the parameters characterizing both output modes are identical. In this example, we fix the spatial separation between the output modes, equal to $\frac{2}{{\cal A}_\Lambda} + D$, and study the amount of entanglement as a function of the proper acceleration of the output modes. It is clear that in order to fix the distance between the modes, the change in proper acceleration has to be compensated by a change in the separation $D$. The resulting dependence of the detected logarithmic negativity is shown in Fig.~\ref{fig_opus1d_Negativity_fixed_distance}. In the upper figure, we plot negativity as a function of the proper acceleration. We choose the range of the plot such that we can study the transition corresponding to the change of sign of the distance $D$. Again we see that the behavior of ${\cal E}_{{\cal N}}$ changes for these two regimes. The plot exhibits oscillations for $D<0$ and it is smooth for $D>0$. In the lower figure, we plot ${\cal E}_{{\cal N}}$ for a larger range of $D>0$. Once again, we discover a sudden death of entanglement that occurs for finite proper acceleration ${\cal A}_\Lambda$ and distance $D$. 

Finally we move on to the scenario in which both output modes are accelerated in the same direction, as schematically shown in Fig.~\ref{fig_opus1d_Para}. Because of the form of Eq. \eqref{opus_b1_N12_D0} it can be shown that for $D=0$ we obtain no entanglement in the regime whereby $N_\Lambda$ and $N^\pm_{\text{I},\text{II}}$ are much smaller than unity. This is the case in most of the physical setups of interest, whereby the average number of Unruh particles is much smaller than unity. There may be some entanglement present, when the noise term is larger, but this is the case only at very high accelerations, much higher than those which the Unruh effect is expected to be observable at. Our framework is suitable to describe those, but we choose not to compute the logarithmic negativity for such a range of parameters, because they have been far from achievable physically as of today. We have numerically evaluated the logarithmic negativity of the output state of the channel for a range of parameters, and found zero negativity also for $D\neq 0$. This strongly suggests that there is no vacuum entanglement within the parallel accelerations setup, in the regime whereby $N_\Lambda$ and $N^\pm_{\text{I},\text{II}}$ are much smaller than unity.

Let us remark that the framework we have introduced, allows one to reliably study the effect of the accelerations of the observers and the distance between them on the observed entanglement. To the best of our knowledge, our results have not been shown previously. We have found that the entanglement witnessed by the accelerated observers in the Minkowski vacuum exhibits an interesting oscillatory behavior, when the two Rindler wedges overlap, i.e., for $D<0$. When the two Rindler wedges are separated by a positive distance, $D>0$, we also find a new phenomenon of sudden death of entanglement, similar to the effect reported in~\cite{dnon0_nick}. We believe that the tools presented in this work can be further used to investigate the rich structure of the vacuum field entanglement.

In our work~\cite{lor_1d} the channel has also been applied to a non-vacuum state, namely a two-mode squeezed thermal state. Here, the results are just briefly outlined. The fidelity of the channel was evaluated numerically and it has been found that the fidelity degrades with increasing accelerations of the observers, but the higher the temperature of the input state is, the weaker the effect becomes. The reason for this degradation is the inevitable mismatch of the inertial and accelerated modes, which increases with increasing accelerations. Also, apart from the modes \eqref{opus_b1_psi_pass}, different output modes have been analyzed, which were chosen depending on the accelerations ${\A}_\Lambda$, such that the coefficients $\alpha_\Lambda$ are maximized. For these modes, the effect is negligible and the fidelity is nearly perfect.

This concludes the chapter on the case of our framework applied to a real scalar field in $1+1$ dimensions. Now we proceed with the development of an analogous framework in $3+1$ dimensions.


\chapter{Real scalar field in 3+1-dimensional spacetime} \label{chb3d}

In this chapter we construct the $3+1$-dimensional case of the framework from the previous chapter. We then use it to investigate the bipartite entanglement of the vacuum. In Sec. \ref{sec_opus_b3_mod_rind}, we discuss the $3+1$-dimensional modified Rindler coordinates. In Sec. \ref{sec_opus_b3_setup} the setup of the problem is formulated and the similarities to the $1+1$-dimensional case are outlined. In Sec. \ref{sec_opus_b3_N} the noise matrix derived. In Sec. \ref{sec_opus_b3_modes} we choose the inertial and accelerating modes, and eventually in Sec. \ref{sec_opus_b3_results} the results for the vacuum entanglement are presented. The chapter concludes with Sec. \ref{sec_opus_b3_skew} which discusses an outlook for extending the framework to observers accelerating at a general angle with respect to each other.


\section{Modified Rindler coordinates}   \label{sec_opus_b3_mod_rind}

In this chapter we develop the framework introduced in Chapter \ref{chb1d}. To allow for the possibility of shifting the Rindler wedges relative to each other, we introduce the $3+1$D modified Rindler coordinates. Their corresponding modified Rindler transformation is as follows:
\begin{equation}      \label{opus_b3_mod_Rindler}
\begin{aligned}
&t = \pm\chi\sinh a\eta,
\\
&x = \pm\chi\cosh a\eta \pm \frac{D}{2},
\\
&{\bf x_\pe} = {\bf x'_\pe},
\end{aligned}
\end{equation}
where $+$ refers to the coordinates covering region I, $-$ refers to the coordinates covering region II and $D$ may be positive or negative. The coordinates have the same form as their $1+1$-dimensional counterparts \eqref{opus_b1_mod_Rindler} but there are two additional dimensions perpendicular to the direction of acceleration.

The mode decomposition \eqref{qft_dec_b_3d} is altered for any $D$ to:
\begin{equation}     \label{opus_dec_b_3d}
\begin{aligned}
\hat{\Phi}=\int_0^{\infty}\text{d}\Omega\int\text{d}^2{\bf k_\perp}
\big(
&\wiOk \biOk + \wiOk\s \biOk\d 
\\
+ &\wiiOk \biiOk + \wiiOk\s \biiOk\d
\big)
+
\hat{\Phi}_\x{III}(D),
\end{aligned}
\end{equation} 
where the operator $\hat{\Phi}_\x{III}(D)=0$ when $D=0$. This decomposition is used in the derivation of the noise matrix elements for $D\neq 0$.

\section{The setup and similarities to the $1+1$-dimensional framework}     \label{sec_opus_b3_setup}

Let us state the assumptions that are made.
We work with a real, $3+1$-dimensional massive Klein-Gordon field $\hat{\Phi}$, with mass $m$. In this case the mass may be greater or equal to zero. We begin with an arbitrary Gaussian state of two localized inertial modes $\phi_\x{I}$ and $\phi_\x{II}$, with associated annihilation operators $\hat{f}_\x{I}$ and $\hat{f}_\x{II}$. The mode $\phi_\x{I}$ at $t=0$ is localized within region I, and $\phi_\x{II}$~at $t=0$ is localized within region II. Similarly, we have two localized accelerating modes $\psi_\x{I}$ and $\psi_\x{II}$, with associated annihilation operators $\hat{d}_\x{I}$ and~$\hat{d}_\x{II}$. The mode $\psi_\x{I}$ is at all times localized within region I, and $\psi_\x{II}$ is at all times localized within region II. We also maintain the assumptions \eqref{opus_b1_ass1}-\eqref{opus_b1_ass6}. The motion of both accelerated observers occurs along the $x$-axis, unless stated otherwise (in Sec. \ref{sec_opus_b3_skew}).

The field decompositions that we consider are \eqref{qft_dec_a}, \eqref{opus_bogo_fd} and \eqref{opus_dec_b_3d}. We assume that the modes have no negative frequency contributions and this leads to the following decompositions of the observers' annihilation operators:
\begin{align}                 \label{opus_b3_dinb}
\hat{d}_\Lambda &=\int\text{d}\Omega\int\text{d}^2{\bf k_\perp}\,\,
\olok \hat{b}_{\Lambda\Omega\bf k_\perp},
\end{align}
where, as previously $\Lambda\in \{\x{I},\x{II}\}$.

The channel is again of the form \eqref{opus_b1_channel_x}-\eqref{opus_b1_channel_s}. Furthermore, the reasoning presented in Sec. \ref{sec_opus_M} does not rely on any preassumptions about the dimensionality of the spacetime thus, it still holds in this case. In particular the matrix $M$ is of the form \eqref{opus_M} with the definitions \eqref{opus_Mbogo_defn}.

Let us proceed with the derivation of the noise matrix $N$.

\section{Derivation of the noise matrix}        \label{sec_opus_b3_N}

The beginning of the derivation follows the steps as the $1+1$D case up to the expressions \eqref{opus_b1_s11}-\eqref{opus_b1_s24}, which are the elements of the covariance matrix $\sdvac$. We further express the terms arising in \eqref{opus_b1_s11}-\eqref{opus_b1_s24} in terms of the modified Rindler creation and annihilation operators via \eqref{opus_b3_dinb} to obtain:
\begin{equation}             \label{opus_b3_d1d1}
\begin{aligned}
{}_M\bra{0} \hat{d}_\x{I} \hat{d}_\x{I} \ket{0}_M = 
\iint\x{d}\Omega\x{d}\Xi 
\iint\text{d}^2{\bf k_\perp}\text{d}^2{\bf l_\perp}
\,\, &\oIok \oIxl
\\ \times \,\,
&{}_M\bra{0} \biOk \biXl \ket{0}_M,
\end{aligned}
\end{equation}
\begin{equation}                        \label{opus_b3_d1d1d}
\begin{aligned}
{}_M\bra{0} \hat{d}_\x{I}\d \hat{d}_\x{I} \ket{0}_M = 
\iint\x{d}\Omega\x{d}\Xi 
\iint\text{d}^2{\bf k_\perp}\text{d}^2{\bf l_\perp}
\,\, &\oIok\s \oIxl
\\ \times \,\,
&{}_M\bra{0} \biOk\d \biXl \ket{0}_M,
\end{aligned}
\end{equation}
\begin{equation}               \label{opus_b3_d1d2}
\begin{aligned}
{}_M\bra{0} \hat{d}_\x{I} \hat{d}_\x{II} \ket{0}_M = 
\iint\x{d}\Omega\x{d}\Xi 
\iint\text{d}^2{\bf k_\perp}\text{d}^2{\bf l_\perp}
\,\, &\oIok \oIIxl
\\ \times \,\,
&{}_M\bra{0} \biOk \biiXl \ket{0}_M,
\end{aligned}
\end{equation}                     
\begin{equation}                       \label{opus_b3_d1d2d}
\begin{aligned}
{}_M\bra{0} \hat{d}_\x{I} \hat{d}_\x{II}\d \ket{0}_M = 
\iint\x{d}\Omega\x{d}\Xi 
\iint\text{d}^2{\bf k_\perp}\text{d}^2{\bf l_\perp}
\,\, &\oIok \oIxl\s
\\ \times \,\,
&{}_M\bra{0} \biOk \biiXl\d \ket{0}_M.
\end{aligned}
\end{equation}
The remaining two expressions follow from replacing subscripts I with II in \eqref{opus_b3_d1d1} and \eqref{opus_b3_d1d1d}. Only the expectation values \eqref{opus_b3_d1d2} and \eqref{opus_b3_d1d2d} are $D$-dependent. The other expectation values may be calculated at $D=0$. They yield~\cite{crispino_higuchi}:
\begin{align}          \label{opus_b3_bb1}
{}_M\bra{0} \biOk \biXl \ket{0}_M
=&\,\,
0,
\\                      \label{opus_b3_bb2}
{}_M\bra{0} \biOk\d \biXl \ket{0}_M
=&\,\,
\frac{1}{e^{\frac{2\pi\Omega}{a}}-1}\,
\delta(\Omega-\Xi)\,
\delta^2({\bf k_\perp}-{\bf l_\perp}).
\end{align}

This allows us to write $\sdvac$ in the form \eqref{opus_sdvac} with $N_\x{I,II}^\pm$ defined as in~\eqref{opus_N12general} and:
\begin{equation}        \label{opus_b3_Nlambda}
N_\Lambda \defn 2
\int\x{d}\Omega \int\text{d}^2{\bf k_\perp} \,\,
\frac{\left| \olok \right|^2}{e^\frac{2\pi\Omega}{a}-1}.
\end{equation}
Let us now move on to the elements appearing in the off-diagonal blocks.

\subsection{Antiparallel accelerations, $D=0$}

This is the case, whereby the Rindler coordinates are unmodified. Here, the results are~\cite{crispino_higuchi}:
\begin{align}        \label{opus_b3_bb3}
{}_M\bra{0} \biOk \biiXl \ket{0}_M
=& \,\,
\frac{1}{2\sinh\left(\frac{\pi\omega}{a}\right)} \,
\delta(\omega-\lambda) \,
\delta^2({\bf k_\perp}+{\bf l_\perp}),
\\                   \label{opus_b3_bb4}
{}_M\bra{0} \biOk \biiXl\d \ket{0}_M
=& \,\,
0.
\end{align}
Combining this with \eqref{opus_N12general}, \eqref{opus_b3_d1d2} and \eqref{opus_b3_d1d2d} we obtain:
\begin{equation}
N_\text{I,II}^\pm
=
\int\text{d}\Omega\int\text{d}^2{\bf k_\perp}
\frac{\oIok \oIIomk}{\sinh\left(\frac{\pi\Omega}{a}\right)}.
\end{equation}

This completes the derivation of $N$ for $D=0$. Now let us proceed with the case of $D\neq 0$.

\subsection{Antiparallel accelerations, $D\neq 0$}     \label{sec_opus_N12_Dn0}

Let us start with the Bogolyubov transformation between the Minkowski and the unmodified Rindler frame, which is of the form \eqref{rqi_bogo_b}:
\begin{equation}
\hat{b}_{\Lambda\Omega{\bf k_\perp}}=
\int\text{d}l_x\int\text{d}^2{\bf l_\perp}
\left(
\alpha_{\Omega{\bf k_\perp}l_x{\bf l_\perp}}^{\Lambda\star}
\hat{a}_{l_x{\bf l_\perp}}
-
\beta_{\Omega{\bf k_\perp}l_x{\bf l_\perp}}^{\Lambda\star}
\hat{a}_{l_x{\bf l_\perp}}\d
\right),
\end{equation}
where the Bogolyubov coefficients are defined in terms analogous overlaps as in \eqref{rqi_u_bogo1d}. The subscripts of the Bogolyubov coefficients include all of the wave vector components that the coefficients depend on. The expressions for them are~\cite{takagi,crispino_higuchi}:
\begin{align}           \label{opus_b3_bogo1}
\alpha_{\Omega{\bf k_\perp}l_x{\bf l_\perp}}^{\text{I}}
&=
\frac{e^\frac{\pi\Omega}{2a}\,\,\delta^2({\bf k_\perp}-{\bf l_\perp})}
{\sqrt{4\pi a\sqrt{l_x^2+l_\perp^2+m^2}\sinh{\frac{\pi\Omega}{a}}}}
\left(
\frac{\sqrt{l_x^2+l_\perp^2+m^2}+l_x}{\sqrt{l_x^2+l_\perp^2+m^2}-l_x}
\right)_,^{-\frac{i\Omega}{2a}}
\\
\beta_{\Omega{\bf k_\perp}l_x{\bf l_\perp}}^{\text{I}}
&=
\frac{-\,\,e^{-\frac{\pi\Omega}{2a}}\,\,\delta^2({\bf k_\perp}+{\bf l_\perp})}
{\sqrt{4\pi a\sqrt{l_x^2+l_\perp^2+m^2}\sinh{\frac{\pi\Omega}{a}}}}
\left(
\frac{\sqrt{l_x^2+l_\perp^2+m^2}+l_x}{\sqrt{l_x^2+l_\perp^2+m^2}-l_x}
\right)_,^{-\frac{i\Omega}{2a}}
\\
\alpha_{\Omega{\bf k_\perp}l_x{\bf l_\perp}}^{\text{II}}
&=
\frac{e^\frac{\pi\Omega}{2a}\,\,\delta^2({\bf k_\perp}-{\bf l_\perp})}
{\sqrt{4\pi a\sqrt{l_x^2+l_\perp^2+m^2}\sinh{\frac{\pi\Omega}{a}}}}
\left(
\frac{\sqrt{l_x^2+l_\perp^2+m^2}+l_x}{\sqrt{l_x^2+l_\perp^2+m^2}-l_x}
\right)_,^{\frac{i\Omega}{2a}}
\\                   \label{opus_b3_bogo4}
\beta_{\Omega{\bf k_\perp}l_x{\bf l_\perp}}^{\text{II}}
&=
\frac{-\,\,e^{-\frac{\pi\Omega}{2a}}\,\,\delta^2({\bf k_\perp}+{\bf l_\perp})}
{\sqrt{4\pi a\sqrt{l_x^2+l_\perp^2+m^2}\sinh{\frac{\pi\Omega}{a}}}}
\left(
\frac{\sqrt{l_x^2+l_\perp^2+m^2}+l_x}{\sqrt{l_x^2+l_\perp^2+m^2}-l_x}
\right)_.^{\frac{i\Omega}{2a}}
\end{align}
Upon a relative shift of the Rindler wedges, these coefficients get modified in the same way as in \eqref{opus_b1_modbogos}. Hence, our expectation values of the modified Rindler creation and annihilation operator products take the form:
\begin{align}
{}_M\bra{0} \bi[\Omega{\bf k_\perp}]\bi[\Xi{\bf l_\perp}]\ket{0}_M
=&
-\iint\text{d}p_x\, \text{d}^2{\bf p_\perp}\,\,
\alpha_{\Omega{\bf k_\perp}p_x{\bf p_\perp}}^{\text{I}\star}
\beta_{\Xi{\bf l_\perp}p_x{\bf p_\perp}}^{\text{I}\star},
\\
{}_M\bra{0} \bi[\Omega{\bf k_\perp}]\bi[\Xi{\bf l_\perp}]\d\ket{0}_M
=&
\iint\text{d}p_x\, \text{d}^2{\bf p_\perp}\,\,
\alpha_{\Omega{\bf k_\perp}p_x{\bf p_\perp}}^{\text{I}\star}
\alpha_{\Xi{\bf l_\perp}p_x{\bf p_\perp}}^{\text{I}},
\\
{}_M\bra{0}\bi[\Omega{\bf k_\perp}]\bii[\Xi{\bf l_\perp}]\ket{0}_M
=&
-\iint\text{d}p_x\, \text{d}^2{\bf p_\perp}\,\,
\alpha_{\Omega{\bf k_\perp}p_x{\bf p_\perp}}^{\text{I}\star}
\beta_{\Xi{\bf l_\perp}p_x{\bf p_\perp}}^{\text{II}\star},
\\
{}_M\bra{0}\bi[\Omega{\bf k_\perp}]\bii[\Xi{\bf l_\perp}]\d\ket{0}_M
=&
\iint\text{d}p_x\, \text{d}^2{\bf p_\perp}\,\,
\alpha_{\Omega{\bf k_\perp}p_x{\bf p_\perp}}^{\text{I}\star}
\alpha_{\Xi{\bf l_\perp}p_x{\bf p_\perp}}^{\text{II}}.
\end{align} 
Now we substitute \eqref{opus_b3_bogo1}-\eqref{opus_b3_bogo4} in the above, and combine it with \eqref{opus_b3_d1d1}-\eqref{opus_b3_d1d2d}. There arises an integral which is evaluated following the steps shown in subsection \ref{sec_opus_N12_Dn0}:
\begin{equation}
\begin{aligned}
\int\frac{\text{d}p_x}{\sqrt{p_x^2+k_\perp^2+m^2}}
&\left(
\frac{\sqrt{p_x^2+k_\perp^2+m^2}+p_x}{\sqrt{p_x^2+k_\perp^2+m^2}-p_x}
\right)^{\frac{i}{2a}(\Omega\pm\Xi)}
e^{iDp_x}
\\=&\,
2\,\cosh\frac{\pi(\Omega\pm\Xi)}{2a}
K_{\frac{i(\Omega\pm\Xi)}{a}}\left(\sqrt{k_\perp^2+m^2}\,|D|\right)
\\
-&
2\,\frac{D}{|D|}\sinh\frac{\pi(\Omega\pm\Xi)}{2a}
K_{\frac{i(\Omega\pm\Xi)}{a}}\left(\sqrt{k_\perp^2+m^2}\,|D|\right).
\end{aligned}
\end{equation}
This altogether, substituted into \eqref{opus_N12general}, leads to the following result:
\begin{equation}       \label{opus_b3_N12_Dn0}
\begin{aligned}
N_\x{I,II}^\pm
=
&\frac{1}{\pi a}
\int\text{d}^2{\bf k_\perp}
\iint\text{d}\Omega\text{d}\Xi
\frac{\oIok}{\sqrt{\sinh\left(\frac{\pi\Omega}{a}\right)\sinh\left(\frac{\pi\Xi}{a}\right)}}
\\ \times
\bigg[&
-\oIIxmk
e^{\frac{\pi}{2a}(\Omega-\Xi)(1-\frac{D}{|D|})}
K_{\frac{i(\Omega-\Xi)}{a}}\left(\sqrt{k_\perp^2+m^2}\,|D|\right)
\\
&+
\oIIxk\s\,
e^{\frac{\pi}{2a}(\Omega+\Xi)(1-\frac{D}{|D|})}
K_{\frac{i(\Omega+\Xi)}{a}}\left(\sqrt{k_\perp^2+m^2}\,|D|\right)
\bigg].
\end{aligned}
\end{equation}

This completes the derivation of the noise matrix for the case of antiparallel-accelerated modes. Now let us proceed with the case of the parallel-accelerated modes.

\subsection{Parallel accelerations}

The derivation in this section follows the steps analogous to the ones shown in the subsection \ref{sec_opus_b1_par}. Here we quote the result:
\begin{equation}       \label{opus_b3_N12_Dn0_par}
\begin{aligned}
&N_\x{I,II}^\pm
=
\frac{1}{\pi a}
\int\text{d}^2{\bf k_\perp}
\iint\text{d}\Omega\text{d}\Xi
\frac{\oIok}{\sqrt{\sinh\left(\frac{\pi\Omega}{a}\right)\sinh\left(\frac{\pi\Xi}{a}\right)}}
\\ \times
\bigg[&
-\oIIxmk
e^{\frac{\pi}{2a}\left[(\Omega-\Xi)-(\Omega+\Xi)\frac{D}{|D|}\right]}
K_{\frac{i(\Omega+\Xi)}{a}}\left(\sqrt{k_\perp^2+m^2}\,|D|\right)
\\
&+
\oIIxk\s\,
e^{\frac{\pi}{2a}\left[(\Omega+\Xi)-(\Omega-\Xi)\frac{D}{|D|}\right]}
K_{\frac{i(\Omega-\Xi)}{a}}\left(\sqrt{k_\perp^2+m^2}\,|D|\right)
\bigg].
\end{aligned}
\end{equation}
For $D=0$ this reduces to:
\begin{equation}    \label{opus_b3_N12_D0_par}
N_\text{I,II}^\pm
=\pm
\int\text{d}\Omega\int\text{d}^2{\bf k_\perp}
\frac{\oIok\oIIok\s}{\sinh\left(\frac{\pi\Omega}{a}\right)}\,\,
e^{\frac{\pi\Omega}{a}}.
\end{equation}
Thus, the noise matrix has been evaluated for all cases. Let us note that in the $3+1$-dimensional case the channel is also $a$-independent, which can be seen analogously to the way presented in Sec. \ref{sec_opus_a}.

\section{The choice of the modes}      \label{sec_opus_b3_modes}

The input modes are chosen such that they simulate modes in an inertial cavity, i.e. \eqref{qft_u_cav_m}, and the output modes such that they simulate modes in an accelerated cavity, i.e. \eqref{qft_w_cav_m}. All of them are assumed to have a smooth envelope and to be localized in all three spatial dimensions. The accelerated modes are at the same position as the stationary ones at $t=0$. Since the $3+1$-dimensional Klein-Gordon equation is a second order differential equation, it is sufficient to specify the values of spatial profiles of the modes and their time derviatives at $t=0$. The motivation for the choice of the exact shape of the mode functions is analogous to the one provided in Sec.~\ref{opus_modes}. For the input modes we choose:
\begin{equation}
\begin{aligned}           \label{opus_b3_phi}
&\phi_{\Lambda}\Big|_{t=0}=
\mathcal{N_\phi}\,\,
e^{-2(\frac{1}{{\A}_\Lambda L_\pa}\ln({\A}_\Lambda x))^2-\frac{2}{L_\pe^2}(y^2+z^2)}
\\ &\quad\quad\times
\sin\left[\sqrt{\Omega_0^2-2\kappa_\pe^2-m^2}\left(x\mp\frac{1}{{\A}_\Lambda}\right)\right]
\sin[\kappa_\pe y]\sin[\kappa_\pe z],
\\
&\partial_t\phi_{\Lambda}\Big|_{t=0}=
-i\Omega_0\phi_{\Lambda}\Big|_{t=0},
\end{aligned}
\end{equation}
where the upper sign refers to $\Lambda=\x{I}$, the lower sign refers to $\Lambda=\x{II}$. Also, ${\A}_\Lambda$ are the proper accelerations of the respective observers, $L_\pa$ is the length of a wavepacket in the direction parallel to the acceleration, $L_\pe$ is the length of a wavepacket in the direction perpendicular to the acceleration (here taken to be the same in $y$ and $z$ direction, but this does need to be the case). Furthermore, $\Omega_0$ is the frequency which the wavepackets are centered around in the frequency space, $\kappa_\pe$ is the wave vector perpendicular to the direction of the acceleration, and $\mathcal{N_\phi}$ is a normalization constant. The accelerating modes are chosen to be:
\begin{align}                      \label{opus_b3_psi}
&\psi_{\Lambda}\Big|_{t=0}=
\mathcal{N_\psi}\,\,
e^{-2(\frac{1}{{\A}_\Lambda L_\pa}\ln({\A}_\Lambda\chi))^2-\frac{2}{L_\pe^2}(y^2+z^2)}
\\ &\quad\times \nonumber
\Im\left\{
I_{-\frac{i\Omega_0}{{\A}_\Lambda}}\left(\frac{\sqrt{2\kappa_\pe^2+m^2}}{{\A}_\Lambda}\right)
I_{\frac{i\Omega_0}{{\A}_\Lambda}}\left(\sqrt{2\kappa_\pe^2+m^2}\,\chi\right)
\right\}
\sin[\kappa_\pe y]\sin[\kappa_\pe z],
\\ \nonumber
&
\partial_\tau\psi_{\Lambda}\Big|_{\tau=0}=
\mp i\Omega_0\psi_{\Lambda}\Big|_{\tau=0},
\end{align}
where the upper sign refers to $\Lambda=\x{I}$, the lower sign refers to $\Lambda=\x{II}$. It should be understood that this mode is defined in the respective wedge and is zero elsewhere.

Similarly to the previous case, we assume that the accelerating modes are localized in the $x$ direction sufficiently, such that one can assign a single proper acceleration to each of them. This leads to the condition $\frac{1}{{\A}_\Lambda} \gg L$. Moreover, in order to make the negative frequency contributions negligible we let $\Omega_0 \gg \frac{1}{L}$. No additional negative frequency cutoff is applied in this case.

\begin{figure}
\centering
\includegraphics[width=0.6\linewidth]{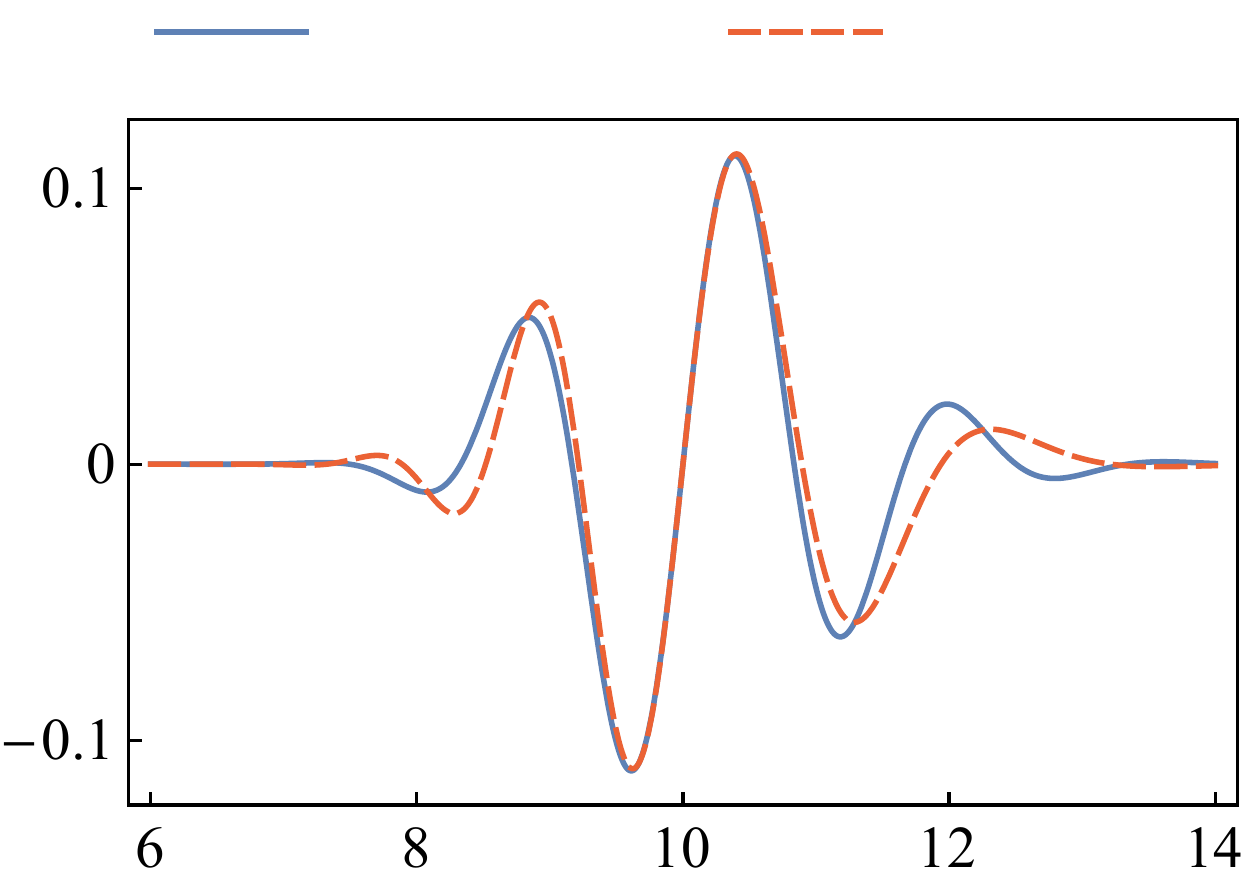}
\put(-167,156){$\phi(x,0)$}
\put(-60,156){{$\psi(\chi,0)$}}
\caption{\small{Comparison between the spatial profiles of modes $\phi_\Lambda$ and $\psi_\Lambda$ for the following choice of parameters: ${\A}_\Lambda=0.1$, $L_{||}=L_\pe=2$, $\Omega_0\approx 5$, $m=0.1$, $\kappa_\pe=2$, along the $x$ axis at $y=z=1$. \label{fig_3d_modes}}}
\end{figure}

Before moving on to the numerical results, in the Fig. \ref{fig_3d_modes} let us compare the spatial profiles of the modes in the $x$ direction. The inevitable mismatch of the modes due to the acceleration is visible. However, there is no mismatch in the directions perpendicular to the acceleration, which can be seen by inspection of the formuale \eqref{opus_b3_phi} and \eqref{opus_b3_psi}. It is a consequence of the fact that the transformation \eqref{opus_b3_mod_Rindler} leaves the coordinates ${\bf x_\pe}$ unchanged.

\section{Results}              \label{sec_opus_b3_results}

Let us now apply our channel to the input Minkowski vacuum state, in order to investigate the vacuum entanglement. Upon this application, the output state is of the form $\vec{X}^{(d)}=\vec{0}$ and \eqref{opus_sdvac}. The results hold also for an input coherent state, with $\vec{X}^{(d)}\neq\vec{0}$.

The numerical calculation is more challenging in the $3+1$-dimensional case than in the $1+1$-dimensional case. 
Now, in the $D=0$ case we have an additional two-dimensional integration over the wave vector perpendicular to the direction of acceleration. Fortunately, one may switch to the polar coordinate system, and perform the integration over the polar angle analytiacally. Therefore, the resultant expression involves a double integral of the overlap which is computed numerically. This is as time-consuming as the integrals from the $1+1$-dimensional case for $D\neq 0$. Furthermore, the expressions for $D\neq 0$ involve quadruple integrals of the overlap, which could not have been computed using the same approach.

To quantify the amount of entanglement, the logarithmic negativity is chosen, as given by the formula \eqref{gqm_neg} with the substitutions: $\boldsymbol\sigma_{AB} = \sdvac\,$ and $\tilde{\Delta} = (1+N_{\text{I}})^2 +  (1+N_{\text{II}})^2 + \left| N_{\text{I,II}} \right|^2$.

\begin{figure}
\centering
\includegraphics[width=0.5\linewidth]{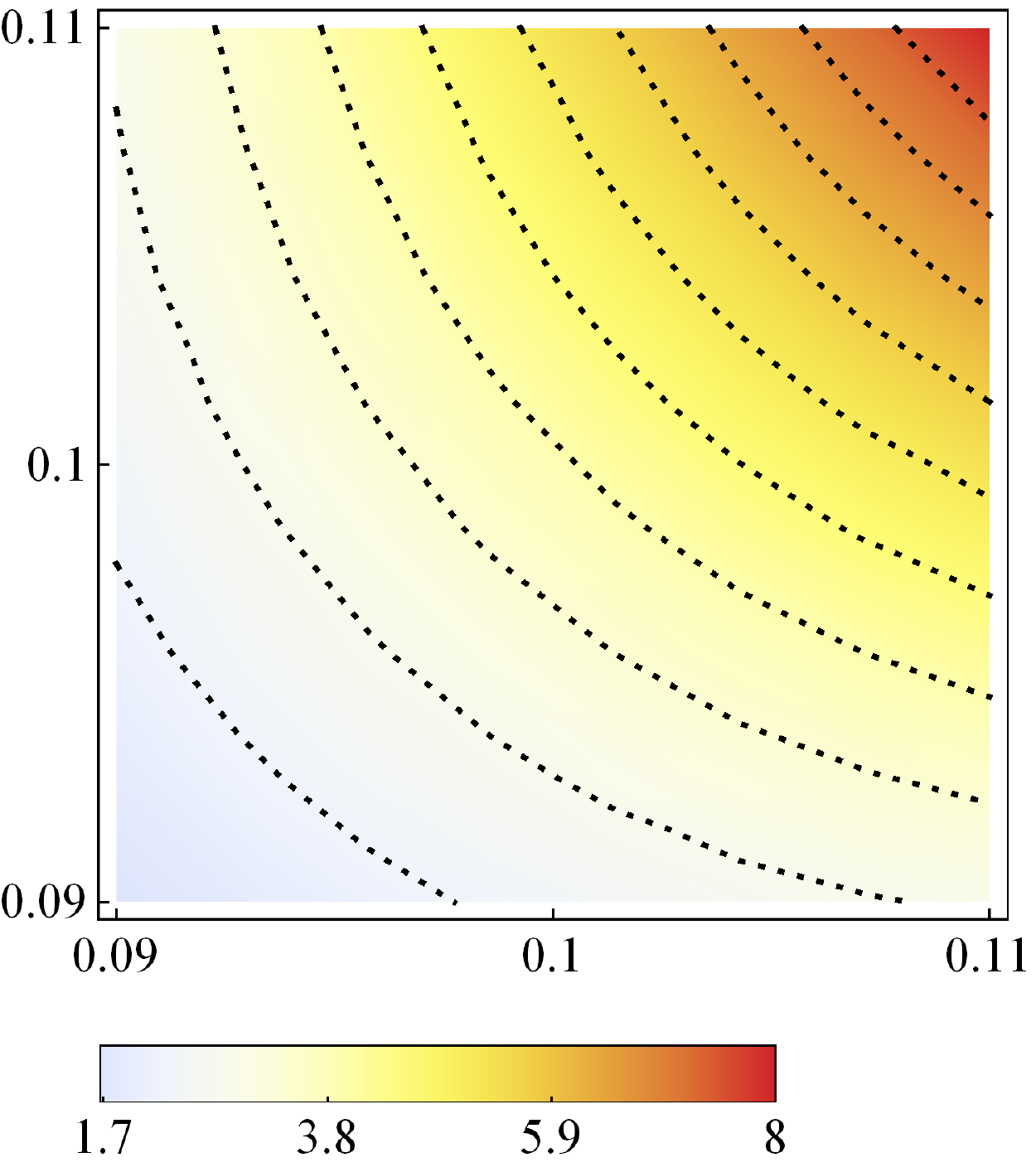}
\put(-195,173){$\A_\text{I}$}
\put(-58,33){$\A_\text{II}$}
\put(-42,15){\footnotesize{$\mathcal{E}_\mathcal{N}\times 10^{-19}$}}
\caption{\small{Logarithmic negativity of the Minkowski vacuum for two counter-accelerated modes, as a function of their proper accelerations for $D=0$. We have chosen $L_{||}=L_\pe=2$, $m=0.1$, $\Omega_0\approx 5.5$ and $\kappa_\pe=2$. \label{fig_3d_plotaa}}}
\end{figure}

Let us begin with the case of the counter-accelerated observers. Fig. \ref{fig_3d_plotaa} shows the dependence of the logarithmic negativity on the proper accelerations of the observers. Similarly to the $1+1$D case, increasing either of the accelerations leads to obtaining more entanglement, because the observers get more affected by the correlated Unruh noise.

\begin{figure}
\centering
\includegraphics[width=0.5\linewidth]{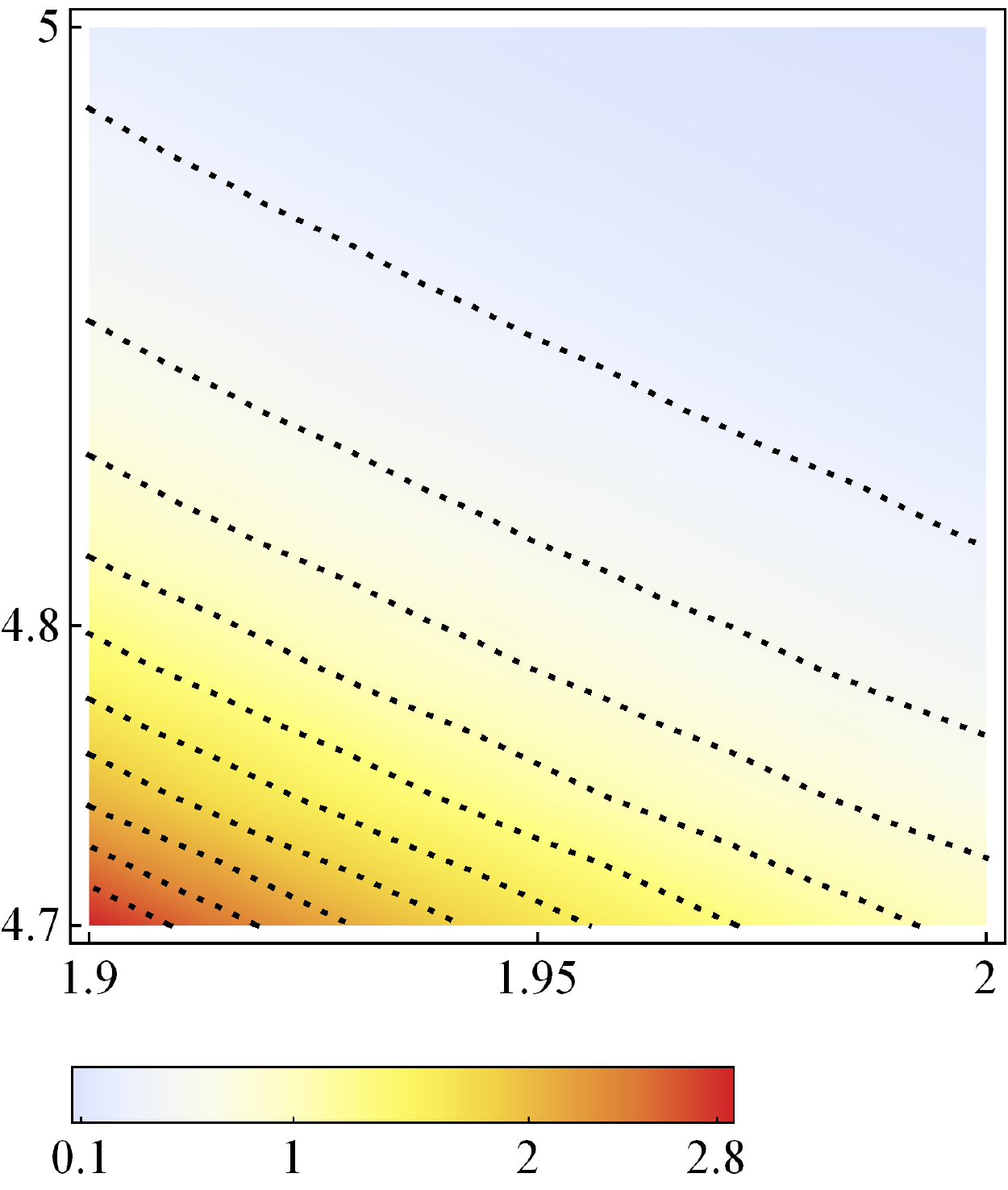}
\put(-200,166){$\Omega_0$}
\put(-50,33){$L_\pa$}
\put(-44,15){\footnotesize{$\mathcal{E}_\mathcal{N}\times 10^{-15}$}}
\caption{\small{Logarithmic negativity of the Minkowski vacuum for two counter-accelerated modes, as a function of $L_{||}$ and $\Omega_0$ for $D=0$. We have chosen ${\A}_\text{I}={\A}_\text{II}=0.1$, $L_\pe=2$, $m=0.1$, and $\kappa_\pe=2$. \label{fig_3d_plotlo}}}
\end{figure}

Fig. \ref{fig_3d_plotlo} shows the dependence of the logarithimc negativity on the sizes of the wavepackets along the $x$ axis, and on their central frequencies. The varying parameters, as well as the proper accelerations, for simplicity have been chosen to be equal for both wavepackets. Again, as in the previous chapter, most entanglement is present, when the sizes and frequencies are smallest

\begin{figure}
\centering
\includegraphics[width=0.5\linewidth]{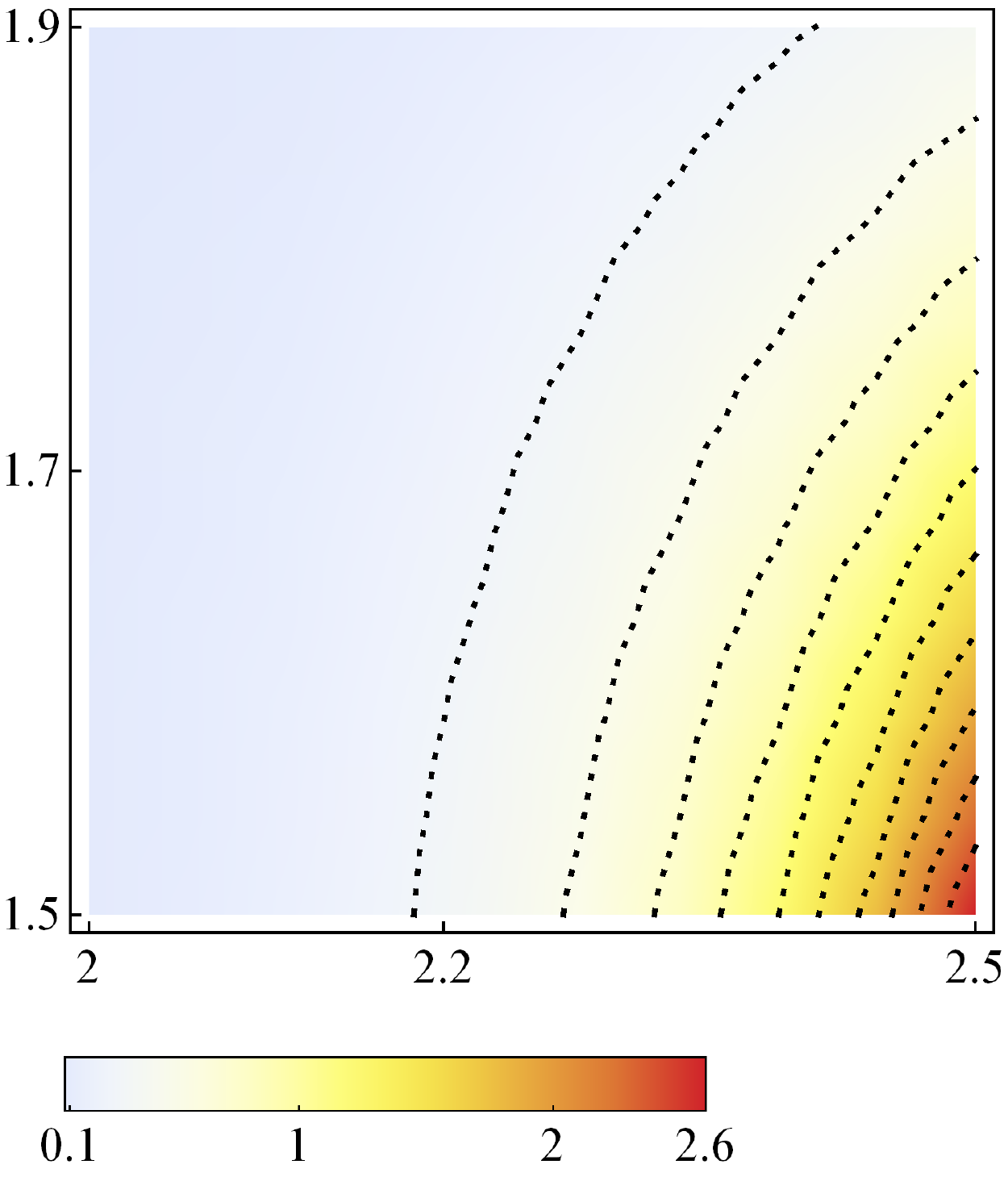}
\put(-202,175){$L_\pe$}
\put(-62,33){$\kappa_\pe$}
\put(-48,15){\footnotesize{$\mathcal{E}_\mathcal{N}\times 10^{-17}$}}
\caption{\small{Logarithmic negativity of the Minkowski vacuum for two counter-accelerated modes, as a function of $L_\pe$ and $\kappa_\pe$ for $D=0$. We have chosen ${\A}_\text{I}={\A}_\text{II}=0.1$, $L_{||}=2$, $m=0.1$, $\Omega_0 \approx 5.5$. \label{fig_3d_plotkl}}}
\end{figure}

We have also investigated the dependence of the logarithmic negativity on the parameters related to the dimensions perpendicular to the motion of the observers, namely on the size and wave vector perpendicular to the $x$ axis. The result is plotted in the Fig. \ref{fig_3d_plotkl}. We see that to obtain more entanglement, one needs to localize the wavepackets more in either dimension. Yet, while it is beneficial to decrease $\Omega_0$, it turns out $\kappa_\pe$ needs to be increased, to obtain more logarithmic negativity.

Moving on to the case of the parallel-accelerated observers, as in the previous chapter, the expression for the logarithmic negativity can be expanded for $|N_\text{I}|$, $|N_\text{II}|$, $|N_\text{I,II}^\pm|$ being much smaller than unity. From its form it is evident that we obtain no entanglement for $D=0$ in the weak Unruh noise regime. 

The existence of the entanglement for counter-accelerating observers and its absence for parallel-accelerating observers is interesting, and the following section provides an outlook at investigating it in more detail.

\section{Skew-accelerating observers}           \label{sec_opus_b3_skew}

Let us now discuss an outlook for a possible extension of our framework.
In the $1+1$-dimensional case we have found that when the output modes accelerate in the same direction there is no entanglement in the output state for $D=0$ when the noise terms are much smaller than unity. Our numerical results suggest that this is true also for $D\neq 0$. The same outcome is seen in the $3+1$-dimensional case, with the caveat that we have been unable to produce numerical results for $D\neq 0$. In the $3+1$-dimensional spacetime, we may also consider observers accelerating at an arbitrary angle with respect to each other.

Let us rotate the modes $\phi_\x{I}$ and $\psi_\x{I}$ around the origin, in the $xz$ plane, by an angle $\Theta$. This setup is depicted in the Fig. \ref{fig_3d_rotn}. It is a general setup which, for $\Theta=0$ reduces to the antiparallel-accelerating modes' case, and for $\Theta=\pi$ reduces to the parallel-accelerating modes' case. The aim of this is to examine the dependence of the extracted logarithmic negativity on $\Theta$ to discover how one case transits into the other.

\definecolor{mzielony}{rgb}{0,0.7,0}
\definecolor{mczerwony}{rgb}{0.922, 0.383, 0.207}
\definecolor{mniebieski}{rgb}{0.367,0.504,0.707}

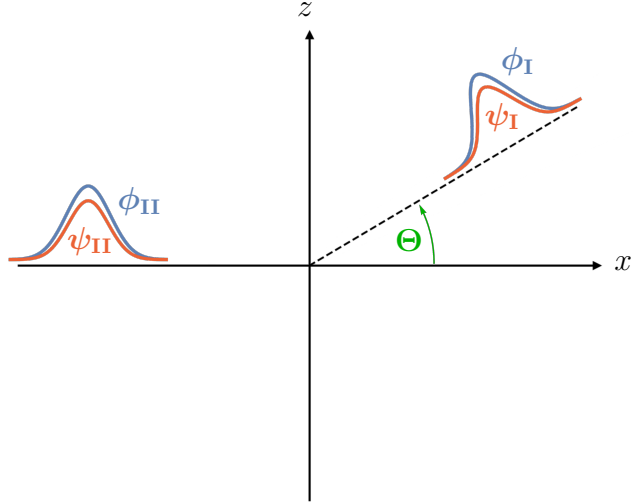
\begin{figure}
\centering
\includegraphics[width=0.7\linewidth]{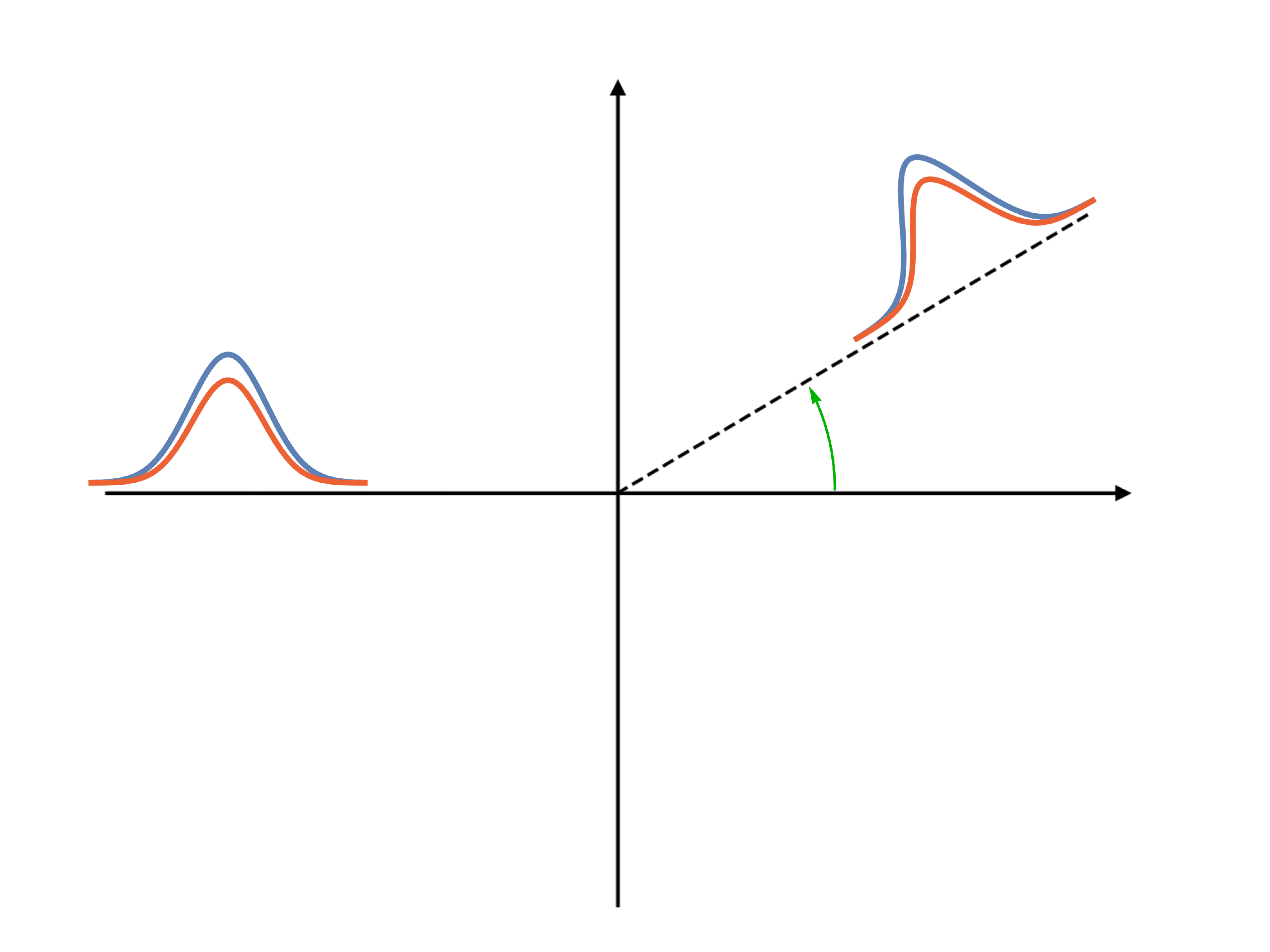}
\put(-145,193){$z$}
\put(-25,97){$x$}
\put(-108,104){\textcolor{mzielony}{$\boldsymbol{\Theta}$}}
\put(-232,105){\textcolor{mczerwony}{$\boldsymbol{\psi_\x{II}}$}}
\put(-213,120){\textcolor{mniebieski}{$\boldsymbol{\phi_\x{II}}$}}
\put(-74,152){\textcolor{mczerwony}{$\boldsymbol{\psi_\x{I}}$}}
\put(-68,173){\textcolor{mniebieski}{$\boldsymbol{\phi_\x{I}}$}}
\caption{\small{Schematic depiction of the skew-accelerating observers' setup. The modes $\phi_\x{I}$ and $\psi_\x{I}$ are rotated around the origin, in the $xz$ plane, by an angle $\Theta$. \label{fig_3d_rotn}}}
\end{figure}

We use unmodified Rindler coordinates, such that $D=0$ at all times. All the assumptions \eqref{opus_b1_ass1}-\eqref{opus_b1_ass6} about the orthogonality of the wavepackets still hold, thus for the wavepackets not to overlap at $\Theta=\pi$, we take ${\A}_\x{I} \neq {\A}_\x{II}$. 

Furthermore, we still assume that the wavepackets have no negative frequency mode overlap. Thus, for $\psi_\x{II}$ we clearly may write:
\begin{equation}        \label{opus_skew_d2}
\hat{d}_\x{II} =\int\text{d}\Omega\int\text{d}^2{\bf k_\perp}\,\,
\oIIok \hat{b}_{\x{II}\Omega\bf k_\perp}.
\end{equation}
When $\psi_\x{I}$ is rotated, it clearly has no negative frequency content with respect to the Rindler coordinate system rotated together with it. It is not obvious however, that its negative frequency content with respect to the unrotated Rindler coordinates should remain zero. Furthermore, for $\Theta\gtrsim\frac{\pi}{2}$ the mode $\psi_\x{I}$ already overlaps with region II of the unrotated Rindler coordinate system. Denoting the rotated terms by a prime, we write in most general:
\begin{equation}       \label{opus_skew_d1}
\begin{aligned}
\hat{d}_\x{I} &=
\int\text{d}\Omega\int\text{d}^2{\bf k'_\perp}\,\,
(\psi'_\x{I}, w'_{\x{I}\Omega\bf k'_\pe}) 
\hat{b}'_{\x{I}\Omega\bf k'_\perp}
\\
&=
\int\text{d}\Omega\int\text{d}^2{\bf k_\perp}\,\,
\big[
(\psi'_\x{I}, w_{\x{I}\Omega\bf k_\pe}) 
\hat{b}_{\x{I}\Omega\bf k_\perp}
+
(\psi'_\x{I}, w_{\x{I}\Omega\bf k_\pe}\s) 
\hat{b}_{\x{I}\Omega\bf k_\perp}\d
\\ &\qquad\qquad\qquad\qquad+
(\psi'_\x{I}, w_{\x{II}\Omega\bf k_\pe}) 
\hat{b}_{\x{II}\Omega\bf k_\perp}
+
(\psi'_\x{I}, w_{\x{II}\Omega\bf k_\pe}\s) 
\hat{b}_{\x{II}\Omega\bf k_\perp}\d
\big].
\end{aligned}
\end{equation}
Using \eqref{opus_skew_d2} and \eqref{opus_skew_d1} one could now carry on with computing expectation values analogous to \eqref{opus_b3_d1d1}-\eqref{opus_b3_d1d2d}. Substituting in \eqref{opus_b3_bb1}, \eqref{opus_b3_bb2}, \eqref{opus_b3_bb3} and \eqref{opus_b3_bb4} one would obtain all the necessary expressions to construct the noise matrix, expressed in terms of overlaps that can be evaluated numerically.

Let us now move on to the discussion of this framework applied to the $1+1$-dimensional Dirac spinor field.


\chapter{Dirac spinor field in 1+1-dimensional spacetime} \label{chf1d}

In this chapter we formulate our framework for the case of the $1+1$-dimensional Dirac spinor field. First, in Sec. \ref{sec_dirac_eqn}, we discuss the basic description of the Dirac spinor field in the Minkowski and Rindler coordinate systems, focussing on the differences with respect to the scalar field. In Sec. \ref{sec_dirac_setup} the setup under investigation is discussed and the derivation of the channel is outlined. In Sec. \ref{sec_dirac_modes} the modes are chosen, and finally in Sec. \ref{sec_dirac_results} the results for the entanglement of the vacuum are obtained. This part of the thesis is concluded in Sec. \ref{chf1conc}.


\section{Dirac spinor field}         \label{sec_dirac_eqn}

Let us begin with a brief discussion of the $1+1$-dimensional Dirac spinor field~\cite{crispino_higuchi}. A massive spinor field is governed by the Dirac equation, which in $1+1$-dimensional Minkowski spacetime has the form:
\begin{equation}
i\partial_t\hat{\Psi}=\left(-i\boldsymbol\alpha_3\partial_x+m\boldsymbol\beta\right)\hat{\Psi},
\end{equation}
where $m$ is the mass of the field and the matrices $\boldsymbol\beta$ and $\boldsymbol\alpha_3$ are defined as:
\begin{equation}\label{matrices}
\boldsymbol\beta \defn \left( \begin{array}{cc}
1 & 0  \\
0& -1   \end{array} \right), \hspace{10mm} 
\boldsymbol\alpha_3 \defn \left( \begin{array}{cc}
0 & 1  \\
1& 0   \end{array}\right).
\end{equation} 
The field $\hat{\Psi}$ is a spinor field, which means that it has two complex components. The scalar product associated to the Dirac equation is of the form:
\begin{equation}  \label{dirac_scalar}
\left(\phi_1,\phi_2\right)=\int \text{d}x\, \phi_1^\dagger\phi_2,
\end{equation}
thus, the properties that it satisfies, are:
\begin{equation} \label{dirac_scalar_prop}
(\phi_1,\phi_2) = (\phi_2,\phi_1)\s = (\phi_2\s , \phi_1\s).
\end{equation}
Comparing these with \eqref{qft_scalar_prop}, one may note that there is a sign difference in the last equality.

The mode solutions of the Dirac equation in the Minkowski coordinates are:
\begin{align}
u_k^\pm=&\frac{1}{\sqrt{4\pi\omega_k}}  \left(
\begin{array}{c}
\sqrt{\omega_k\pm m}\\
\pm\sqrt{\omega_k\mp m}
\end{array}
\right)e^{\mp i\omega_k t+ikx}  ,
\end{align} 
where $\omega_k=\sqrt{k^2+m^2}$. The upper sign corresponds to positive frequency modes. These modes are called {\it particle} modes. The lower sign corresponds to negative frequencies. These modes are called {\it antiparticle} modes. Both particle and antiparticle modes are normalized to a delta function using the scalar product \eqref{dirac_scalar}.

In Rindler coordinates, the Dirac equation takes the form:
\begin{equation}         
i \frac{1}{a}\partial_\eta\hat{\Psi}=\left(m\chi \beta-\frac{i}{2} \alpha_3-i\chi\alpha_3\partial_{\chi}\right)\hat{\Psi},
\end{equation}
with the scalar product:
\begin{equation}  
\left(\phi_1,\phi_2\right)=\int \text{d}\chi\, \phi_1^\dagger\phi_2,
\end{equation}
and the mode solutions:
\begin{align}
w^\pm_{\text{I}\Omega}=\sqrt{\frac{m \cosh(\frac{\pi \Omega}{a})}{2\pi^2 a}} \left(
\begin{array}{c}
K_{\pm i\frac{\Omega}{a}+\frac{1}{2}}(m\chi)+i K_{\pm i\frac{\Omega}{a}-\frac{1}{2}}(m\chi)\\
- K_{\pm i\frac{\Omega}{a}+\frac{1}{2}}(m\chi)+i K_{\pm i\frac{\Omega}{a}-\frac{1}{2}}(m\chi)
\end{array}
\right)  e^{\mp i\Omega\eta},
\\
w^\pm_{\text{II}\Omega}=\sqrt{\frac{m \cosh(\frac{\pi \Omega}{a})}{2\pi^2 a}}  \left(
\begin{array}{c}
K_{\pm i\frac{\Omega}{a}+\frac{1}{2}}(-m\chi)+i K_{\pm i\frac{\Omega}{a}-\frac{1}{2}}(-m\chi)\\
- K_{\pm i\frac{\Omega}{a}+\frac{1}{2}}(-m\chi)+i K_{\pm i\frac{\Omega}{a}-\frac{1}{2}}(-m\chi)
\end{array}
\right) e^{\pm i\Omega\eta},
\end{align}
where the former solution holds in wedge I, and the latter in wedge II. The creation and annihilation operators associated with particles are named $\hat{b}_{\x{I}\Omega}^+$, and those associated with antipaticles are named $\hat{b}_{\x{I}\Omega}^-$.

The transformation between the Minkowski and the Rindler vacuum state is of the form:
\begin{equation}
|0\rangle_\text{M}=\hat{S}_\x{I,II} |0\rangle_\text{R},
\end{equation}
as in Eq. \eqref{rqi_s12}. However, the relations analogous to \eqref{rqi_sbs} are as follows~\cite{fer_sbs}:
\begin{equation}    \label{dirac_sbs}
\begin{aligned}
\hat{S}_\x{I,II}^\dagger \hat{b}_{\text{I}\Omega}^+ \hat{S}_\x{I,II}=\,\,& 
 \hat{b}_{\text{I}\Omega}^+\cos r_\Omega - \hat{b}_{\text{II}\Omega}^{-\dagger} \sin r_\Omega,\\
\hat{S}_\x{I,II}^\dagger \hat{b}_{\text{II}\Omega}^- \hat{S}_\x{I,II}=\,\,& 
 \hat{b}_{\text{II}\Omega}^-\cos r_\Omega + \hat{b}_{\text{I}\Omega}^{+\dagger} \sin r_\Omega,
\end{aligned}
\end{equation}
where $r_\Omega=\x{atan}\,e^{-\frac{\pi\Omega}{a}}$. The transformations of the operators $\hat{b}_{\text{II}\Omega}^+$ and $\hat{b}_{\text{I}\Omega}^-$ are obtained by interchanging $\hat{b}^-_{\Lambda\Omega}\leftrightarrow \hat{b}^+_{\Lambda\Omega}$ in (\ref{dirac_sbs}), where $\Lambda\in\{\x{I},\x{II}\}$.

\section{The setup and comparison with the case of the Klein-Gordon field}   \label{sec_dirac_setup}

The setup of the problem is analogous to the one presented in chapter \ref{chb1d}. We consider two localized stationary wavepacket modes $\phi_\Lambda^\pm$, each one localized in its respective wedge at $t=0$. Similarly, we have two uniformly accelerated modes $\psi_\Lambda^\pm$, each one localized in its respective wedge at all times. The modes now have two components. Moreover, the $+$ or $-$ signs refer to the particle or antiparticle modes respectively, both of which are chosen in Sec. \ref{sec_dirac_modes}. The annihilation operators corresponding to the inertial modes are denoted as $\hat{f}_\Lambda^\pm$, whereas the annihilation operators corresponding to the accelerating modes are denoted $\hat{d}_\Lambda^\pm$. All the assumptions from Sec. \ref{sec_opus_b1_setup} hold for particle and antiparticle modes. Let us stress that even though the creation and annihilation operators satisfy anticommutation relations, e.g. $\{\hat{d}_\x{I}^+ , \hat{d}_\x{I}^{+\dagger}\}=1$, the assumption that the two modes do not overlap is still expressed with the commutation relations \eqref{opus_b1_comm_dd}.

Let us write a decomposition of the field, analogous to \eqref{opus_dec_wavepackets}:
\begin{equation}       
\hat{\Phi} =
\sum\limits_n \phi_n^+ \hat{f}_n^+ + \phi_n^{-\star} \hat{f}_n^{-\dagger} =
\sum\limits_n \psi_n^+ \hat{d}_n^+ + \psi_n^{-\star} \hat{d}_n^{-\dagger},
\end{equation}
where the index $n$ labels the modes in the wavepacket basis and it may be I,~II or different.

We are interested in applying the channel \eqref{opus_b1_channel_x}-\eqref{opus_b1_channel_s} to the input Minkowski vacuum state at $D=0$, in order to investigate the entanglement of the vacuum and compare the results to those presented in Sec. \ref{sec_opus_b1_results}. The input Minkowski vacuum state for a spinor field, has the form $\vec{X}^{(f)}=\vec{0}$ and:
\begin{align}
\boldsymbol\sigma^{(f)}=&\left(\begin{array}{cccccccc}
0 & 1& 0 & 0  & 0 & 0& 0 & 0     \\
-1 & 0& 0 & 0  & 0 & 0& 0 & 0    \\
0 & 0& 0 & 1 & 0 & 0& 0 & 0     \\
0 & 0 & -1 & 0 & 0 & 0& 0 & 0 \\
0 & 0 & 0 & 0 & 0 & 1& 0 & 0 \\
0 & 0 & 0 & 0 & -1 & 0& 0 & 0 \\
0 & 0 & 0 & 0 & 0 & 0& 0 &1 \\
0 & 0 & 0 & 0 & 0 & 0& -1 & 0 \\
\end{array}\right).
\end{align}
The output state is evaluated in an analogous way to the one presented in chapter \ref{chb1d} for $D=0$. The procedure leads to $\vec{X}^{(d)}=\vec{0}$ and the covariance matrix:
\begin{equation}
\sdvac=\left(\begin{array}{cc}
\boldsymbol\sigma_{+}^{(d)} & \boldsymbol\sigma_{c}^{(d)}    \\
\boldsymbol\sigma_{c}^{(d)}  & \boldsymbol\sigma_{-}^{(d)}  
\end{array}\right),
\end{equation}
where the block $\boldsymbol\sigma_{+}^{(d)}$ captures the effect of acceleration on the particle modes. Similarly, $\boldsymbol\sigma_{-}^{(d)}$ describes antiparticle modes. $\boldsymbol\sigma_{c}^{(d)} $ captures the correlations between particles and anti-particles that arise due to the acceleration. For a spinor field the full covariance matrix is antisymmetric. Explicitly, the block matrices take the form:
\begin{equation}
\boldsymbol\sigma_{\pm}^{(d)}=\left(\begin{array}{cccc}
0 & 1-N_\text{I}^\pm& 0 & 0     \\
-1+N_\text{I}^\pm & 0& 0 & 0    \\
0 & 0 & 0 & 1-N_\text{II}^\pm   \\
0& 0& -1+N_\text{II}^\pm & 0  
\end{array}\right),
\end{equation}
and:
\begin{equation}
\boldsymbol\sigma_{c}^{(d)}=\left(\begin{array}{cccc}
0 & 0& \Im\,N_\text{I,II}^- & \Re\,N_\text{I,II}^-  \\
0 & 0& \Re\,N_\text{I,II}^- & -\Im\,N_\text{I,II}^- \\
-\Im\,N_\text{I,II}^+ & -\Re\,N_\text{I,II}^+       \\
-\Re\,N_\text{I,II}^+ & \Im\,N_\text{I,II}^+  
\end{array}\right).
\end{equation}
The expressions appearing in the output covariance matrix are:
\begin{align}          \label{dirac_NLambda}
N_\Lambda^\pm=& \,\,2\int \text{d}\Omega\, \frac{|(\psi_\Lambda^\pm , w^\pm_{\Lambda\Omega})|^2}{e^{\frac{2\pi\Omega}{a}}+1} \, ,\\   \label{dirac_N12}
N_\text{I,II}^\pm=&- \int \text{d}\Omega\, \frac{(\psi_\text{I}^\pm, w^\pm_{\text{I}\Omega}) (\psi_\text{II}^\mp, w^\mp_{ \text{II}\Omega})}{\cosh\left(\frac{\pi\Omega}{a}\right)}.
\end{align}
Comparing \eqref{dirac_NLambda} with \eqref{opus_b1_Nlambda}, one may note that where in the case of the Klein-Gordon equation the Bose-Einstein distribution appears, for fermions we obtain the Fermi-Dirac distribution.

For the technical details the Reader is referred to the publication~\cite{lor_fer}. Let us now discuss the modes chosen and then move on to the presentation of the results.

\section{The choice of the modes}             \label{sec_dirac_modes}

\begin{figure}
\centering
\includegraphics[scale=1]{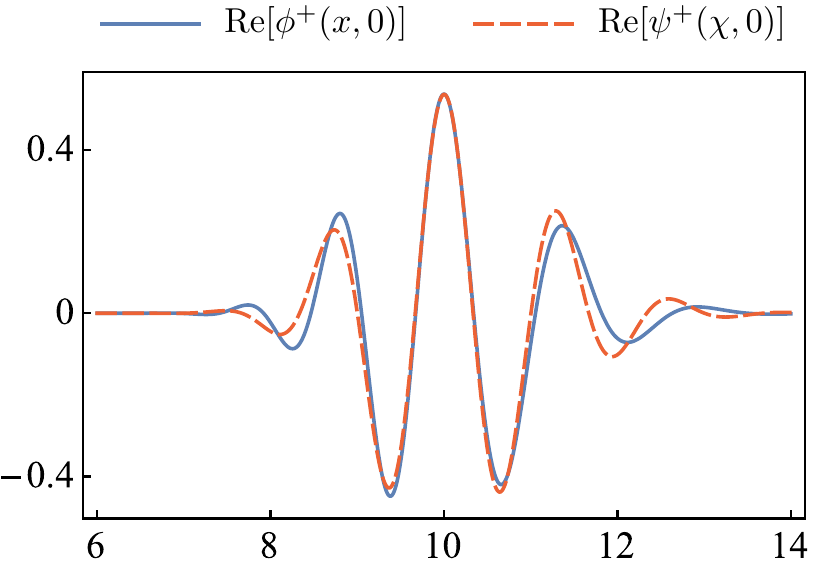}
\put(-150,151){\scriptsize{$\Lambda$}}
\put(-41,151){\scriptsize{$\Lambda$}}
\caption{\small{Comparison between the real parts of the spatial profiles of modes $\phi^+_\Lambda$ and $\psi^+_\Lambda$ for the following choice of parameters: ${\cal A}_\Lambda=0.1$, $L=2$, $\Omega_0\approx 5$, $m=0.1$.}}
\label{fig_fer_picmodes}
\end{figure}

Following the reasoning from the previous chapters, for the inertial modes we choose stationary cavity modes mutliplied by a smooth envelope. Similarly, for the accelerated modes we choose modes of an analogous accelerating cavity. We assume that our modes have a purely positive frequency mode content, or in other words are purely particle modes. Since the Dirac equation is a first order differential equation, it suffices to specify only the value of the mode at $t=0$. Motivated by the solutions obtained in~\cite{modes_fermions}, the choice is:
\begin{equation} 
\begin{aligned}
\phi_\Lambda^+(x,0)= \mathcal{N}_\phi \, &\left(
\begin{array}{c}
\cos\left(\frac{\kappa_\phi}{2}\right)+\sin\left(\frac{\kappa_\phi}{2}\right)\\
\cos\left(\frac{\kappa_\phi}{2}\right)-\sin\left(\frac{\kappa_\phi}{2}\right)
\end{array}
\right) 
\\
\times \,
 &e^{-2\left( \frac{1}{{\A}_\Lambda L}\log({\A}_\Lambda x) \right)^2+i\,\sqrt{\Omega_0^2-m^2}\,(x-\frac{1}{{\A}_\Lambda}) - \frac{i}{2}\kappa_\phi},
\end{aligned}
\end{equation}
where $L$ corresponds to the size of the wave packet that is centered at $\frac{1}{{\A}_\Lambda}$, where ${\A}_\Lambda$ are the proper accelerations of the accelerated wavepackets. Furthermore, $\mathcal{N}_\phi$ is a normalization constant, and $\kappa_\phi=\x{atan}\left(\frac{m}{\sqrt{\Omega_0^2-m^2}}\right)$ where $\Omega_0$ is the frequency which the wavepacket is centered around in the frequency space.
For the accelerated modes, the choice is:
\begin{equation}
\begin{aligned}
\psi_\Lambda^+(\chi,0)=& \,\,\mathcal{N}_\psi \,
\left( 
I_{-i\frac{ \Omega_0 }{a}-\frac{1}{2}}\left(m\chi_0\right)-
I_{-i\frac{ \Omega_0 }{a}+\frac{1}{2}}\left(m\chi_0\right)\right) 
\\ \times \,
&\left(
\begin{array}{c}
I_{i\frac{ \Omega_0 }{a}-\frac{1}{2}}(m \chi )+i I_{i\frac{\Omega_0 }{a}+\frac{1}{2}}(m \chi )\\
I_{i\frac{ \Omega_0 }{a}-\frac{1}{2}}(m \chi )-i I_{i\frac{ \Omega_0 }{a}+\frac{1}{2}}(m \chi )
\end{array}
\right)
 \, e^{-2\left(\frac{1}{{\A}_\Lambda L}\log({\A}_\Lambda\chi)\right)^2},
\end{aligned}
\end{equation}
with the constant of normalization $\mathcal{N}_\psi$.
The spatial profiles of the real parts of these modes are compared in Fig. \ref{fig_fer_picmodes}. The inevitable mismatch due to the acceleration is visible.
  
In both cases the corresponding antiparticle modes are obtained by performing charge conjugation, which is of the form:
\begin{equation}
\phi_\Lambda^-
=
\hat{C} {\phi_\Lambda^+}\s,
\end{equation}
and similarly for $\psi_\Lambda^-$, where the charge conjugation operator in our case takes the form:
\begin{equation}
\hat{C}=\begin{pmatrix}
0 & -1 \\ -1 & 0
\end{pmatrix}.
\end{equation}
This way we construct all the modes necessary to proceed with computing the results.

\section{Results}             \label{sec_dirac_results}

Let us move on to the results obtained for input Minkowski vacuum state. The amount of entanglement is studied for the same ranges of parameters as in Sec. \ref{sec_opus_b1_results} for $D=0$.

In \cite{fermions_ln1}, it was shown that the partial transpose of a fermionic Gaussian state is, in general, not Gaussian and therefore it is difficult to calculate the logarithmic negativity exactly. However, it was shown that a lower bound can be obtained~\cite{fermions_ln1, fermions_ln2}. Building on this construction, the respective bound  $\tilde{\mathcal{E}}_\mathcal{N}$ is:
\begin{equation}
\begin{aligned}
\tilde{\mathcal{E}}_\mathcal{N}=
\ln\bigg[1 \,\,+\,\, &\frac{1}{2}\bigg(-N_\text{I}^+-N_\text{II}^-+N_\text{I}^+N_\text{II}^-+|N_\text{I,II}^+|^2  \\
+\,\,& \Re\sqrt{(N_\text{I}^+-N_\text{II}^-)^2-4|N_\text{I,II}^+|^2}\\
+\,\,& \Im\sqrt{(N_\text{I}^+-N_\text{II}^-)^2-4|N_\text{I,II}^+|^2} \bigg)\bigg],
\end{aligned}
\end{equation}
where the arising noise matrix terms are given in \eqref{dirac_NLambda} and \eqref{dirac_N12}. It is important to emphasize that $\tilde{\mathcal{E}}_\mathcal{N}$ is a lower bound for the negativity and therefore it is difficult to make strong quantitative statements. However, it is reasonable to assume that the actual value of the negativity is close to the lower bound $\tilde{\mathcal{E}}_\mathcal{N}$. Therefore, in the following, we refer to $\tilde{\mathcal{E}}_\mathcal{N}$ as the logarithmic negativity, keeping this subtlety in mind.

\begin{figure}
\centering
\includegraphics[width=0.5\linewidth]{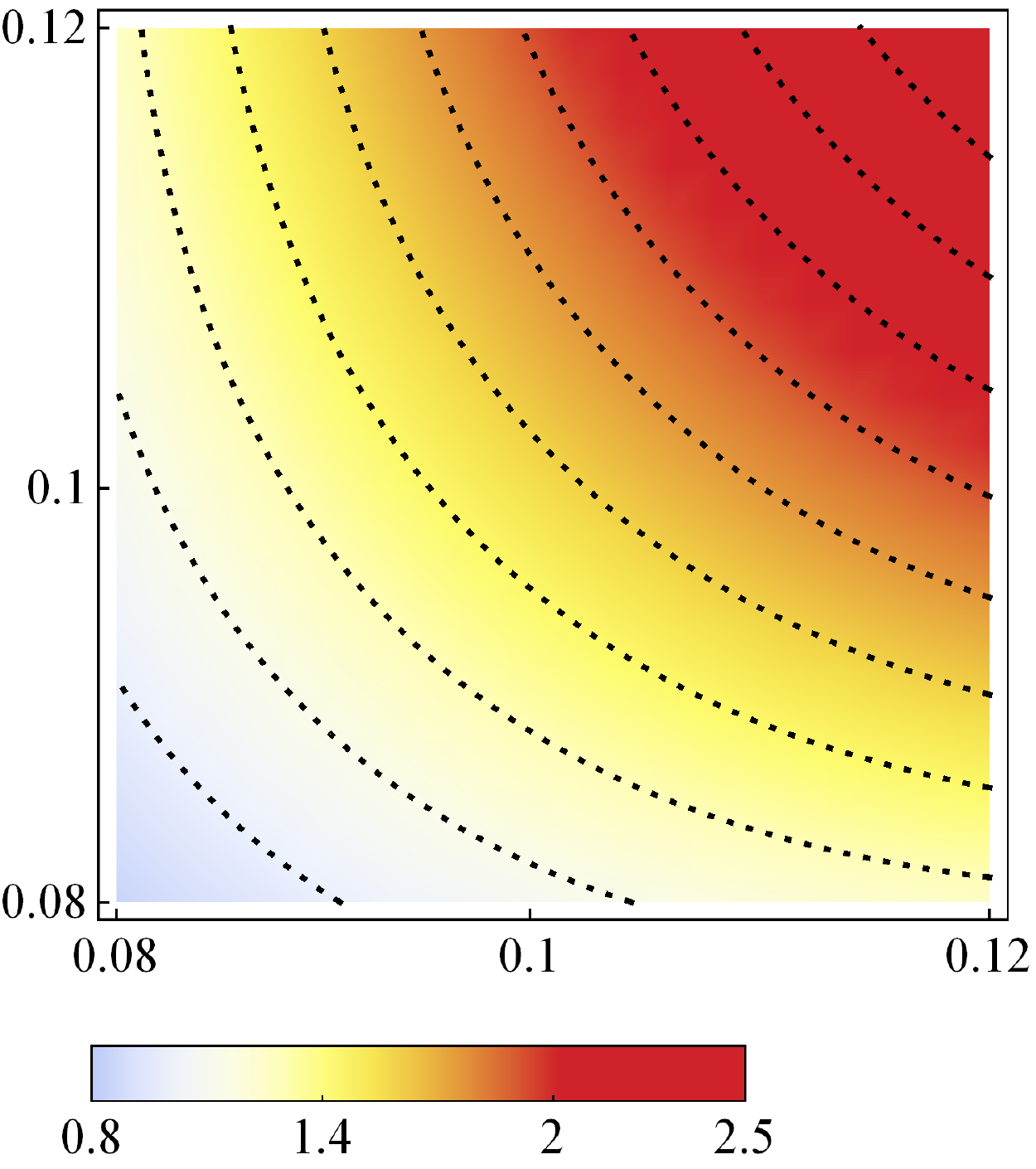}
\put(-195,170){${\A}_\x{I}$}
\put(-58,32){${\A}_\x{II}$}
\put(-50,14){{\footnotesize{$\tilde{\mathcal{E}}_\mathcal{N}\times 10^{-12}$}}}
\caption{\small{Logarithmic negativity $\tilde{\mathcal{E}}_\mathcal{N}$ of the Minkowski vacuum for two counter-accelerated modes, as a function of their proper accelerations. We have chosen $m=0.1$, $L=2$ and $\Omega_0 \approx 5$. 
}}
\label{fig_fer_picnegvac}
\end{figure}

\begin{figure}
\centering
\includegraphics[width=0.5\linewidth]{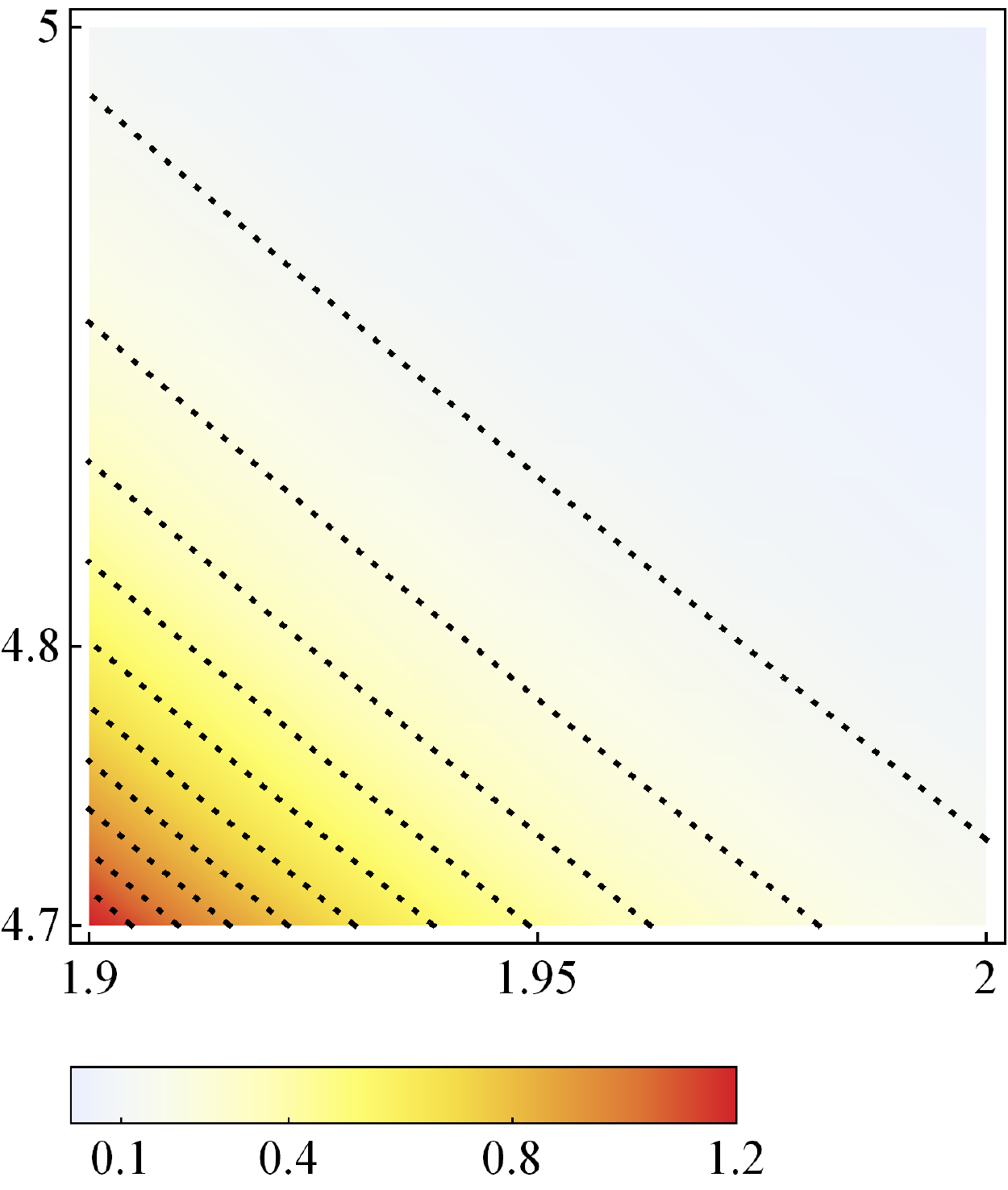}
\put(-200,165){$\Omega_0$}
\put(-48,32){$L$}
\put(-49,14){{\footnotesize{$\tilde{\mathcal{E}}_\mathcal{N}\times 10^{-11}$}}}
\caption{\small{Logarithmic negativity $\tilde{\mathcal{E}}_\mathcal{N}$ of the Minkowski vacuum for two counter-accelerated modes characterized by the same proper acceleration ${\cal A}_\Lambda=0.1$ as a function of $L$ and $\Omega_0$. We have chosen $m=0.1$. 
}}
\label{fig_fer_lo}
\end{figure}

Fig. \ref{fig_fer_picnegvac} shows the negativity as a function of the two proper accelerations. The higher the accelerations are, the more entanglement is obtained in the output state. Furthermore, Fig. \ref{fig_fer_lo} shows the negativity as a function of the proper length of the wavepackets and their central frequency. Most entanglement exists when the length and the central frequency are smallest. The results are in a qualitative agreement with those obtained in Sec. \ref{sec_opus_b1_results}. 

Let us also note that quantitatively, the amount of vacuum entanglement for fermions tends to be lower than for bosons by approximately one order of magnitude. On the one hand one could explain this with the fact that the Fermi-Dirac distribution itself has lower values than the Bose-Einstein distribution, but on the other hand what has been computed is the lower bound for the logarithmic negativity, therefore it is difficult to say whether the effect arises physically.

In the publication of this work~\cite{lor_fer} the channel was also applied to a non-vacuum state, namely a Bell state. The negativity was numerically evaluated as a function of the proper accelerations of the observers, and it was found that it degrades with increasing accelerations. It is in agreement with the result for the squeezed thermal state in the case of the Klein-Gordon field, which was described in Sec. \ref{sec_opus_b1_results}. The reason for this effect is the increasing inevitable mismatch between the inertial and the accelerated modes.

The $D\neq 0$ case of this setup is currently under development~\cite{grzesiek}. It is interesting, whether the outcomes of this case are in agreement with those presented in Sec. \ref{sec_opus_b1_results} or perhaps new effects will be discovered.

\section{Conclusions of Part \ref{part3}}   \label{chf1conc}

This concludes the part of the thesis devoted to the investigation of the bipartite vacuum entanglement using a two-mode Gaussian channel between two stationary and two accelerating modes. The results for the case of $D=0$ agree with the previous literature on the topic. Also, certain new effects were seen. In particular, with the use of the modified Rindler coordinates it was possible to disentangle the variables ${\A}_\Lambda$ and $D$ and obtain results for $D\neq 0$ in $1+1$ dimensions, including the sudden death of the entanglement, discussed in Sec. \ref{sec_opus_b1_results}. Also, another new idea was proposed, of analyzing skew-accelerating observers in $3+1$ dimensions to gain an insight into the behaviour of antiparallel and parallel accelerating observers. 
Furthermore, we believe this framework may have broad applications. It has already been applied to  continuous variable qauntum teleportation and continuous varaible dense coding~\cite{piotrek}. Also, extensions of it are developed, including the multimode channel~\cite{channel_kacper} and the $D\neq 0$ channel in $1+1$ dimensions for the Dirac spinor field~\cite{grzesiek}. Let us now move on to the part of the thesis devoted to  investigation of another constituent of the structure of the vacuum entanglement, i.e. the tripartite entanglement.



\part{Tripartite entanglement}
\label{part4}


\chapter{Extraction of genuine tripartite entanglement from the vacuum} \label{chtri}

In this chapter we study the bipartite and tripartite entanglement contained in the vacuum state of a $1+1$-dimensional scalar quantum field. The entanglement is harvested by three particle detectors, initially in a separable state. The detectors couple to the field for a finite time. After this interaction their final state is entangled. If the interaction time is sufficiently short, such that no subluminal signal can propagate between the detectors while they are switched on, it is clear that this entanglement originates from the vacuum. The calculations are performed non-perturbatively, using the formalism of Gaussian quantum mechanics. We begin with Sec. \ref{sec_tri_mot}, which motivates the chapter. In Sec. \ref{sec_tri_setup} the details of the setup and the model are discussed. The calculation of the final state of the detectors is performed in Sec. \ref{sec_tri_calc}. Further, in Sec.~\ref{sec_tri_tri} we discuss how various types of entanglement are quantified or estimated. In Sec. \ref{sec_tri_results} detailed maps are presented and analyzed, which show the regions of the parameter space, wherein each type of entanglement exists. Sec. \ref{sec_tri_concl} concludes the chapter.


\section{Motivation}      \label{sec_tri_mot}

The Minkowski vacuum state contains quantum entanglement. This was the subject of investigation in the previous parts of the thesis. This entanglement can be accessed e.g. from a uniformly accelerated frame. As discussed in~Sec.~\ref{sec_vac_ent}, the origin of this entanglement is the non-trivial form of the Minkowski vacuum state, as seen from the Rindler frame \eqref{rqi_s12}. It is natural to ask questions about the properties and the structure of this entanglement. In Part \ref{part3} properties of the vacuum bipartite entanglement have been investigated using the framework of the two-mode Gaussian quantum channel. In this part of the thesis, the vacuum bipartite and tripartite entanglement is examined with a particle detector.

In relativistic quantum information the Unruh-DeWitt detector~\cite{birrell_davies,unruh_effect,dewitt} is a commonly used particle detector model. We consider a situation whereby the field and the detector interact with each other locally for a finite time. This may also be referred to as turning the detector on, and later, turning it off. Afterwards, the state of the detector is analyzed. During such an interaction the state of the quantum field is also altered, and the resultant field fluctuations cannot propagate faster than the speed of light.

Let us consider two stationary detectors, positioned at different locations in space and switched on for a finite time. They both get excited because of the interaction with the field, and in general their final state may be entangled. If the interaction time is sufficiently long, this fact is not surprising, because a field fluctuation arising at one detector propagates and might reach the other detector and excite it. Hence, the excitations of the detectors might be correlated. It is perhaps more surprising that the detectors may get entangled even when the interaction time is purposely made so short, that fluctuations from one detector may not reach the other one by the time it gets turned off. In such a case no subluminal signal can be exchanged between the detectors. The entanglement in the final state of the detectors in such a case originates from the fact that they both interact with the same quantum field, whose state is already entangled. This entanglement can be swapped over onto the detectors. This way the detectors can extract a certain amount of entanglement from the Minkowski vacuum state.

In~\cite{reznik1,reznik2} two qubit detectors in a $3+1$-dimensional scalar field are considered. The detectors are at rest and the interaction is turned on for a finite time, short enough to exclude any causal contact between them. It is found that their state after the interaction violates Bell's inequalities, and the amount of violation decreases with increasing distance between detectors. On a technical side, the calculations are performed using perturbation theory.

The idea is further developed in~\cite{eric_dets_for_probing} and~\cite{dragan_fuentes}, wherein two inertial harmonic oscillator detectors, are placed in a cavity. Again, the interaction is turned on for a short time, preventing any causal contact between the detectors. Afterwards the state of the detectors is found to be entangled. Furthermore, the formalism of Gaussian quantum mechanics is used, the calculations are non-perturbative and the obtained results are exact. Further literature concerning such an extraction of entanglement, called {\it vacuum entanglement harvesting}, includes~\cite{farming}.

In this part of thesis we extend the scenario with detectors in a cavity, by considering three detectors. This allows us to study and compare the amounts of extracted bipartite and tripartite entanglement. Furthermore, in our work we also investigate the influence of different types of cavity boundary conditions on the amount of harvested entanglement. The results are interesting from the theoretical point of view, because they provide an insight into the structure of the entanglement of the vacuum. Also, they might have practical applications, since an experimental verification of these phenomena is likely to be realised in a cavity system similar to e.g.~\cite{intro_dcasimir_conf}. The knowledge of the most efficient way to harvest entanglement may also be useful in situations whereby the entanglement can be used as a resource e.g. in quantum information protocols. Let us now discuss the details of our work.

\section{The setup}         \label{sec_tri_setup}

We work in $1+1$-dimensional spacetime with a massless scalar field $\hat{\Phi}$ enclosed in a cavity extending between $x=0$ and $x=L$. Inside the cavity there are three detectors at rest, whose positions $x_\Lambda$ will be specified later. Here $\Lambda\in\{1,2,3\}$ is an index labelling the detectors, which are modelled by harmonic oscillators with associated annihilation operators $\hat{d}_\Lambda$. Initially the detectors and the field are in their ground states, thus they are not entangled. At time $t=0$ the interaction is turned on and at $t=T$ it is turned off. Afterwards, the entanglement across various partitions of the system is examined.

We consider two cases, wherein the field is subject to different types of boundary conditions, introduced in Sec.~\ref{sec_qft_m}. The first type is the periodic boundary conditions, which yield mode solutions \eqref{qft_u_cav_m_per}. The second type is the hard wall Dirichlet boundary conditions, which yield mode solutions \eqref{qft_u_cav_m}. The mode decomposition of the field $\hat{\Phi}$ is given by \eqref{qft_dec_A}.


The full Hamiltonian governing the evolution of the system, can be written as:
\begin{equation}           \label{tri_Hfull}
\hat{H} = \hat{H}_\x{free} + \hat{H}_\x{int}.
\end{equation}
The free Hamiltonian is given by:
\begin{equation}           \label{tri_Hfree}
\hat{H}_\x{free} = 
\sum\limits_\Lambda \Omega_d \,\hat{d}\d_\Lambda \hat{d}_\Lambda +
\sum\limits_n \omega_n \,\hat{A}_n\d \hat{A}_n,
\end{equation}
where $\Omega_d$ is the size of the gap between the detectors' energy levels (in natural units), also referred to as the frequency of the detectors. Furthermore, the interaction is assumed to be of the Unruh-DeWitt type:
\begin{equation}          \label{tri_Hint}
\hat{H}_\x{int} =
\sum\limits_\Lambda \lambda \,(\hat{d}_\Lambda + \hat{d}_\Lambda\d)\, \hat{\Phi}(x=x_\Lambda),
\end{equation}
where $\lambda$ is the coupling strength, and it is understood that the field \eqref{qft_dec_A} is evaluated at the positions of the detectors only ($\hat{H}_\x{int}$ is vanishing elsewhere). 

The detectors interact with an infinite number of field modes, but the high energy modes have a negligible impact on the evolution of the state of the detectors. It is thus safe to apply a UV cutoff and consider the interaction only with $n_c$ lowest energy modes. 

Let us now describe the evolution of the state of the detectors using the formalism of Gaussian quantum mechanics.

\section{Calculation of the final state of the detectors}        \label{sec_tri_calc}

Let us now calculate the final state of the detectors in the formalism of Gaussian quantum mechanics. The vectors and matrices have different elements depending on the chosen boundary conditions, thus we discuss each case separately.

Starting with the periodic boundary conditions, let us define a vector containing the quadrature operators \eqref{gqm_quad} of the detectors and the field modes:
\begin{equation}  
\hat{\mathbf{x}}=(\hat{q}_{d1},\hat{p}_{d1},...\,,
\hat{q}_{d3},\hat{p}_{d3},\hat{q}_{-\frac{n_c}{2}},\hat{p}_{-\frac{n_c}{2}},...\,,\hat{q}_{\frac{n_c}{2}},\hat{p}_{\frac{n_c}{2}})^T,
\end{equation}
where the vector has $2n_c+6$ elements and the first $6$ correspond to the detectors, whereas the remaining $2n_c$ correspond to the field modes (with the zero mode $n=0$ excluded).

We may write any Hamiltonian that is quadratic in the quadrature operators, as:
\begin{equation}    \label{tri_xFx}
\hat{H}(t)=\hat{\mathbf{x}}^T\mathbf{F}(t)\hat{\mathbf{x}},
\end{equation}
where $\mathbf F$ is a matrix with dimensions $(2n_c+6) \times (2n_c+6)$. It is a phase-space matrix that fully characterizes the Hamiltonian. It is found~\cite{eric_dets_for_probing} that only the symmetric part of $\mathbf F$ impacts the evolution, hence we further define $\mathbf{F}^\x{sym}\defn\mathbf{F}+\mathbf{F}^T$.

Similarly to \eqref{tri_Hfull}, the phase-space Hamiltonian matrix is divided into $\mathbf{F}^\x{sym}=\mathbf{F}^\x{sym}_\x{free}+\mathbf{F}^\x{sym}_\x{int}$. From \eqref{tri_Hfree} it follows that:
\begin{equation}  \label{tri_Ffree_per}
\mathbf{F}_\text{free}^\text{sym}=\textrm{diag}(\Omega_d,\Omega_d,\Omega_d,\Omega_d,\Omega_d,\Omega_d,\omega_{-\frac{n_c}{2}},\omega_{-\frac{n_c}{2}},
...\,,\omega_{\frac{n_c}{2}},\omega_{\frac{n_c}{2}}),
\end{equation}
and furthermore, from \eqref{tri_Hint} we obtain:
\begin{equation} \label{tri_Fint_per}
\mathbf{F}_\text{int}^\text{sym}=2\lambda 
\begin{pmatrix}
\mathbf{0}_6 & \mathbf{G}^T \\
\mathbf{G} & \mathbf{0}_{2n_c}
\end{pmatrix}
,
\end{equation}
where $\mathbf{0}_n$ is a zero matrix with dimensions $n\times n$, and the matrix $\mathbf G$ takes the form:
\begin{eqnarray}   
	\mathbf{G} \defn
\begin{pmatrix} 
\frac{\cos\left( k_{-n_c/2}x_1\right)}{\sqrt{2 \pi n_c}} & 0 &
\frac{\cos\left( k_{-n_c/2}x_2\right)}{\sqrt{2 \pi n_c}} & 0 &
\frac{\cos\left( k_{-n_c/2}x_3\right)}{\sqrt{2 \pi n_c}} & 0 \\
\frac{-\sin( k_{-n_c/2}x_1)}{\sqrt{2\pi n_c}} & 0 &
\frac{-\sin( k_{-n_c/2}x_2)}{\sqrt{2\pi n_c}} & 0 &
\frac{-\sin( k_{-n_c/2}x_3)}{\sqrt{2\pi n_c}} & 0 \\
\frac{\cos (k_{1-n_c/2} x_1)}{\sqrt{2 \pi(N-2)}} & 0 &
\frac{\cos (k_{1-n_c/2} x_2)}{\sqrt{2 \pi(N-2)}} & 0 &
\frac{\cos (k_{1-n_c/2} x_3)}{\sqrt{2 \pi(N-2)}} & 0 \\
\frac{-\sin (k_{1-n_c/2}x_1)}{\sqrt{2 \pi (n_c-2)}} & 0 &
\frac{-\sin (k_{1-n_c/2}x_2)}{\sqrt{2 \pi (n_c-2)}} & 0 &
\frac{-\sin (k_{1-n_c/2}x_3)}{\sqrt{2 \pi (n_c-2)}} & 0 \\
\vdots & \vdots & \vdots & \vdots & \vdots & \vdots \\
\frac{\cos (k_{n_c/2} x_1)}{\sqrt{2\pi n_c}} & 0 &
\frac{\cos (k_{n_c/2} x_2)}{\sqrt{2\pi n_c}} & 0 &
\frac{\cos (k_{n_c/2} x_3)}{\sqrt{2\pi n_c}} & 0 \\
\frac{-\sin (k_{n_c/2} x_1)}{\sqrt{2 \pi n_c}}  & 0 &
\frac{-\sin (k_{n_c/2} x_2)}{\sqrt{2 \pi n_c}} & 0 &
\frac{-\sin (k_{n_c/2} x_3)}{\sqrt{2 \pi n_c}} & 0 
\end{pmatrix}.
\end{eqnarray}

The matrices $\mathbf{F}$ and $\mathbf{F}^\x{sym}$ may be also constructed for the case of the Dirichlet boundary conditions. We start with defining the vector of quadrature operators:
\begin{equation} 
\hat{\mathbf{x}}=(\hat{q}_{d1},\hat{p}_{d1},...\,,
\hat{q}_{d3},\hat{p}_{d3},\hat{q}_{1},\hat{p}_{1},...\,,\hat{q}_{n_c},\hat{p}_{n_c})^T,
\end{equation} 
where the vector has $2n_c+6$ elements and the first $6$ elements correspond to the detectors, whereas the remaining $2n_c$ elements correspond to the field modes.

Hence the phase space matrix corresponding to the free Hamiltonian is of the form:
\begin{equation}  \label{tri_Ffree_dir}
\mathbf{F}_\text{free}^\text{sym}=\textrm{diag}(\Omega_d,\Omega_d,\Omega_d,\Omega_d,\Omega_d,\Omega_d,\omega_1,\omega_1,
...\,,\omega_{n_c},\omega_{n_c}).
\end{equation}
The phase space matrix corresponding to the interaction Hamiltonian still has the block form given by \eqref{tri_Fint_per}, where the matrix $\mathbf G$ has the following form:
\begin{eqnarray}   
	\mathbf{G} \defn
\begin{pmatrix} 
\frac{\textrm{sin} k_1x_1}{\sqrt{\pi}} & 0 &
\frac{\textrm{sin} k_1x_2}{\sqrt{\pi}} & 0 &
\frac{\textrm{sin} k_1x_3}{\sqrt{\pi}} & 0 \\
0 & 0 & 0 & 0 & 0 & 0 \\
\frac{\textrm{sin} k_2x_1}{\sqrt{2\pi}} & 0 &
\frac{\textrm{sin} k_2x_2}{\sqrt{2\pi}} & 0 &
\frac{\textrm{sin} k_2x_3}{\sqrt{2\pi}} & 0 \\
0 & 0 & 0 & 0 & 0 & 0 \\
\vdots & \vdots & \vdots & \vdots & \vdots & \vdots \\
\frac{\textrm{sin} k_{n_c} x_1}{\sqrt{n_c\pi}} & 0 &
\frac{\textrm{sin} k_{n_c} x_2}{\sqrt{n_c\pi}} & 0 &
\frac{\textrm{sin} k_{n_c} x_3}{\sqrt{n_c\pi}} & 0 \\
0 & 0 & 0 & 0 & 0 & 0 
\end{pmatrix}.
\end{eqnarray}

Regardless of the type of boundary conditions that the field $\hat{\Phi}$ is subject to, the time evolution of the covariance matrix is governed by the equation \eqref{gqm_sympl_ev}, where the symplectic matrix obeys the equation~\cite{eric_dets_for_probing}:
\begin{align}  
	\partial_t \,S(t)= \Omega_d \, \boldsymbol F^\text{sym}(t)\, S(t).
\end{align}
In our work $\boldsymbol F^\text{sym}$ is time-independent, hence the solution is as follows:
\begin{equation}        \label{tri_S_soln}
S(t)=\mathrm{exp}(\Omega_d\,\mathbf{ F}^\text{sym}t).
\end{equation}

The initial state of the system is that all the detectors and the field is in the vacuum state. Therefore the initial covariance matrix is $\boldsymbol\sigma(0)=\I$. This together with \eqref{tri_S_soln} can be substituted into \eqref{gqm_sympl_ev} and hence the final covariance matrix of the system may be obtained. Further, we isolate the covariance matrix of the detectors only, by simply truncating the resultant matrix to its upper-left $6\times 6$ block. All of these operations are performed numerically to yield a matrix of the form:
\begin{equation}   \label{tri_final_state}
\boldsymbol\sigma_{123}=\left( \begin{array}{ccc}
              \boldsymbol\sigma_1 & \boldsymbol\gamma_{12} & \boldsymbol\gamma_{13} \\
              \boldsymbol\gamma_{12} & \boldsymbol\sigma_2 & \boldsymbol\gamma_{23} \\
              \boldsymbol\gamma_{13} & \boldsymbol\gamma_{23} & \boldsymbol\sigma_3 \end{array} \right),
\end{equation}
where the interpretation of each block matrix is discussed in Sec. \ref{sec_gqm_comp_sys} and the subscripts denote the number of the corresponding detector. It should be stressed that the obtained solution is exact in the sense that it is calculated without resorting to perturbation theory.

Let us now discuss the alignement of the detectors, which is most suitable for a proper comparison of the results for different types of boundary conditions. In both cases the positions of the detectors are chosen to be $(x_1,x_2,x_3)=(\frac{1}{6}L,\frac{1}{2}L,\frac{5}{6}L)$. For periodic boundary conditions this alignment is symmetric, meaning that each detector is placed at the distance~$\frac{L}{3}$ away from either of the remaining two, and hence the state of the system is invariant under the exchange of any two detectors. This property allows us to easily determine whether or not the trio is tripartitely entangled. For Dirichlet boundary conditions this alignment leaves the middle detector at the distance \nolinebreak $\frac{L}{3}$ away from the detectors on the sides. In this case also the exchange of the detectors on the sides leaves the state of the system unchanged. However, the asymmetry under the exchange of the middle detector with one on a side is unavoidable. For this reason we are unable to compute the extracted tripartite entanglement in the case of a Dirichlet cavity. We rather only compare the extracted bipartite entanglement with that of the periodic cavity.

\section{Tripartite entanglement}        \label{sec_tri_tri}


We intend to study the bipartite and tripartite entanglement in the state \eqref{tri_final_state}. Sec. \ref{sec_gqm_ent} shows how to quantify bipartite entanglement using logarithmic negativity. This method can be explicitly applied to calculate the amount of entanglement (which from now on may be briefly referred to as ``entanglement'') between given two detectors, e.g. ${\cal E_N}(1|2)$. For the purposes of estimation of the tripartite entanglement, we also need to compute the bipartite entanglement across the bipartitions of the trio of detectors, e.g.~${\cal E_N}(3|12)$. This poses a problem, because bipartitions of the system have a different number of degrees of freedom. The solution to this problem is the so-called {\it unitary localization}~\cite{adesso1,adesso2}. It involves applying an appropriate local unitary operation, which does not change the amount of entanglement across a given bipartition, but swaps the entire entanglement onto two detectors. Then it suffices to compute the entanglement between these two detectors.

In the case of periodic boundary conditions, the system is symmetric under the exchange of any two detectors. Hence the covariance matrix of the system \eqref{tri_final_state} can be written as:
\begin{align}
	\boldsymbol \sigma_\text{periodic}=\begin{pmatrix}
		\boldsymbol \sigma_s & \boldsymbol \gamma & \boldsymbol \gamma \\
		\boldsymbol \gamma & \boldsymbol \sigma_s & \boldsymbol \gamma \\
		\boldsymbol \gamma & \boldsymbol \gamma &\boldsymbol \sigma_s
	\end{pmatrix}.
\end{align}
In order to compute the entanglement across the bipartition $3|12$, we consider applying a beam-splitter operation to the $12$ mode subsystem, which globally is given by
\begin{equation}
\mathbf{S}_\text{BS}=\begin{pmatrix}
              \I/\sqrt{2} & - \I/\sqrt{2} & \boldsymbol 0  \\
              \I/\sqrt{2} &  \I/\sqrt{2} & \boldsymbol 0 \\
	\boldsymbol 0 & \boldsymbol 0 & \I
	\end{pmatrix}.
\end{equation}
Being a unitary operation within the subsystem $1$ and $2$, this does not affect the entanglement across the $3|12$ bipartition. Furthermore, with mode symmetry this operation is seen to isolate all correlations solely between modes $2$ and $3$, as shown below:
\begin{align}
	\mat{S}_\text{BS} \,\mat \sigma_\text{periodic} \,\mat S_\text{BS}^T=\begin{pmatrix}
		\mat \sigma_s-\mat \gamma & \mat 0 & \mat 0 \\
		\mat 0 & \mat \sigma_s+\mat \gamma & \sqrt{2} \mat \,\gamma \\
		\mat 0 & \sqrt{2} \,\mat\gamma & \mat \sigma_s
	\end{pmatrix}.
\end{align}
From here, we can apply the formula in \eqref{gqm_neg} to compute the entanglement between modes $2$ and $3$ of this transformed state. This is equivalent to the bipartite entanglement across the $3|12$ partition.

In the case of the Dirichlet boundary conditions it is very complex to find a local unitary operation that would swap all the entanglement onto two detectors. We have successfully found it only for the bipartition $2|13$, which is easier due to the symmetry under the exchange of detectors $1$ and $3$. This has been accomplished numerically, by solving a cumbersome system of equations following from demanding appropriate elements of the transformed covariance matrix to vanish.

Recalling the discussion from Sec. \ref{sec_qm_multi}, by definition, a tripartite system contains genuine tripartite entanglement if the state is inseparable across all three of the possible bipartitions \cite{horodecki_entanglement, adesso_ent}. What we focus on, in regard to the tripartite entanglement, is its existence or absence rather than its amount. Thus, we estimate the amount of tripartite entanglement using the geometrical average of the logarithmic negativities across all the bipartitions:
\begin{equation}      \label{tri_tri_est}
\tilde{\mathcal{E}}_\mathcal{N}(123)=
\sqrt[3]{\mathcal{E_N}(1|23)\,\, \mathcal{E_N}(2|13)\,\, \mathcal{E_N}(3|12)}.
\end{equation}
This quantity does not constitute a proper entanglement measure, but certainly it provides an answer to whether or not there is tripartite entanglement in the system. In the case of the periodic cavity, the fact that the three detector state is symmetric under detector permutation means that to determine the presence of tripartite entanglement we need only consider a single bipartition, since the other two are equivalent. In the case of the Dirichlet boundary conditions we forgo this calculation because of the inability of calculating the bipartite entanglement across certain bipartitions.

\section{Results}           \label{sec_tri_results}

In this section we present the results that are obtained for the alignment of the detectors described in Sec.~\ref{sec_tri_calc}. The amounts of entanglement between detectors are computed for both types of boundary conditions. We show two results. First, that for a periodic cavity one can harvest tripartite entanglement without causal contact between detectors, and that this is in fact easier to do so than for the bipartite entanglement. Second, we compare the harvesting of bipartite entanglement between the cases of a periodic and a Dirichlet cavity, finding that it is considerably easier to harvest entanglement in the periodic cavity.

The convergence of the obtained results with increasing $n_c$ was tested, and it was found that $n_c=50$ was a sufficient number of modes in all cases.
All the results are evaluated at a fixed value of the coupling constant $\lambda=0.01$.
The time of interaction $T$ is presented here in the units of $r$, where $r=\frac{L}{3}$ is the light-crossing time between neighbouring detectors. Hence, for times $T>r$ neighbouring detectors may communicate with each other, and for $T<r$ they may not and all the entanglement generated in the system must be due to the extraction from the field.

\subsection{Periodic boundary conditions}

\begin{figure*}[]
    \centering
    \includegraphics[width=1.05\textwidth]{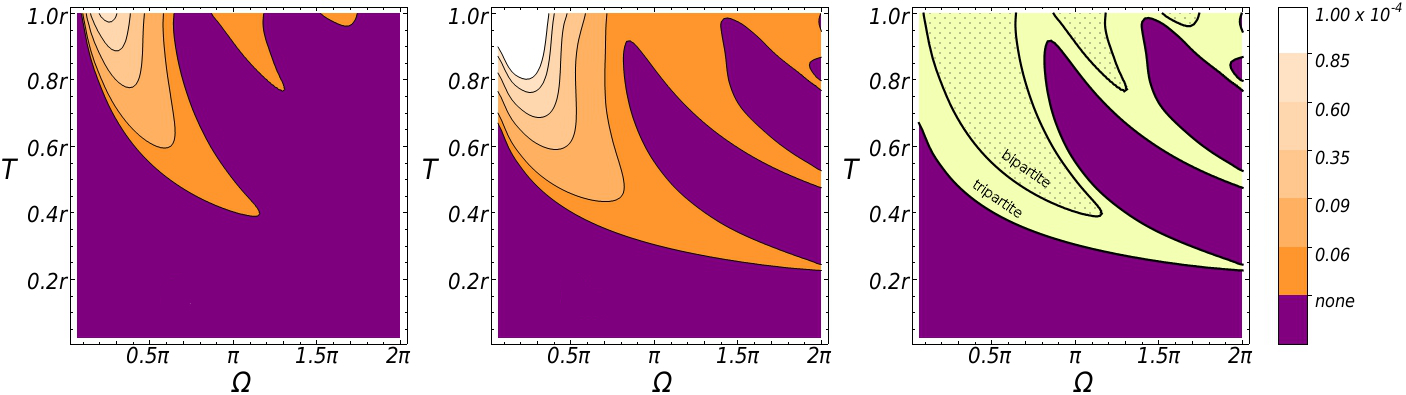}
    \put(-337,-2){{\tiny\it d}}
    \put(-214,-2){{\tiny\it d}}
    \put(-91,-2){{\tiny\it d}}
    \put(-380,26){\twh{$\boldsymbol{\mathcal{E_N}(s|s)}$}}
    \put(-257,26){\twh{$\boldsymbol{\mathcal{\tilde{E}_N}(sss)}$}}
    \caption{\small{Entanglements $\mathcal{E_N}(s|s)$, $\mathcal{\tilde{E}_N}(sss)$ in the system subject to 
             periodic boundary conditions, as a function of $T$ and
             $\Omega_d$ at $L=10$ and $\lambda=0.01$; On the right: regions of existence
             of $\mathcal{E_N}(s|s)$ and $\mathcal{\tilde{E}_N}(sss)$ plotted together.}}
    \label{fig_tri_periodic}
\end{figure*}

\begin{figure*}[]
    \centering
    \includegraphics[width=1.05\textwidth]{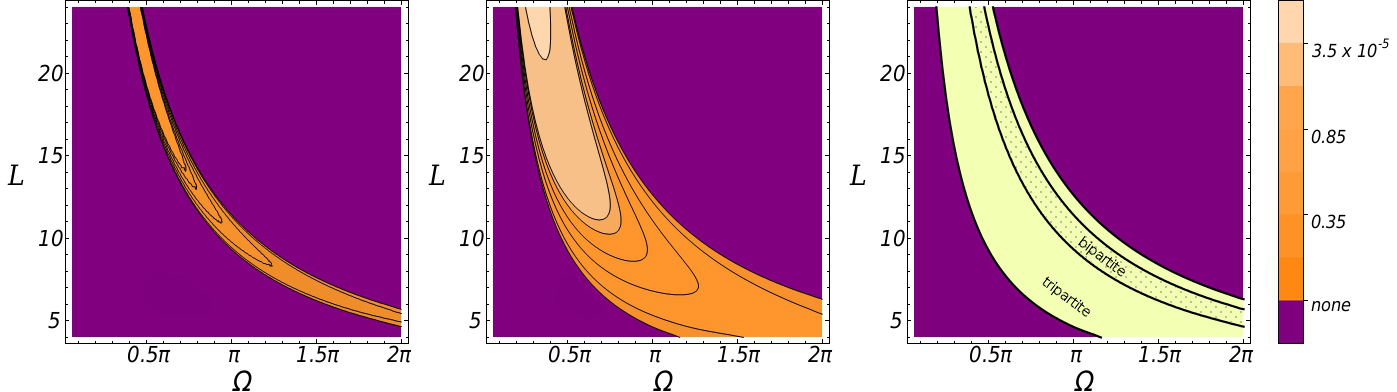}
    \put(-336,-2){{\tiny\it d}}
    \put(-212,-2){{\tiny\it d}}
    \put(-88.5,-2){{\tiny\it d}}
    \put(-341,96){\twh{$\boldsymbol{\mathcal{E_N}(s|s)}$}}
    \put(-220,96){\twh{$\boldsymbol{\mathcal{\tilde{E}_N}(sss)}$}}
    \caption{\small{Entanglements $\mathcal{E_N}(s|s)$, $\mathcal{\tilde{E}_N}(sss)$ in the system subject to 
             periodic boundary conditions, as a function of $L$ and
             $\Omega_d$ at $T=0.4r$ and $\lambda=0.01$; On the right: regions of existence
            of $\mathcal{E_N}(s|s)$ and $\mathcal{\tilde{E}_N}(sss)$ plotted together.}}
    \label{fig_tri_lo}
\end{figure*}
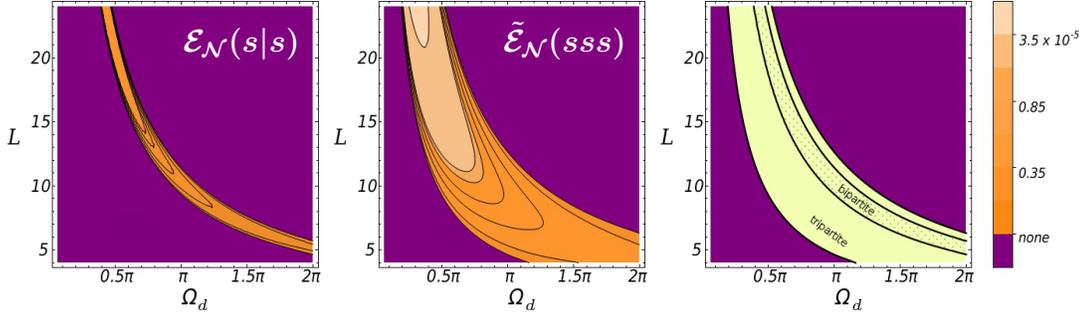

In this subsection due to symmetry we denote all the detectors by $s$, hence $\mathcal{E_N}(s|s)$
is the bipartite entanglement between a pair of detectors, which is identical for each pair. $\mathcal{\tilde{E}_N}(sss)$ is the estimate for tripartite entanglement~\eqref{tri_tri_est}, which by symmetry is equal to $\mathcal{E_N}(s|ss)$. For simplicity the former is referred
to as ``bipartite entanglement'' and the latter is referred to as ``tripartite entanglement''.

For the case of periodic boundary conditions the plots of the amount of entanglement as a function of $T$ and $\Omega_d$ (the frequency of the detectors), at a fixed value of $L=10$, are shown in Fig. \ref{fig_tri_periodic}, where we plot both $\mathcal{E_N}(s|s)$ and $\mathcal{\tilde{E}_N}(sss)$. In agreement with intuition, most entanglement is produced after the light-crossing time of the neighbouring detectors, which equals $r$. There are however regions in the $T\Omega_d$ plane, along which both types of entanglement exist far in the $T<r$ regime. They do not diminish under increasing the number of field modes $n_c$, hence this is not an artifact of the imposed UV cutoff but rather a true physical effect. Moreover, the regions where tripartite entanglement is 
produced are certainly broader than for those of the bipartite entanglement, meaning that the
tripartite entanglement emerges earlier and is therefore easier to be harvested. This however is not surprising if we recall that $\mathcal{\tilde{E}_N}(sss)=\mathcal{E_N}(s|ss)$ 
and we obtain $\mathcal{E_N}(s|s)$ from $\mathcal{E_N}(s|ss)$
by tracing out one of the detectors. This is a local operation and so can only decrease the amount of entanglement, which implies that $\mathcal{E_N}(s|s)$ always has to be less or equal $\mathcal{\tilde{E}_N}(sss)$, as can be seen from our results.

Fig. \ref{fig_tri_lo} shows a plot of entanglement versus
$L$ and $\Omega_d$ produced at fixed $T=0.4r$. In the $L\Omega_d$ plane we find a hyperbolic-like region in which both bipartite and tripartite entanglement get extracted by this time. Again, the region of non-zero tripartite entanglement is larger than that of bipartite entanglement. We also note that, within this scenario, the longer the cavity is and the smaller the detector frequency is the more entanglement can be acausally extracted.

\subsection{Dirichlet boundary conditions}           

\begin{figure*}[]
    \centering
    \includegraphics[width=\textwidth]{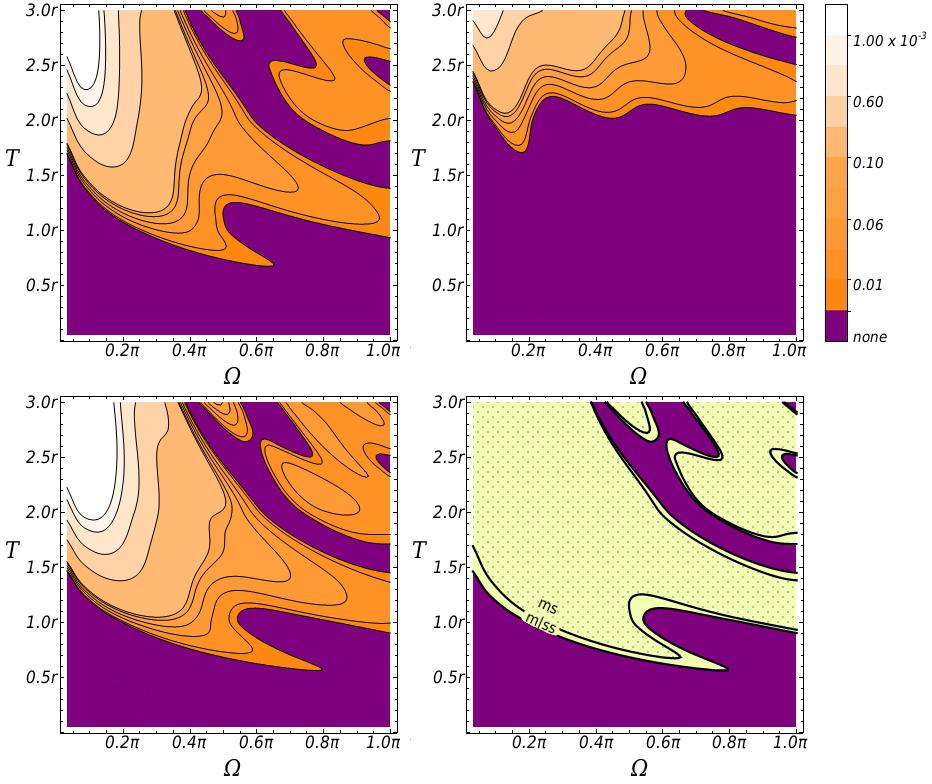}
    \put(-289.5,0.5){{\tiny\it d}}
    \put(-119,0.5){{\tiny\it d}}
    \put(-289.5,165){{\tiny\it d}}
    \put(-119.2,165){{\tiny\it d}}
    \put(-350,198){\twh{$\boldsymbol{\mathcal{E_N}(m|s)}$}}
    \put(-180,198){\twh{$\boldsymbol{\mathcal{E_N}(s|s)}$}}
    \put(-351,33){\twh{$\boldsymbol{\mathcal{E_N}(m|ss)}$}}
    \caption{\small{Entanglements $\mathcal{E_N}(m|s)$, $\mathcal{E_N}(s|s)$ and $\mathcal{E_N}(m|ss)$ in the system 
             subject to 
             Dirichlet boundary conditions, as a function of $T$ and
             $\Omega_d$ at $L=10$ and $\lambda=0.01$; Bottom right: regions of existence
             of $\mathcal{E_N}(m|s)$ and $\mathcal{E_N}(m|ss)$ plotted together.}}
    \label{fig_tri_dirichlet}
\end{figure*}

\begin{figure}
    \centering
    \includegraphics[width=0.5\textwidth]{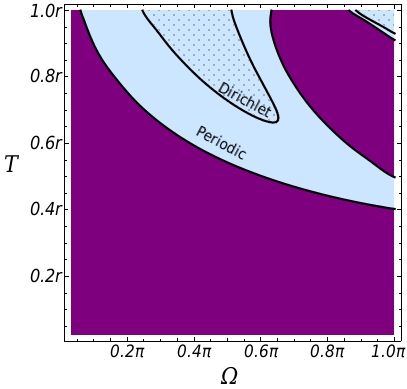}
    \put(-81,0){{\scriptsize\it d}}
    \caption{\small{A comparison, between the cases of periodic and Dirichlet boundary conditions, of the regions in which bipartite entanglement is harvested between neighboring detectors. Here $L=10$ and $\lambda=0.01$.}}
    \label{fig_tri_both}
\end{figure}

Throughout this subsection we denote the middle detector as $m$ and the detectors on the sides as $s$.

The behaviour of the entanglement under Dirichlet boundary conditions is found to be different to some extent from the periodic cavity case. The quantities which are involved, include: the bipartite entanglement $\mathcal{E_N}(m|s)$, the bipartite entanglement $\mathcal{E_N}(s|s)$, and the bipartite entanglement $\mathcal{E_N}(m|ss)$ (which can be computed using the method of Sec. \ref{sec_tri_tri}). We plot them in Fig. \ref{fig_tri_dirichlet} as a function of $T$ and $\Omega_d$, at fixed value of $L=10$. Again we observe that most entanglement is produced after the corresponding light-crossing time, which for the case of neighbouring detectors is $r$, whereas for the detectors on the sides it is $2r$. Under these boundary conditions there exist regimes in the parameter space for which the entanglement $\mathcal{E_N}(m|s)$, which is more broadly available, persists in the spacelike regime down to $T\sim 0.6r$. This is a considerably longer time than in the case of periodic boundary conditions. This can be seen directly in Fig. \ref{fig_tri_both}. We clearly see that it is easier to extract bipartite entanglement in the periodic cavity than in one with Dirichlet boundary conditions.

The tripartite entanglement $\mathcal{\tilde{E}_N}(mss)$ has not been calculated for this case, for reasons described in Sec. \ref{sec_tri_tri}, however the bipartite entanglement $\mathcal{E_N}(m|ss)$ has been evaluated, which allows us to draw certain conclusions. The entanglements $\mathcal{E_N}(m|s)$ and $\mathcal{E_N}(m|ss)$ are plotted together in Fig. \ref{fig_tri_dirichlet} for comparison. The region of existence of $\mathcal{E_N}(m|s)$ is enclosed in the region of existence of $\mathcal{E_N}(m|ss)$, which is in agreement with the argument posed in the previous subsection. The region of non-zero $\mathcal{E_N}(m|ss)$ is slightly broader than that of non-zero $\mathcal{E_N}(m|s)$ similarly to the case of a periodic cavity.

The existence of $\mathcal{E_N}(m|ss)$ is a necessary (but not sufficient) condition for the existence of tripartite entanglement, hence no tripartite entanglement can exist in the system if $\mathcal{E_N}(m|ss)=0$. We therefore conclude that, similarly to bipartite entanglement, the extraction of tripartite entanglement is considerably easier to achieve using a periodic cavity rather than a Dirichlet one.

Therefore we see from the plot in Fig. \ref{fig_tri_dirichlet} that tripartite entanglement may not emerge earlier than $T\sim 0.55r$. This is considerably later than in the case of the periodic boundary conditions.

\section{Conclusions}         \label{sec_tri_concl}

In summary, we have used non-perturbative methods of Gaussian quantum mechanics to examine the harvesting of both bipartite and tripartite entanglement from the vacuum state of a field in a cavity. That is, a set of detectors interacting with a common quantum field can become entangled without causal contact by means of swapping the spatial entanglement present in the field.

There are two primary conclusions that we have made. First, from the vacuum state of a periodically-identified cavity field it appears that genuine tripartite entanglement can be harvested. In fact, it is considerably easier to obtain tripartite entanglement than bipartite entanglement between any two of the three detectors. Indeed, we have been able to obtain tripartite entanglement after a time of interaction considerably smaller than the light-crossing time between pairs of detector. Specifically, we have seen that a time as small as $t=0.21 r$, where $r$ is the distance between detectors, can be sufficient. We have provided detailed maps of the regions in parameter space in which bipartite and tripartite entanglement can be harvested.

Second, we have demonstrated that there is a significant difference between a periodic cavity field and one subject to Dirichlet boundary conditions for the harvesting of bipartite entanglement. Our finding is that such harvesting is considerably easier to achieve in a periodic cavity than in one subject to Dirichlet boundary conditions. Although we have not fully considered tripartite entanglement in the case of a Dirichlet cavity, the same conclusion can be drawn, that harvesting of tripartite entanglement is easier in a periodic cavity.

Both of these results may be experimentally tested as well as have applicability to a possible utilization in entanglement harvesting scenarios. They furthermore may be relevant to more general system-bath setups than those considered in this chapter, helping to point the way towards optimal strategies for generating quantum correlations utilizing such systems.




\chapter*{Conclusions}         \label{chconc}
\addcontentsline{toc}{part}{Conclusions}

This thesis covered a number of topics from the intersection of relativity and quantum information. In particular, a focus was put on uniformly accelerated systems and the impact of this acceleration on quantum states, and on the quantum entanglement that can be detected and exctracted from the vacuum.

In Part \ref{part2} we investigated the behaviour of a uniformly accelerating quantum clock. The mechanism of the clock was based on a rate of the decay of an unstable particle. This rate was calculated for an accelerating clock and compared with the one obtained for a stationary clock. We discovered that the discrepancy in the rates of ticking of the clocks, differs from the one predicted by special relativity. This is a consequence of the Unruh effect. The existence of the Unruh thermal bath of particles inevitably alters the ticking rate of the accelerating clock. The existence of the effect was demonstrated for a toy model of a clock, and further work has been recently developed~\cite{clk_roberto}, which involves analyzing a realistic decay of a muon in a particle accelerator. Its results support the existence of the effect. Our work shows that the concept of an ideal clock, widely used in relativity, is fundamentally not achievable physically. Furthermore, an observation of this effect in an experiment, would be of great importance. Not only would it be a confirmation of our claims, but also a confirmation of the existence of the Unruh effect itself, which is greatly sought-after nowadays.

In Part \ref{part3} we moved on to considering how a general two-mode Gaussian state of two inertial observers is perceived by a pair of uniformly accelerated observers. The framework developed in this work involved the construction of a quantum channel transforming the state of two localized stationary modes into the state of two localized accelerated modes. The task was performed for a $1+1$-dimensional and $3+1$-dimensional Klein-Gordon field, as well as a $1+1$-dimensional Dirac field. A major strength of this framework was that via usage of the modified Rindler coordinates, it was possible to treat independently the accelerations of the observers and their relative distance at the time of observation of the inertial modes. While this framework may be applied to any Gaussian state, a focus was put on the Minkowski vacuum state, because this allows for the properties of the vacuum entanglement to be studied. The analysis of each type of the field yields similar dependences of the observed bipartite entanglement on the parameters of the system. In general, when the Rindler wedges of the two observers share a common apex, there exisits more entanglement when the accelerations of the observers are greater, and the sizes and central frequencies of the corresponding modes are smaller. Also, some interesting properties of the entanglement of the vacuum were uncovered in the case of the $1+1$-dimensional Klein-Gordon field, where we moved beyond the standard Rindler chart and observed examples of sudden death of the entanglement. This framework can be extended in various ways, including an arbitrary number of observers~\cite{channel_kacper} or a $3+1$-dimensional Dirac field~\cite{grzesiek} and applied to the study of many effects such as e.g. certain relativistic quantum information protocols~\cite{piotrek}.

\thispagestyle{concl}

In Part \ref{part4} we continued the study of the entanglement of the vacuum, and also extended our interest to the tripartite entanglement. A cavity with three particle detectors was analyzed using the Gaussian quantum mechanics formalism, which greatly simplifies the calculations and removes the necessity of resorting to perturbation theory. The detectors were allowed to interact with a quantum field enclosed in a cavity for a certain amount of time. This time was chosen to be short enough such that they could not get entangled by quantum fluctuaions travelling between them, but only by swapping the entanglement over from the field. Maps of the regions in parameter space, wherein various entanglement types are seen, were presented for two types of boundary conditions at the walls of the cavity. It was found that it is easier to extract entanglement in a cavity subject to periodic boundary conditions than in one subject to Dirichlet boundary conditions. Furthermore, the tripartite entanglement was discovered to be more easily accessible than the bipartite entanglement. This work provides new insights into the structure of the vacuum entanglement, as well as suggests the most suitable conditions for an experimental observation of these phenomena and an efficient extraction of the entanglement for further practical purposes.

The questions posed and the conclusions drawn from the work presented in this thesis are of fundamental importance in relativistic quantum information and related disciplines, which combine relativity and quantum physics. Also, the effects discussed, have direct experimental implications and may be tested in the future, given the rate of development of the current technology. Thus, the importance of this research ranges from providing helpful insights into the properties of quantum states in the presence of acceleration, to discovering effects that may become significant for the technology of tomorrow, which may involve e.g. quantum communication with satellites~\cite{ivette_satellite, satelity}, quantum accelerometers~\cite{andrzej_mody_besseli} and vacuum entanglement extraction machines. The Author of this thesis is proud to have been able to perform research presented here, which does constitute a tiny brick in the ever-growing building of human knowledge. 

\thispagestyle{concl}




\appendix
\part*{\thispagestyle{empty}}

\phantomsection
\addcontentsline{toc}{part}{Appendices}

\chapter{Squared sinc function representation of the Dirac delta}   \label{app_sec_delta}

In this appendix we prove the formula:
\begin{equation}  \label{app_delta}
\lim_{t\rightarrow\infty} \frac{1}{\pi t} \frac{\sin^2\left[\left(x-x_0\right)t\right]}{\left(x-x_0\right)^2}
=
\delta(x-x_0),
\end{equation}
where $x$ and $x_0$ are real.
First let us observe that:
\begin{equation}
\lim_{\epsilon\rightarrow 0}\frac{\epsilon}{\pi} \frac{\sin^2\left(\frac{x}{\epsilon}\right)}{x^2}
\end{equation}
does not exist.
Now consider:
\begin{equation}
\lim_{\epsilon\rightarrow 0}\frac{\epsilon}{\pi} 
{\int\limits_{x_-<0}^{x_+>0}}\text{d}x \,f(x)\,
\frac{\sin^2\left(\frac{x}{\epsilon}\right)}{x^2}.
\end{equation}
Changing variables such that $y=\frac{x}{\epsilon}$ yields:
\begin{equation}
\lim_{\epsilon\rightarrow 0}\frac{1}{\pi} 
{\int\limits_{\frac{x_-}{\epsilon}}^{\frac{x_+}{\epsilon}}}\text{d}y \,f(\epsilon y)\,
\frac{\sin^2\left(y\right)}{y^2}
=
\frac{f(0)}{\pi} 
{\int\limits_{-\infty}^{+\infty}}\text{d}y\,
\frac{\sin^2\left(y\right)}{y^2}
=
f(0).
\end{equation}
This suffices to show that:
\begin{equation}
\lim_{\epsilon\rightarrow 0}\frac{\epsilon}{\pi} \frac{\sin^2\left(\frac{x}{\epsilon}\right)}{x^2}
=
\delta(x).
\end{equation}
Substituting in $t=\frac{1}{\epsilon}$ and shifting $x\rightarrow x-x_0$ proves the formula \eqref{app_delta}.

\chapter{Asymptotic properties of the modified Bessel function} \label{App_N12limits}
In this section, we study the limiting behavior of the modified Bessel function that is useful in finding the asymptotic behavior of the off-diagonal blocks of the noise matrix $N$ as $D\to0$. 

Let us begin by invoking the asymptotic expression for the modified Bessel function of small positive arguments $\epsilon$~\cite{crispino_higuchi}:
\begin{align}\label{MBK}
K_{i\nu}(2\epsilon)&\approx \frac{i\pi}{2\sinh(\pi\nu)}\left[\frac{\epsilon^{i\nu}}{\Gamma(1+i\nu)}-\frac{\epsilon^{-i\nu}}{\Gamma(1-i\nu)} \right],
\end{align}
where $\Gamma(z)$ is the Euler's Gamma function, defined as $\Gamma(z)=\int_{0}^{\infty}\mathrm{d}t\,e^{-t} t^{z-1}$, for complex numbers $z$ such that $\Re\,z>0$. When $\Re\,z\leq 0$, it is defined by analytic continuation and it has simple poles at nonpositive integer arguments. 

Using the identity $\Gamma(z)\Gamma(1-z)=\frac{\pi}{\sin(\pi z)}$, we can rewrite the asymptotic form  of $K_{i\nu}(2\epsilon)$, given in Eq.~\eqref{MBK}, as:
\begin{align}
K_{i\nu}(2\epsilon)&\approx \Re \left[\epsilon^{i\nu}\Gamma (-i\nu)\right].\label{Kapprox}
\end{align}
Note that $\Im\,\Gamma(ix)\propto-\frac{1}{x}$, which is divergent at $x=0$, can only be treated as a distribution, but $\Re\,\Gamma(ix)$ is regular at $x=0$. Let us substitute $\epsilon = e^{-\lambda}$ in Eq.~(\ref{Kapprox}) and look for the limit $\lambda\to\infty$:
\begin{align}\label{BKPV}
K_{i\nu}(2\epsilon)&\approx \cos(\lambda\nu)\,\Re\,\Gamma(-i\nu)+\sin(\lambda\nu)\,\Im\,\Gamma(-i\nu).
\end{align}
We can now take the limit of $\lambda\to\infty$. In doing so we use the \textsl{Riemann-Lebesgue lemma}\footnote{Let $f$ be a Riemann-integrable function defined on an interval $a\leq x\leq b$ of the real line, the Riemann-Lebesgue lemma states that $\lim_{|\alpha|\rightarrow\infty}\int_{a}^{b}f(x)\sin(\alpha x)\mathrm{d}x=\lim_{|\alpha|\rightarrow\infty}\int_{a}^{b}f(x)\cos(\alpha x)\mathrm{d}x=0$.}. According to it the limit of the first term in Eq.~\eqref{BKPV} vanishes. The limit of the second term can only be given in a distributional sense and it is non-zero only at the pole of the Gamma function at $\nu=0$:
\begin{align}
\lim_{\epsilon\to 0^+} K_{i\nu}(2\epsilon) &= \lim_{\lambda\to\infty }\Im\,\Gamma(-i\nu)\sin(\lambda\nu) = \lim_{\lambda\to\infty }\frac{\sin(\lambda\nu)}{\nu} \nonumber \\
&= \pi \delta(\nu),
\end{align}
where we have used one of the representations of the Dirac delta function.

\bibliographystyle{aipnum4-1}
\bibliography{bibfile}

\addcontentsline{toc}{part}{Bibliography}


\end{document}